\documentclass[acmtog, authorversion]{acmartarxiv}
\pdfoutput=1

\usepackage{booktabs} %

\usepackage{enumitem}
\setitemize{noitemsep,topsep=0pt,parsep=0pt,partopsep=0pt,leftmargin=*}

\usepackage[ruled]{algorithm2e} %

\SetAlFnt{\small}
\SetAlCapFnt{\small}
\SetAlCapNameFnt{\small}
\SetAlCapHSkip{0pt}

\usepackage{cleveref}
\usepackage{siunitx}
\usepackage{tabularx}
\usepackage[export]{adjustbox}
\usepackage{braket}
\usepackage{url}
\usepackage{subfigure}

\setcopyright{rightsretained}

\definecolor{lightgray}{rgb}{0.8,0.8,0.8}
\definecolor{darkblue}{rgb}{0,0,0.7}
\definecolor{darkgreen}{rgb}{0,0.5,0}
\definecolor{darkpurple}{rgb}{0.5,0.0,0.4}
\definecolor{midred}{rgb}{0.8,0,0}
\definecolor{midgreen}{rgb}{0,0.8,0}
\definecolor{midorange}{rgb}{0.8,0.6,0}
\definecolor{shadecolor}{RGB}{255,255,255}

\newcommand{\urlcolor}[1]{\textcolor{darkblue}{#1}}

\newcommand{\changed}[1]{\textcolor{darkgreen}{#1}}
\renewcommand{\changed}[1]{#1}

\crefname{figure}{Figure}{Figures}

\newcommand{\Fig}[1]{Figure~\ref{#1}}

\newcommand{\Eq}[1]{Equation~\ref{#1}}

\newcommand{\Sec}[1]{Section~\ref{#1}}
\newcommand{\Tab}[1]{Table~\ref{#1}}
\newcommand{\hide}[1]{}

\renewcommand{\vec}[1]{\mathbf{#1}}
\DeclareMathOperator*{\argmin}{arg\,min}

\newcommand{\eg}{e.g.\@ }
\newcommand{\ie}{i.e.\@ }

\begin{document}
\begin{anonsuppress}
\thanks{This work was supported by the German Research Foundation (DFG) under grant (HU-2273/2-1), the
X-Rite Chair for Digital Material Appearance, and the ERC starting grant ``ECHO''. We also gratefully acknowledge the support of NVIDIA Corporation with the donation a Titan Xp GPU that was used in this research.}
\end{anonsuppress}
\title{Computational Parquetry: Fabricated Style Transfer with Wood Pixels}

\author{Julian Iseringhausen}
\email{iseringhausen@cs.uni-bonn.de}
\orcid{0000-0002-6240-5105}
\author{Michael Weinmann}
\email{mw@cs.uni-bonn.de}
\orcid{0000-0003-3634-0093}
\author{Weizhen Huang}
\email{whuang@cs.uni-bonn.de}
\orcid{0000-0002-5068-8339}
\author{Matthias B. Hullin}
\email{hullin@cs.uni-bonn.de}
\affiliation{
  \institution{\\University of Bonn}
  \department{Institute of Computer Science II}
  \city{Bonn}
  \postcode{53115}
  \country{Germany}
}

\renewcommand\shortauthors{Iseringhausen, J. et al}

\begin{abstract}
Parquetry is the art and craft of decorating a surface with a pattern of differently colored veneers of wood, stone or other materials. Traditionally, the process of designing and making parquetry has been driven by color, using the texture found in real wood only for stylization or as a decorative effect. Here, we introduce a computational pipeline that draws from the rich natural structure of strongly textured real-world veneers as a source of detail in order to approximate a target image as faithfully as possible using a manageable number of parts.
This challenge is closely related to the established problems of patch-based image synthesis and stylization in some ways, but fundamentally different in others. 
Most importantly, the limited availability of resources (any piece of wood can only be used once) turns the relatively simple problem of finding the right piece for the target location into the combinatorial problem of finding optimal parts while avoiding resource collisions. We introduce an algorithm that allows to efficiently solve an approximation to the problem. It further addresses challenges like gamut mapping, feature characterization and the search for fabricable cuts. We demonstrate the effectiveness of the system by fabricating a selection of pieces of parquetry from different kinds of unstained wood veneer.
\end{abstract}

\ccsdesc[500]{Computing methodologies~Image-based rendering}
\ccsdesc[300]{Computing methodologies~Image processing}
\ccsdesc[500]{Applied computing~Fine arts}
\ccsdesc[300]{Applied computing~Computer-aided manufacturing}

\maketitle
\thispagestyle{empty}
\section{Motivation}\label{sec:motivation}
The use of differently colored and structured woods and other materials to form inlay and intarsia has been known at least since ancient Roman and Greek times. In the modern interpretation of this principle, pieces of veneer form a continuous thin layer that covers the surface of an object (\emph{marquetry} or \emph{parquetry}) \cite{jackson1996complete}. The techniques denoted by these two terms share many similarities but are not identical. Marquetry usually refers to a process similar to ``painting by numbers'', where a target image is segmented into mostly homogeneous pieces which are then cut from more or less uniformly colored veneer and assembled to form the final ornament or picture. Parquetry, on the other hand, denotes the (ornamental) covering of a surface using a regular geometric arrangement of differently colored pieces. While most artists in their work embrace the grain and texture found in their source materials, they mostly use it as a decorative effect. Nevertheless, the resulting artworks can attain high levels of detail, depending on the amount of labor and care devoted to the task (\Fig{fig:examples}).

\begin{figure}[t]%
\includegraphics[width=0.9\columnwidth]{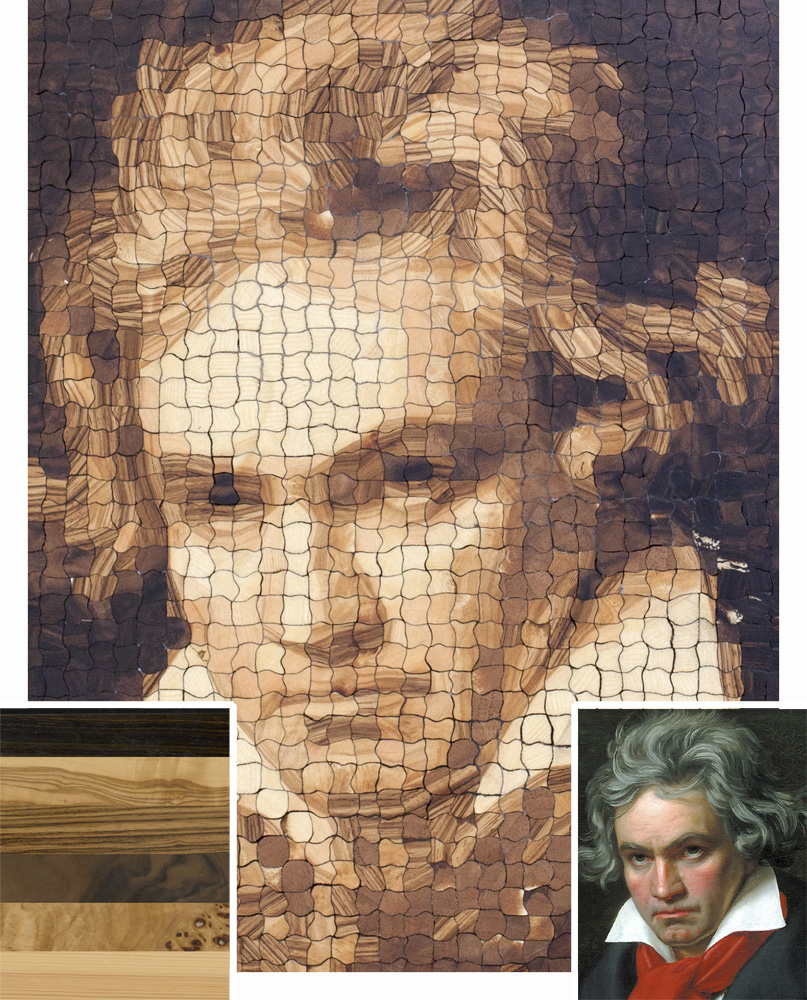}\hfill%
\\%
\caption{\changed{\textbf{Fabricated}} A fabricated piece of wood parquetry, produced using our pipeline. The inputs are a set of six different wood veneers (bottom left corner: poplar burl, walnut burl, santos rosewood, quartersawn zebrawood, olive, fir), and a target image (bottom right corner). The total size of the parquetry is approx.\@ \SI{27 x 34}{cm}. By combining the different appearance profiles (including color and grain structures) of multiple wood types, we are able to produce results with high contrast and fine structural details.}
\label{fig:RealResults_beethoven}%
\end{figure}

To overcome the ``posterized'' look of existing woodworking techniques, make use of fine-grained wood structures, and obtain results that are properly shaded, we introduce \emph{computational parquetry}. Our technique can be considered a novel hybrid of both methods and is vitally driven by a computational design process. The goal of computational parquetry is to make deliberate use of the rich structure present in real woods, using heterogeneities such as knots, grain or other texture as a source of detail for recreating more faithful renditions of target images in wood, using a moderate number of pieces, see e.g.\ \changed{\Fig{fig:RealResults_beethoven}}. Since this goal can only be achieved by exhaustively searching suitable pieces of source material to represent small regions of the target image, the task is absolutely intractable to solve by hand.
In the computer graphics world, our technique is closely related to patch-based image synthesis \cite{barnes2017survey}, texture synthesis \cite{wei2009state} and photo mosaics \cite{Battiato:2006}, well-explored families of problems for which a multitude of very elaborate and advanced solutions exist today. \changed{To our knowledge, none of these solutions are prepared to deal with the fabrication-specific challenges that are inherent to our problem.}

\begingroup
\urlstyle{tt}
Our end-to-end system for fabricated style transfer only uses commonly available real-world materials and can be implemented on hobby-grade hardware (laser cutter and flatbed scanner). \changed{To make this new type of computational art accessible to a wide user base, the source code will be made available at \url{https://github.com/isering/WoodPixel}}.
\endgroup

\section{Related work} 
\begin{figure}[t]%
\includegraphics[width=0.5\columnwidth]{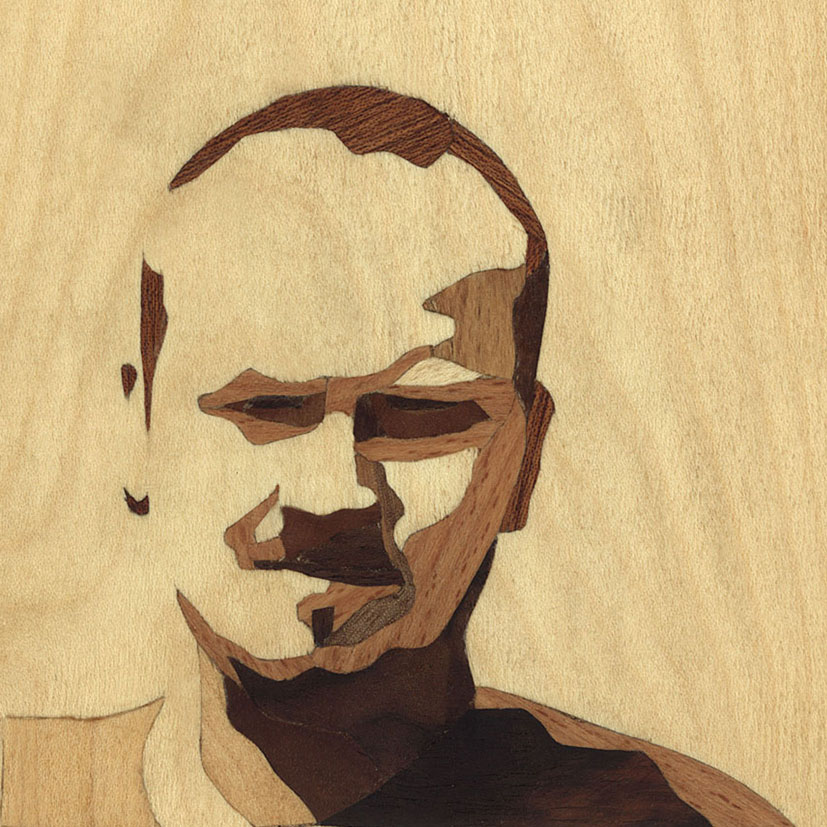}%
\includegraphics[width=0.5\columnwidth]{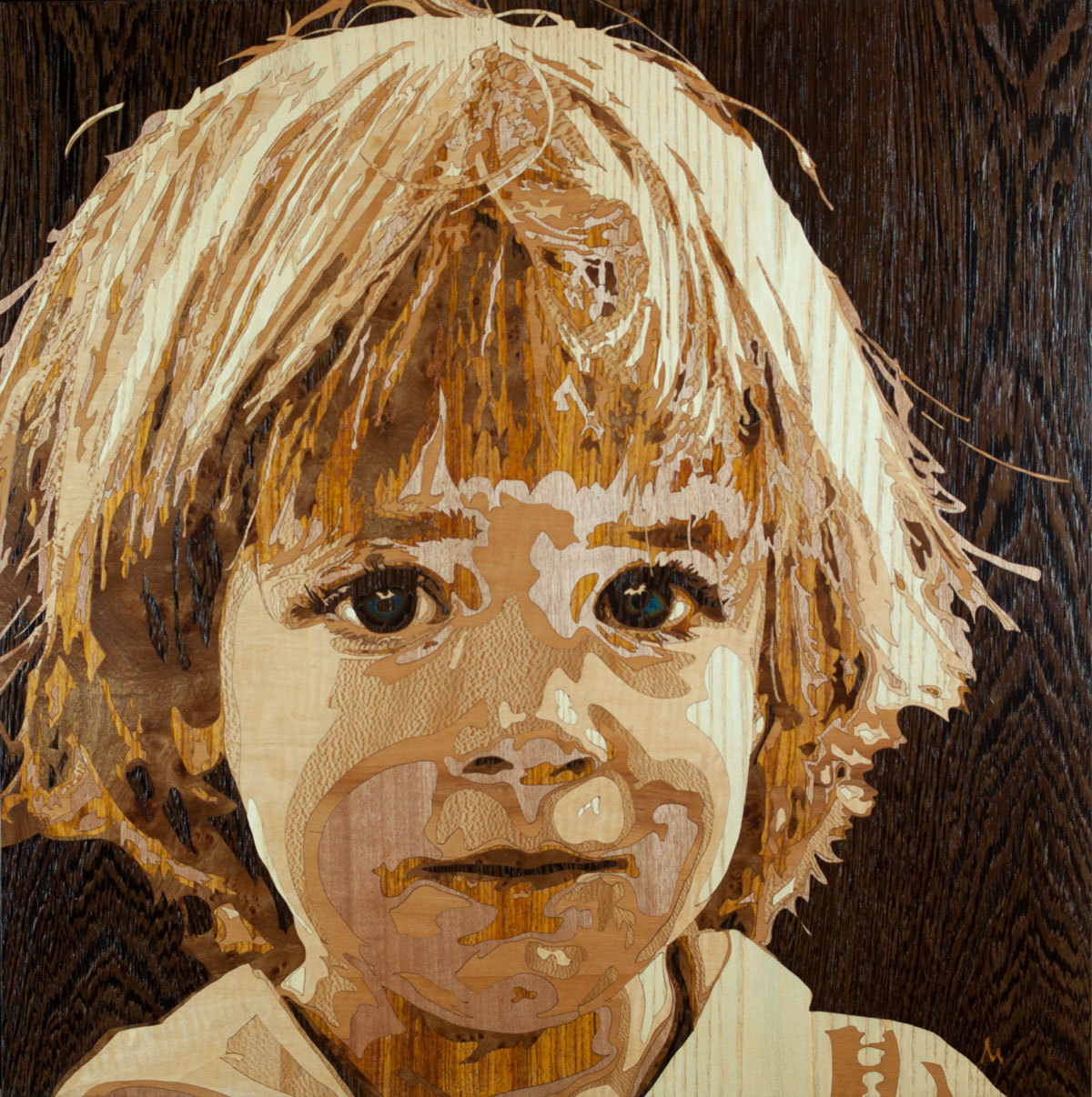}\caption{Two modern examples of marquetry portraits of different complexity. Left: Self-portrait by Laszlo Sandor (using two maple specimens, brown and black walnut, beech, Indian rosewood, okoume and sapele; original size approx.\@ \SI{10 x 10}{cm}). Right: Portrait of a girl by Rob Milam (using wenge, Carpathian elm burl, Honduran rosewood, lauan, pear, plaintree, maple and ash; original size approx.\@ \SI{53 x 53}{cm}).}%
\label{fig:examples}%
\end{figure}

\begin{figure}[ht]%
\includegraphics[height=2.7cm]{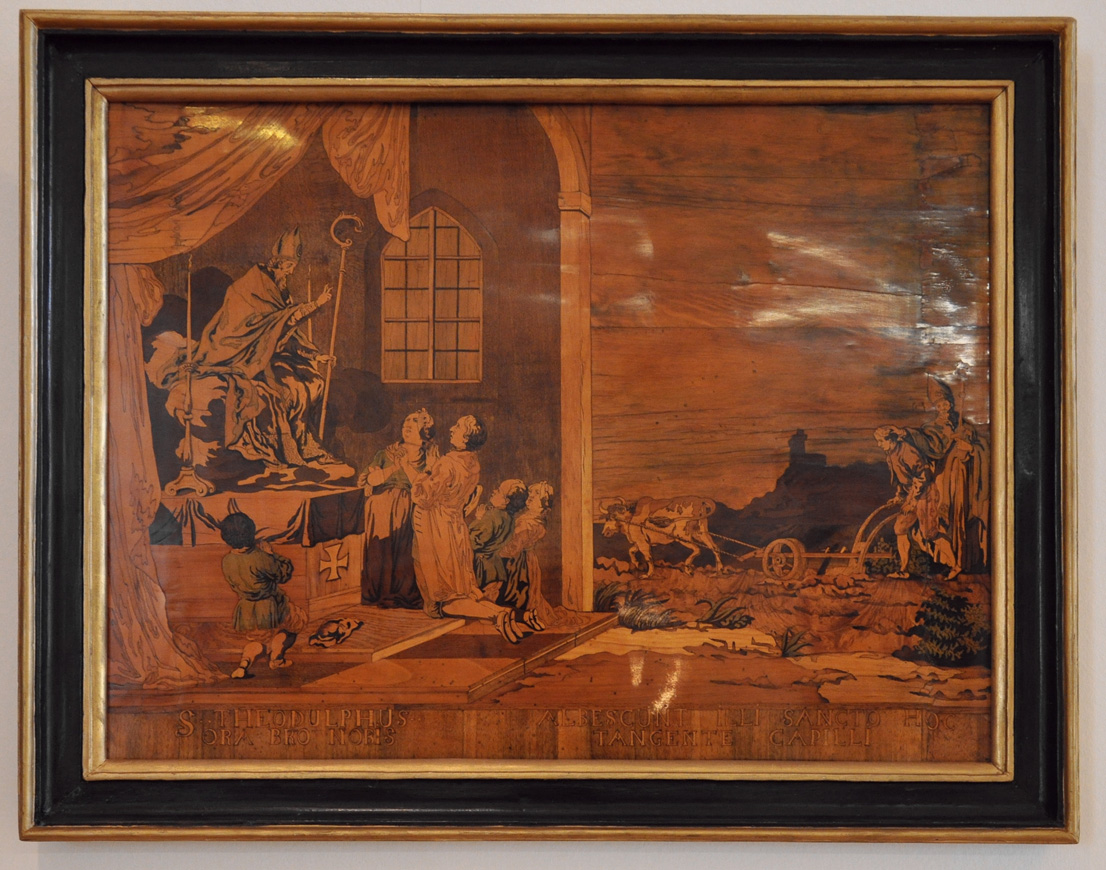}%
\includegraphics[height=2.7cm]{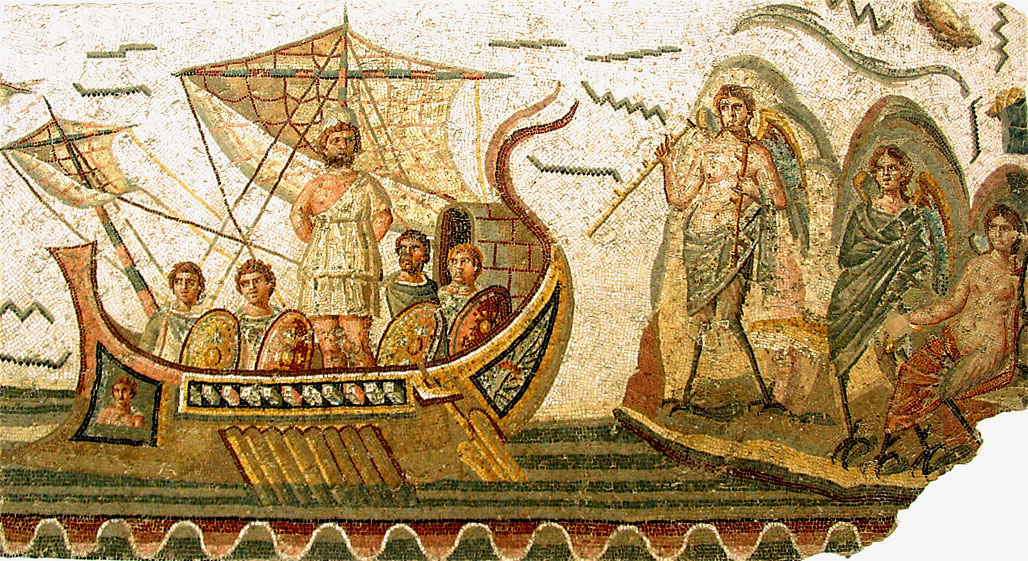}\caption{Examples for intarsia and ancient mosaics: The intarsia from the year 1776 depicts the adoration of St. Theodulf of Trier and a landscape with plowing farmers and St. Theodulf (left). The mosaic from the 2nd century AD depicts a scene from the Odyssey (right).}%
\label{fig:examples2}%
\end{figure}

\paragraph{History of the craft.} 
History knows a rich tradition of techniques that use patches of material for the purpose of composing images. Ancient Roman and Greek mosaics are probably the best-known early instances of this idea. An exemplary mosaic from the second century AD is shown in \Fig{fig:examples2}.
Often, such mosaics consist of largely uniformly shaped primitive shapes (e.g., square tiles) that are aligned with important structures, such as object boundaries, found in the target image. 
A modern counterpart of mosaics is pixel artwork, which has played a similarly ubiquitous role predominantly through video games in the 80s and 90s. Here, the design pattern is generally aligned with a Cartesian grid.

Marquetry can be considered a generalization of mosaic. This art of forming decorative images by covering object surfaces with fragments of materials such as wood, bone, ivory, mother of pearl or metal, has also been known at least since Roman times~\cite{ulrich2007roman}, see \Fig{fig:examples2}. The appearance of the resulting image, however, is mostly dominated by the choice of materials and the shape of fragments.
The closely related term parquetry refers to the assembly of wooden pieces to obtain decorative floor coverings. Either technique can be implemented either by carving and filling a wood surface (inlay) or covering the entire surface with a continuous layer of thin veneer pieces. The materials can be altered in appearance, for instance by staining, painting or carving.

In this paper, we use the term parquetry more restrictively to refer to two-dimensional arrangements of wood veneer that are unaltered in color (except for a final layer of clear varnish that is applied to the entire design). While some artists use computational tools, such as posterization, to find image segmentations (\Fig{fig:examples}), we believe that our method marks the first time that a measured texture of the source material has been used to drive the design process, explicitly making use of features present in the wood.

\paragraph{Stylization.}
With the goal of non-photorealistic rendering, numerous techniques have been proposed to transform 2D inputs into artistically stylized renderings~\cite{Kyprianidis:2013}.
This includes approaches for the simulation of different painting media such as paints, charcoal and watercolor~\cite{Chen:2015,Lu:2013,Panotopoulou:2018}.
In recent years, the potential of deep learning has been revealed for rendering a given content image in different artistic styles~\cite{Jing2017NeuralST}.
Inspired by the ancient mosaics and the application of mosaics for arts (see e.g. Salvador Dal\'i's lithograph \emph{Lincoln in Dalivision}~\shortcite{Dali:1996} or \emph{Self Portrait I} by Chuck Close~\shortcite{Close:1995}), a lot of effort has been spent on non-photorealistic rendering in mosaic-style.
The original photo mosaic approach~\cite{Silvers:1997} creates a mosaic by matching and stitching images from a database.
Further work focused on the application to non-rectangular grids and color correction~\cite{Finkelstein:1998} and tiles of arbitrary shape (jigsaw image mosaics or puzzle image mosaics)~\cite{Kim:2002,diBlasi:2005b,Pavic:2009}, the adjustment of the tiles in order to emphasize image features within the resulting mosaic~\cite{Hausner:2001,Elber:2003,Liu:2010,Battiato:2012} as well as speed-ups of the involved search process~\cite{diBlasi:2005,diBlasi:2005b,Kang:2011}.
More recently, texture mosaics have also been generated with the aid of deep learning techniques (e.g.~\cite{Jetchev:2017}).
We refer to respective surveys~\cite{Battiato:2006,battiatosurvey2007} for a more detailed discussion of the underlying principles.
Furthermore, panoramic image mosaics~\cite{Szeliski:1997} have been introduced where photos taken from different views are stitched based on correspondences within the individual images and a final image blending.

\begin{figure*}[ht]
\centering
\includegraphics[width=\textwidth]{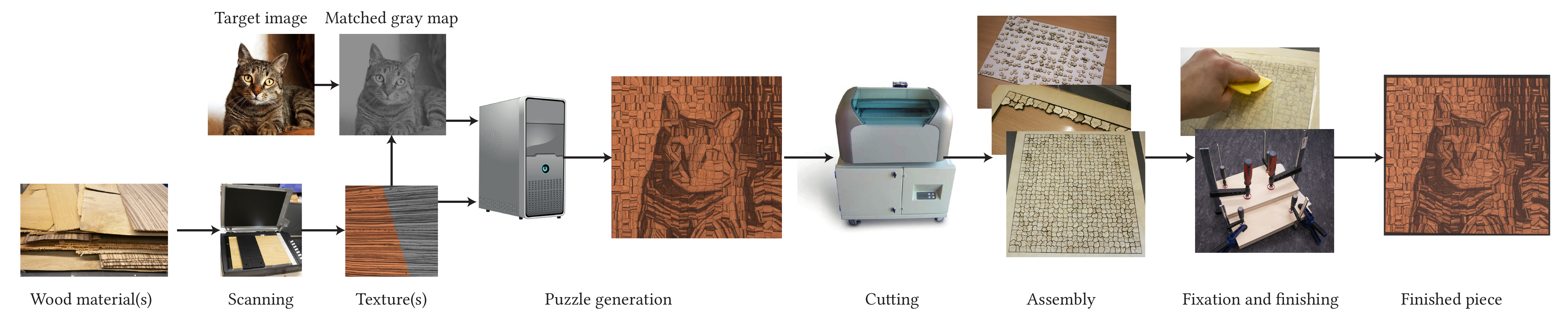}
\caption{The proposed end-to-end pipeline for creating faithful renditions of target images based on exploiting the rich structure present in input wood samples as a source of detail. The involved major steps are data acquisition, cut pattern optimization and the final fabrication of the real-world rendition of the target image.}
\label{fig:pipeline}
\end{figure*}

\paragraph{Example-based synthesis.}
\emph{Pixel-based} synthesis techniques~\cite{Paget:1998,Efros:1999,Wei:2000,Hertzmann:2001} rely on copying single pixels from an exemplar to the desired output image while matching neighborhood constraints.

In contrast, \emph{patch-based or stitching-based} texture synthesis approaches ~\cite{Praun:2000,Efros01,Kwatra:2003} involve copying entire patches from given exemplars.
One major challenge of these approaches is the generation of correspondences between locations in the exemplar image and locations in the generated output image to copy the locally most suitable patches from the exemplar to the output image.
For this purpose, common strategies include arranging patches in raster scan order and subsequently selecting several patch candidates that best fit to the already copied patches.
As this matching process becomes computationally challenging for larger images, several investigations focused on improving matching efficiency~\cite{Ashikhmin:2001,Tong:2002,Barnes:2009,Barnes:2010,Barnes:2011,Datar:2004,Liang:2001,Simakov:2008,He:2012,Olonetsky:2012,Wei:2000}.
In addition, finding an adequate composition and blending of the copied patches has been addressed based on simple compositions of irregularly shaped patches~\cite{Praun:2000}, the blending of overlapping patches within the overlap region~\cite{Liang:2001}, the specification of seams within the overlap region using dynamic programming or graph cuts~\cite{Efros01,Kwatra:2003}, or the application of a weighted averaging for several overlapping regions~\cite{Wexler:2007,Simakov:2008,Barnes:2009}.
Furthermore, \emph{optimization-based} techniques~\cite{Portilla:2000,Han:2006,Kwatra:2005,Kopf:2007,Darabi:2012,Kaspar:2015} are based on the formulation of texture synthesis in terms of an optimization problem which is solved by minimizing an energy function and combines pixel-based and patch-based techniques.
\changed{The global statistics of source patch usage and arrangement can be guided based on histogram matching~\cite{Kopf:2007} or, in the context of general image editing (including tasks such as re-organization of objects in the image, image retargeting, inpainting, image fusion/composition), based on image statistics, saliency information, semantics, and user constraints~\cite{Cho:2008,Pritch:2009}.
The latter applications are related to our work in the sense that they rely on re-arranging patches/regions of the input images, where Cho et al.~\shortcite{Cho:2008} even only use each patch once as also done in our work.
In contrast to these methods, our approach involves a cross-domain analysis between an input target image and images of wooden veneers, where we use the patches of the wooden veneers to compose a \emph{stylized} version of the target image.}

Recently, the potential of deep learning has also been demonstrated in the context of optimization-based texture synthesis (see e.g.~\cite{Gatys:2015,Gatys:2016,Li:2016a,Li:2016b}).
For a more detailed review, we refer to the surveys provided by Wei et al.~\shortcite{wei2009state} and Barnes and Zhang~\shortcite{barnes2017survey}.

Patch-based synthesis in the real world, as described and performed in this work, is characterized by fundamental constraints that are inherent to the task of parquetry and other forms of real-world collage. Any piece of input material can only be used once without being stained, scaled, stretched, copied, blended or filtered. Our synthesis algorithm therefore restricts itself to cutting operations and rigid transformations. More importantly, it must keep track of resource use in order to prevent source patches from colliding with each other. On the output side, the cuts must be fabricable, i.e., the individual fragments must be connected (no isolated pixels) and they may not expose too thin protruding structures. We are not aware of prior work that addressed these specific challenges.

\paragraph{Computational fabrication.}

Developments in the context of stylized fabrication~\cite{Bickel:2018} took benefit from the rapid progress in fabrication technology.
In the context of 2D arts, the computational fabrication of paintings has been approached based on robotic arms to paint strokes for a given input image (e.g.~\cite{Deussen:2012,Lindemeier:2013,Tresset:2012}).
The fabrication of artistic renditions of images has been approached based on a computational non-photorealistic rendering pipeline, the generation of respective woodblocks and a final woodblock printing process~\cite{Panotopoulou:2018}.
Further work addressed mosaic rendering using colored paper~\cite{Gi:2006b}, where computational approaches have been used for tile generation and tile arrangement.
This is followed by the respective generation of colored paper tiles and their arrangement according to the energy optimization.

Most works in computational fabrication aim at obtaining constant results despite possible variations in the material used. In contrast, we embrace the ``personality'' of the input and use it to create artworks that are inherently unique.

\section{Method}
\label{sec:method}
The main objective of this work is the development of a computational pipeline for creating faithful renditions of a target image $\vec{I}_T$ from wood samples by exploiting the rich structures in wood as a source of detail.
The pipeline devised in this work takes $n_\text{samples}$ physical, wooden material samples and a target image as inputs and consists of three major steps: data acquisition, data analysis and cut pattern generation (i.e.\@ tile generation, arrangement, and boundary shape optimization), and the final fabrication of the real-world counterpart (\Fig{fig:pipeline}).

In the first step, the wooden samples are prepared before they can be scanned with a flatbed scanner.
This is followed by extracting local features in the input images and by detecting corresponding patches between the source textures and the target image, yielding a stylized, digital wood parquetry of the target image.
Finally, the patches are converted to cut instructions (taking into account that the cuts have to be fabricable by a laser cutter), specified pieces are cut with a laser cutter, and assembled to a physical sample of parquetry.
We discuss details in the following sections.

\subsection{Data acquisition}
Before the scanning can be conducted, we first prepare the wood samples.
Whereas thicker veneers can be utilized directly, standard veneers (\SIrange{0.6}{0.8}{mm} thick) are glued to a substrate of \SI{1.5}{mm} birch plywood in order to improve stability and minimize waviness. Especially burl veneers tend to be very brittle and assume strongly warped shapes; in contrast, the bending of the substrate is relatively easy to counter by screwing it to a rigid substrate.
We enhance the contrast of the wood veneers (and consequently the contrast of the final parquetry) by sanding and applying a thin layer of clear coat or oil finish.
After letting the finishing layer dry, the specimens are placed on a flatbed scanner and scanned at \SI{300}{dpi}.
The scans are aligned in order to get a common coordinate frame and a mask is generated to separate usable veneer from empty background and screw holes.
We note that for larger scale productions, this step could easily be automated using machine vision techniques.
The output of this step is a set of source textures $\vec{I}_{S} = \{\vec{I}_{S,i}: i \in \{1,\ldots,n_\text{samples}\}\}$, one for each physical wood sample.

\subsection{Feature extraction}
\label{sec:feature_extraction}
\begin{figure}[t]
\includegraphics[width=0.24\columnwidth]{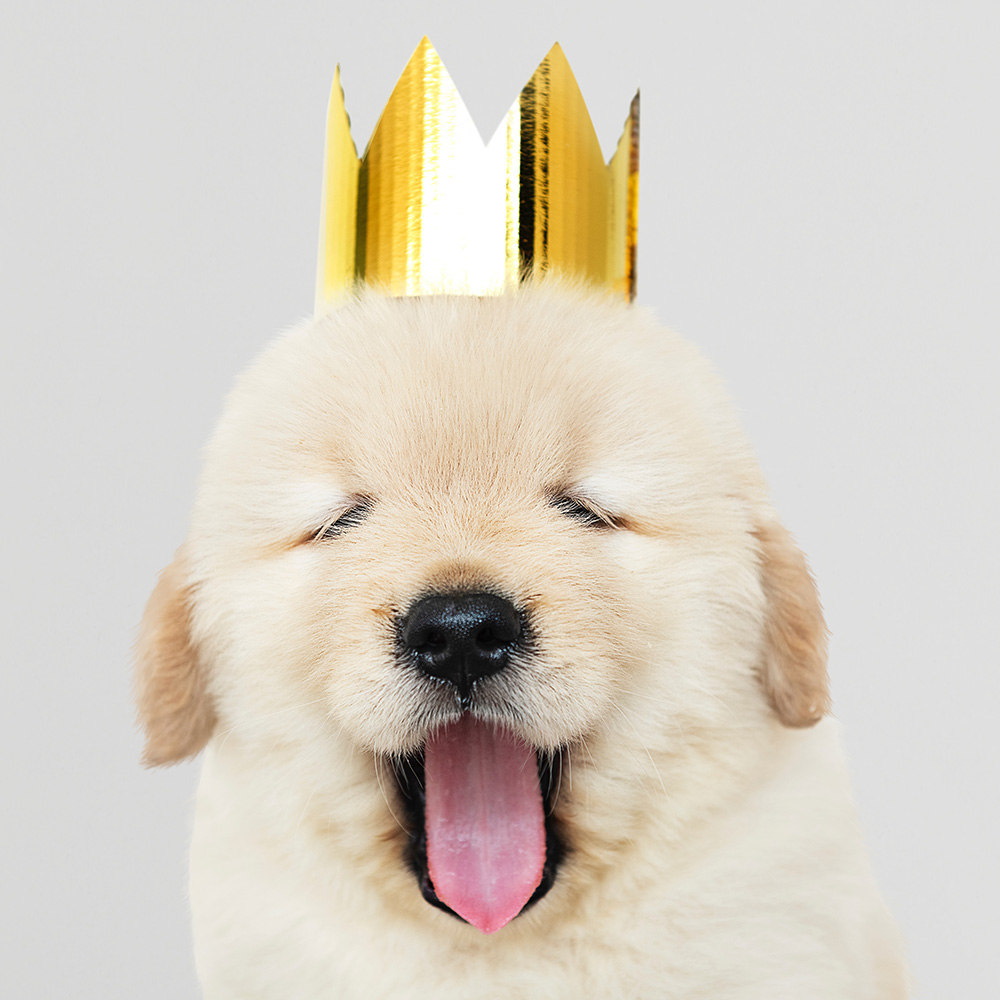}
\hfill
\includegraphics[width=0.24\columnwidth]{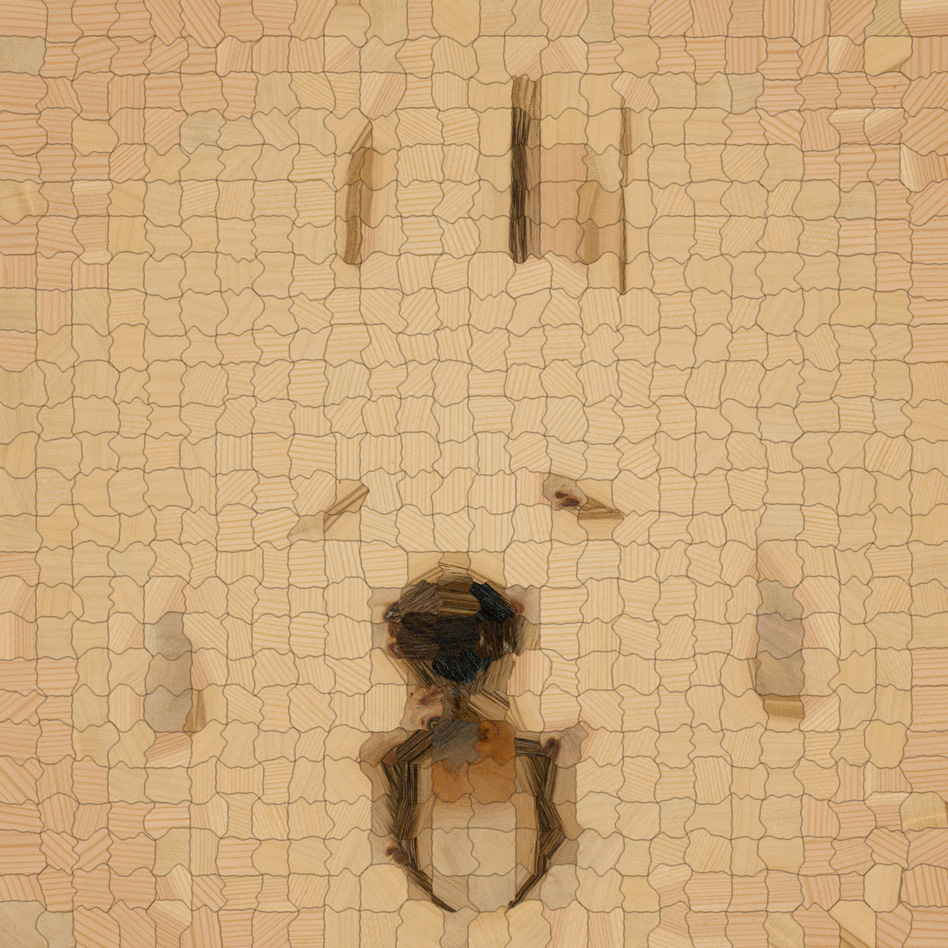}
\hfill
\includegraphics[width=0.24\columnwidth]{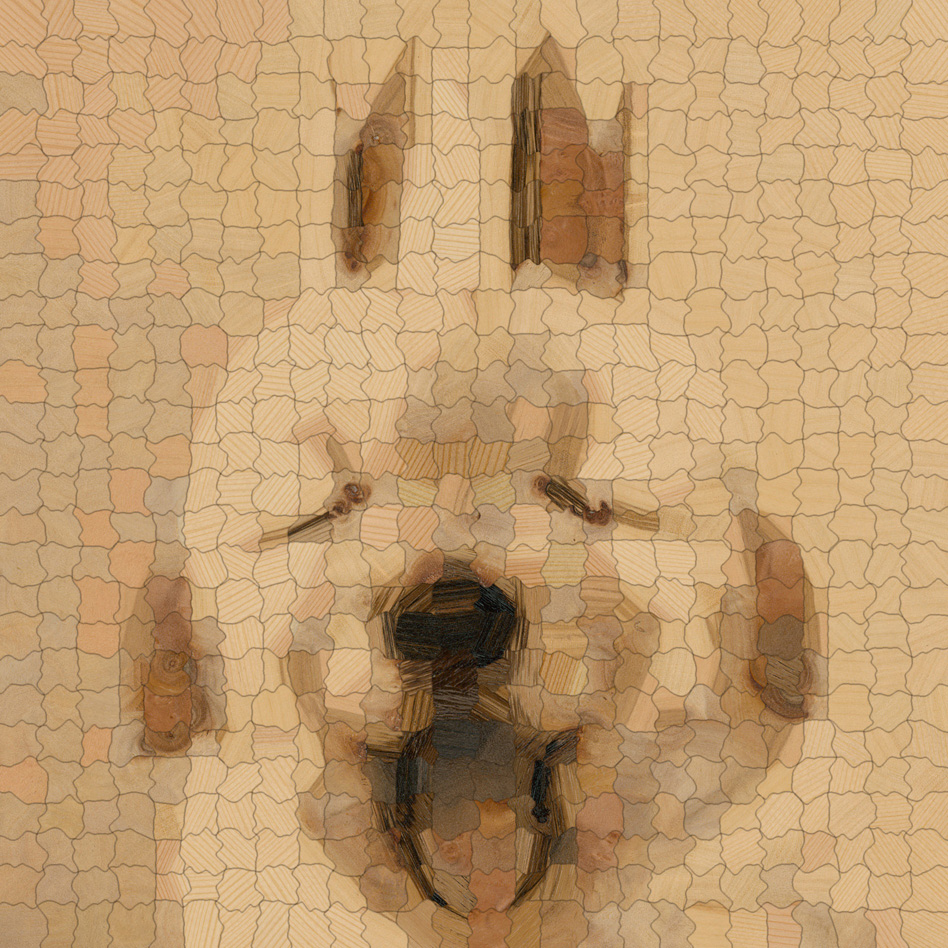}
\hfill
\includegraphics[width=0.24\columnwidth]{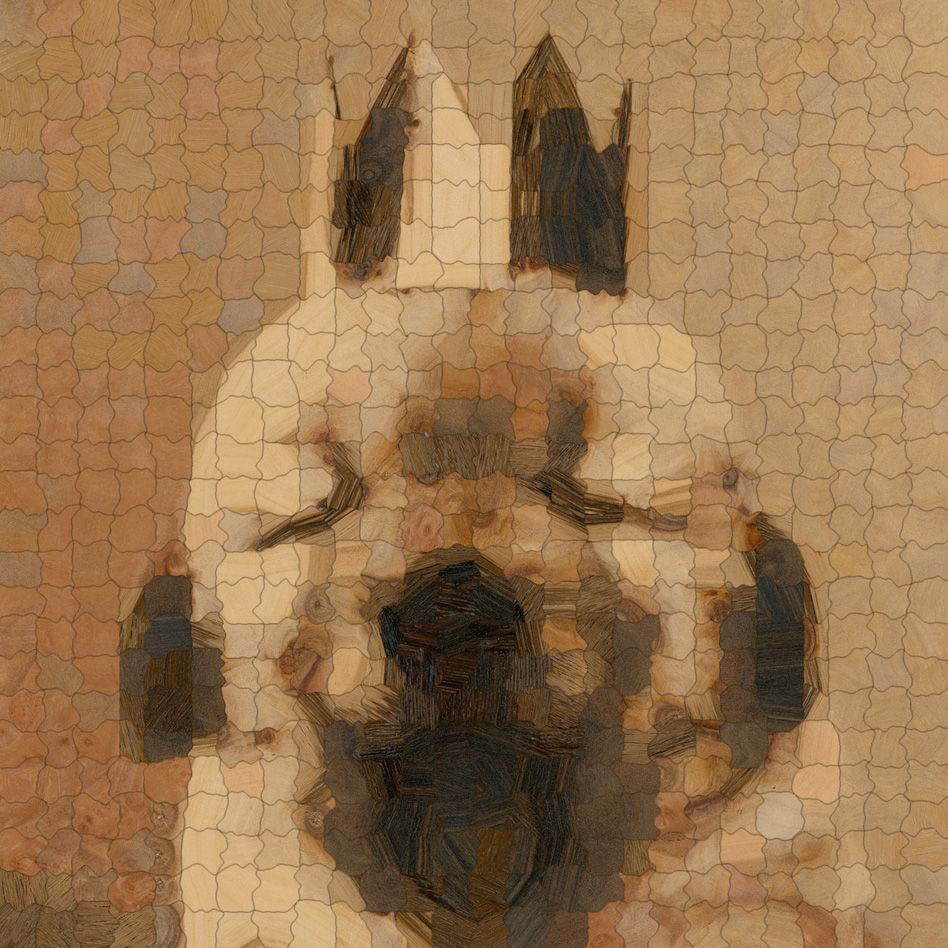}\\[0.5mm]
\includegraphics[width=0.24\columnwidth]{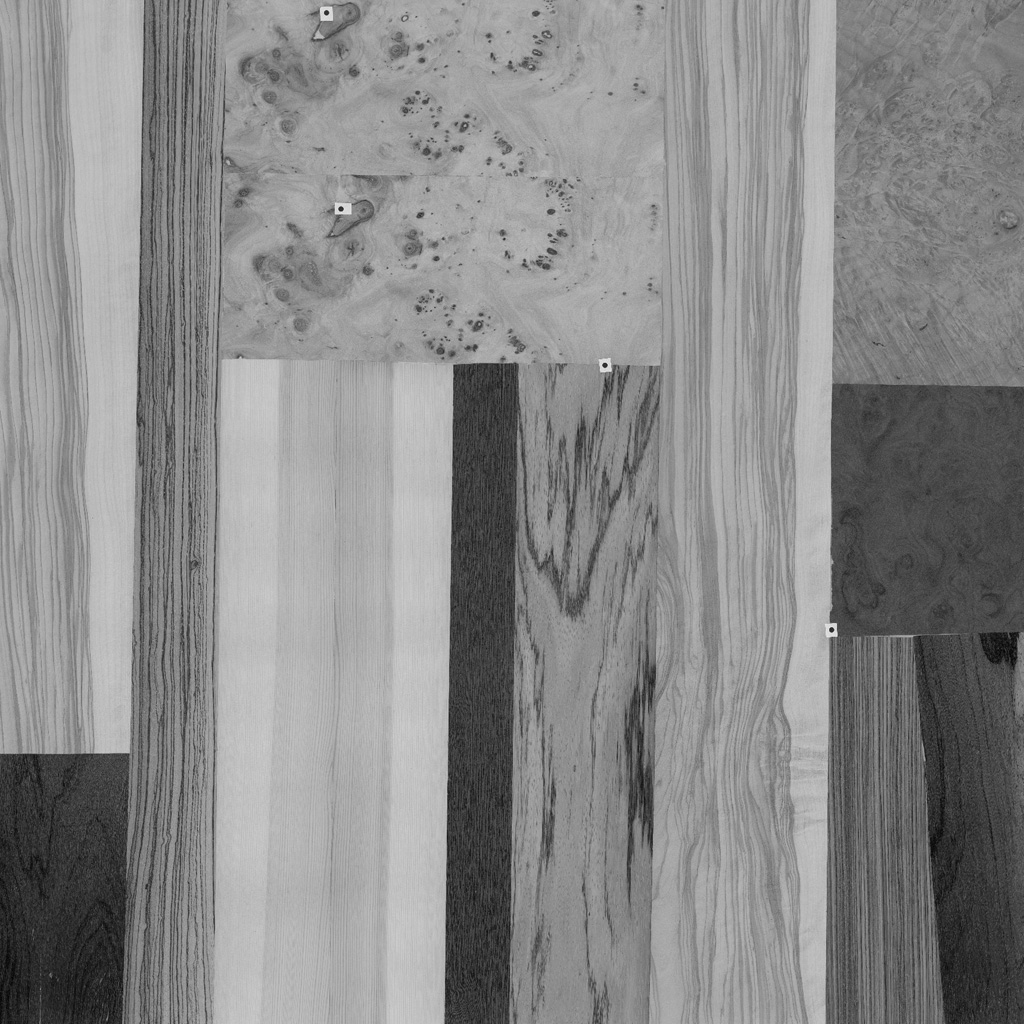}
\hfill
\includegraphics[width=0.24\columnwidth]{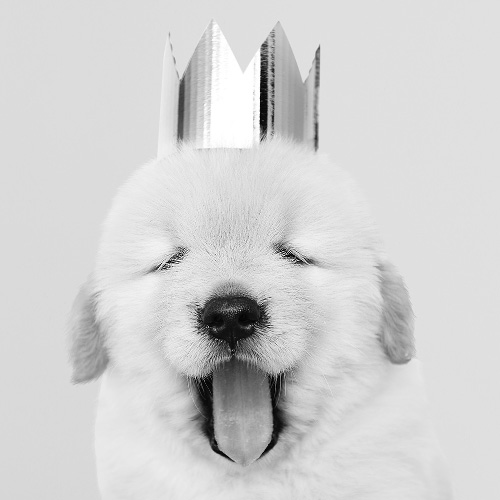}
\hfill
\includegraphics[width=0.24\columnwidth]{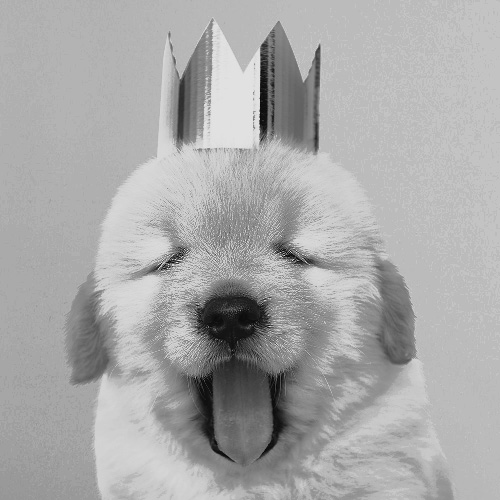}
\hfill
\includegraphics[width=0.24\columnwidth]{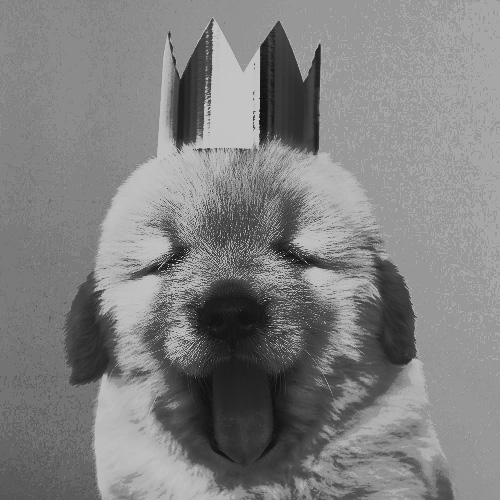}
\caption{\changed{\textbf{Synthetic}} Top row: For a target image exhibiting low contrast and a bad foreground separation (left) the generated wood puzzle shows the same, undesirable effects when discarding histogram matching (middle left). In contrast, applying histogram matching (right) allows to exploit the whole wood texture gamut, which yields a high contrast at the cost of a strong change in appearance compared to the target. By interpolating between the intensity filter responses obtained with and without histogram matching, we generate a parquetry with medium contrast (middle right). Bottom row: The images depict the intensity filter response of a wooden veneer panel (left) and the respective responses obtained for the target image without histogram matching (middle left) and with histogram matching (right), as well as their interpolation (middle right).}
\label{fig:histogram_equalization}
\end{figure}
In order to find patch correspondences between the target image and the source images, we define a suitable representation for textural structures within the individual patches.
We densely evaluate texture features using a filter bank consisting of $2$ image filters, an intensity filter and a Sobel edge filter.
We have experimented with higher-dimensional filter banks similar to the {{Leung--Malik}} filter bank \cite{Leung2001} and found that the potential increase in reconstruction quality does not offset the additional computational cost induced by the higher-dimensional feature space.
Applying the filter bank to an image $\vec{I}$ results in the $2$-dimensional feature response maps
\begin{equation}
\vec{F}(\vec{I}) = \left(w_\text{intens} \cdot \vec{F}_\text{intens}(\vec{I}),\  w_\text{edge} \cdot \vec{F}_\text{edge}(\vec{I}) \right)^{\top},
\end{equation}
where $\vec{F}_x$ are the particular image filters, $w_x \in [0,1]$ are the feature weights, $x \in \{\text{intens}, \text{edge} \}$. The weights allow artistic control over the emphasis on overall intensity matching ($w_\text{intens}$) and fine scale gradient features ($w_\text{edge}$).
Please note that this approach can easily be expanded to different feature vectors, allowing additional artistic control.
We increase the probability of finding good matches by taking $n_\text{rot}$ rotated versions of the wood source textures into account.

Source and target textures may exhibit highly different gamuts and filter response distributions, so we apply a histogram matching step in order to achieve a meaningful matching between target and source patches.
We use a CDF-based histogram equalization \cite[ch.4]{Russ:2002:IPH:581172} to transform the intensity distribution of the target image to that of the available source textures.
As the wood samples generally span a smaller gamut than the target image, gamut mapping is of great importance to allow for the representation of the target image based on sampling the whole range of available wood patch intensities so that characteristic image structures can be emphasized.
For challenging target images with low foreground contrast or bad foreground separation (\Fig{fig:histogram_equalization}) we found that the histogram equalization tends to overshoot.
We alleviate this by interpolating between equalized and original target intensity,
\begin{equation}
\vec{F}_\text{intens}^\prime(\vec{I}_T) = (1-w_\text{hist}) \vec{F}_\text{intens}(\vec{I}_T) + w_\text{hist} \vec{F}_\text{equalize}(\vec{I}_T, \vec{I}_S),
\end{equation}
where $w_\text{hist}$ denotes the interpolation weight, $\vec{F}_\text{equalize}$ the histogram equalization operator, and $\vec{I}_S$ the set of all wood textures.

The output of this step is a set of $n_\text{samples} \cdot n_\text{rot}$ filter responses
\begin{equation}
\vec{F}(\vec{I}_S) = \left\{ \vec{F}(\vec{I}_{S,i,\phi_j}) : i \in \{1, \ldots, n_\text{samples} \}, j \in \{1, \ldots, n_\text{rot} \} \right\}
\end{equation}
for the source textures, and one filter response $\vec{F}(\vec{I}_T)$ for the target image. We typically used $n_\text{rot}=15$ source texture rotations for our experiments.

\hide{
\begin{figure*}[ht]
\centering
\includegraphics[width=\textwidth]{figures/bobbyflow-paper.pdf}\\%
~\hfill(a)\hfill\hfill(b)\hfill\hfill(c)\hfill\hfill(d)\hfill\,%
\caption{Starting with the initial square patch boundaries (a), we use dynamic programming to obtain optimal seams between adjacent patches (b). The pixelwise results are then fitted using cubic splines (c) and re-rendered for validation (d).}
\label{fig:cuts}
\end{figure*}
}

\subsection{Cut pattern optimization}
After evaluating the filter responses, the next step is to find corresponding patches between target image and source textures.
\changed{Given a target patch $\vec{P}_T \subset \vec{I}_T$ consisting of $n_P$ simply connected pixels $\vec{p}_k$, we determine a corresponding source patch $\vec{P}_S$ by a dense template matching using the sum of squared differences between the respective feature maps $\vec{F}$:
\begin{equation}
\begin{aligned}
D_{i, j}\left(\vec{x}\right) &= \sum_{k = 1}^{n_P} \left( \vec{F}(\vec{P}_T(\vec{p}_k)) - \vec{F}(\vec{I}_{S, i,\phi_j}(\vec{x}+\vec{p}_k)) \right)^2,\\
\vec{P}_S &= \argmin_{i, j, \vec{x}} D_{i, j}(\vec{x}). 
\end{aligned}
\label{eq:template_matching}
\end{equation}
}

\begin{figure}[t]
\includegraphics[width=0.49\linewidth]{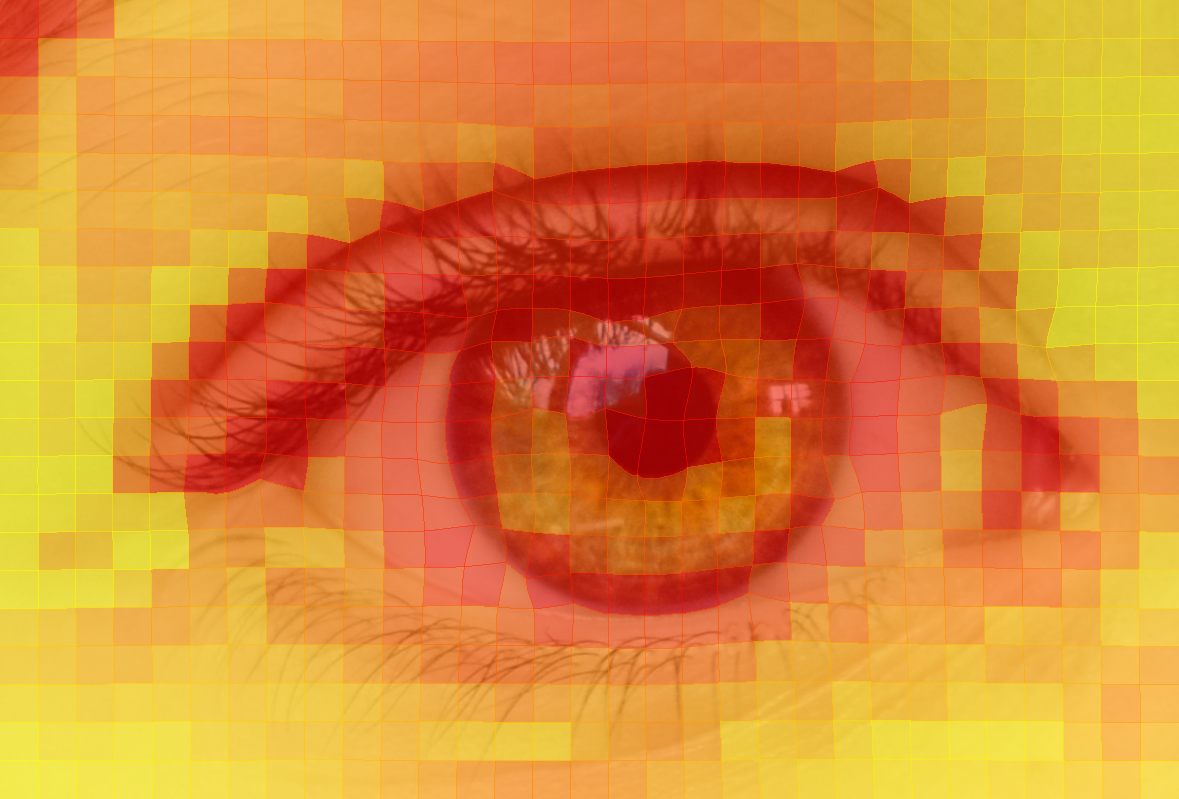}%
\hfill%
\includegraphics[width=0.49\linewidth]{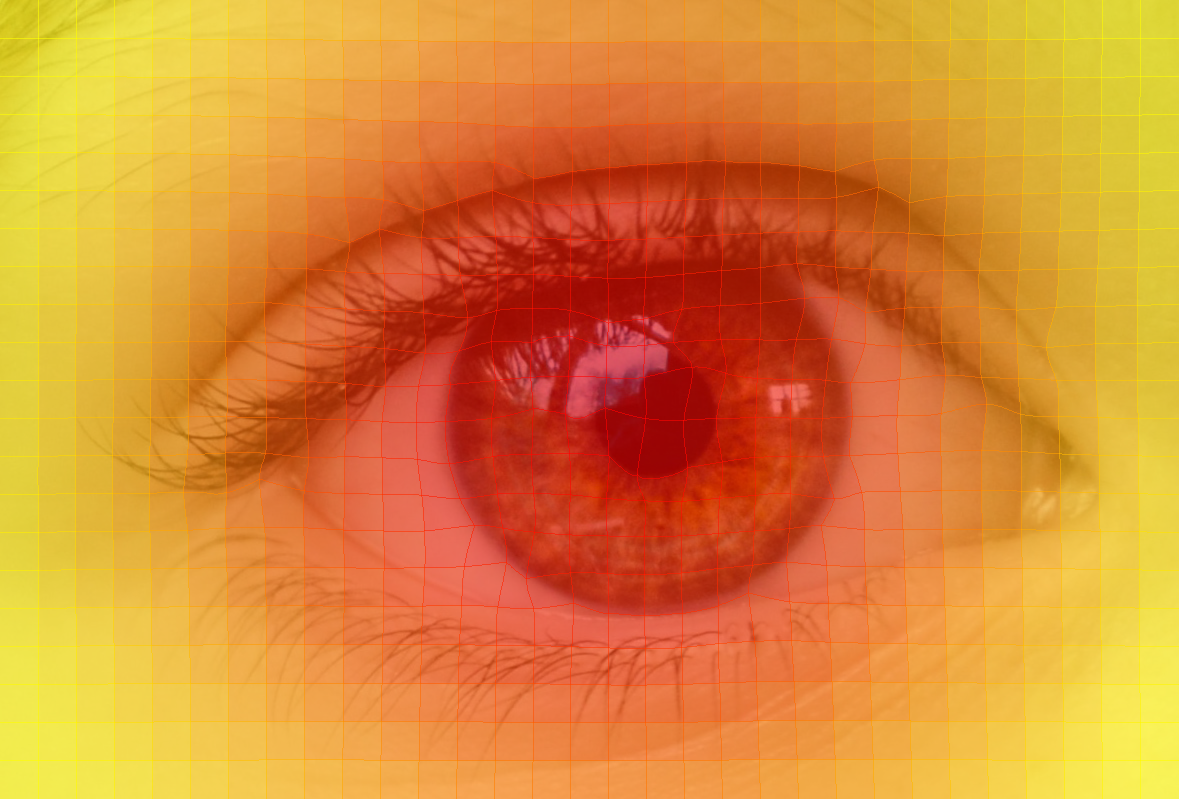}%
\caption{\changed{Before cut pattern optimization, target patches are sorted either by their accumulated saliency score (left) or by their distance to the image center (right). Patches overlaid with red color are reconstructed first, yellow patches are reconstructed last.}}
\label{fig:sort_saliency}
\end{figure}
\changed{We avoid the multiple usage of already matched veneer sample regions by carrying along a binary mask for each source texture.
As more and more wood area is consumed, the probability of finding good patch correspondences decreases as the algorithm advances.
The most effective way to remedy this effect is to ensure that a sufficient amount of panel space for each utilized wood type is available at all times during reconstruction.
Since this might not always be feasible, we store the target patches $\vec{P}_T$ in a priority queue, such that salient patches are reconstructed first.
The patches are sorted either based on their accumulated image saliency score \cite{Montabone:2010:HDU:1660175.1660558} or by their distance to the image center, see \cref{fig:sort_saliency}.
While the former method assures that salient image regions containing high-frequency features will be reconstructed first (and thus be of potentially higher quality), the latter approach creates a bokeh-like effect, which produces more visually pleasing results under a mild shortage of patches with correct low-frequency (intensity) features.
The usage of a priority queue also allows us to reconstruct multiple target images using the same wood panel, either sequentially or interleaved.
The latter allows us to reconstruct multiple target images simultaneously by inserting their respective patches into the queue and sorting them by their image saliency score.}

Target image regions with less salient features can be represented by larger patches.
To exploit this, we implemented an adaptive patch matching step, where a patch is broken into four smaller patches if their combined matching cost is lower than the cost of the larger patch multiplied by a factor $w_\text{adaptive}$.
The factor $w_\text{adaptive}$ can be used to control the artistic balance between larger and smaller patches.
We apply this step $n_\text{adaptive}$ (typ.\@ $0$ to $2$) times.

\changed{At the end of this step, we have covered the target image plane with patches drawn from the source textures $\vec{I}_S$.}

\subsection{Patch shape optimization}
\label{sec:pre_shape_optimization}
\changed{In the previous section we have defined how, given a set of target image patches, correspondences between the target image and the source textures are determined.
In this section, we will discuss the segmentation of the target image into patches, which has an enormous influence on the appearance and fabricability of the final piece of computational wood parquetry art.}

\changed{Ideally, the shapes and placement of the wood patches would be determined using a global scheme that jointly performs template matching and shape optimization. 
The computational costs for solving such a multi-constrained problem, however, would be prohibitively high.
Instead, we explore approaches for decoupling the segmentation step from the matching problem, in order to make both computationally tractable and to enable detailed artistic control.}

\changed{Our approach is to start with a Cartesian grid that is not aligned to image features.
Reconstructions obtained from such grids have a strongly stylized look that resembles pixel art and can be attractive for certain resolutions.
However, it may fail to convey sufficient detail for very coarse grids, in particular when high-contrast areas have to be reconstructed using a single patch and subsequently a single type of wood.
Consequently, we have implemented two different refinement strategies.}

\begin{figure*}[t]
\subfigure[Input image]{\label{subfig:a}\includegraphics[width=0.195\textwidth]{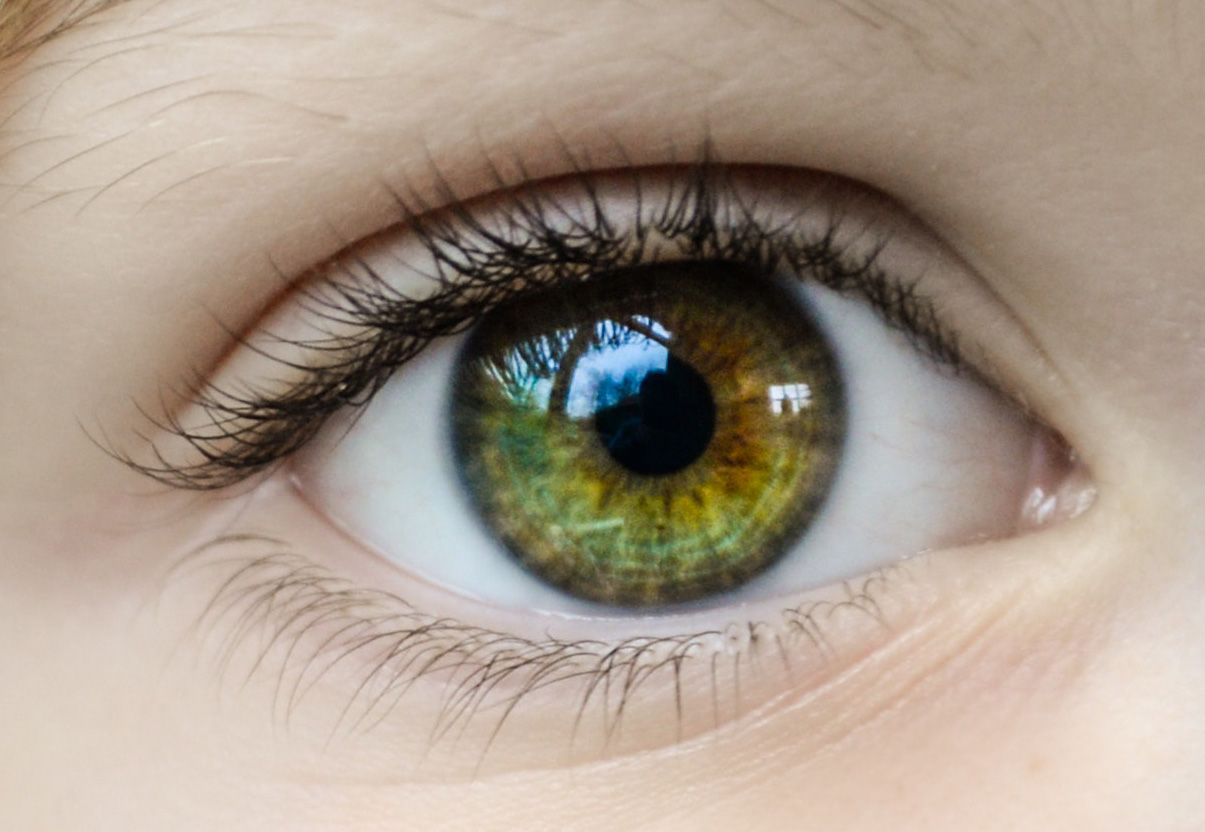}}
\subfigure[Rolling guidance filter]{\label{subfig:b}\includegraphics[width=0.195\textwidth]{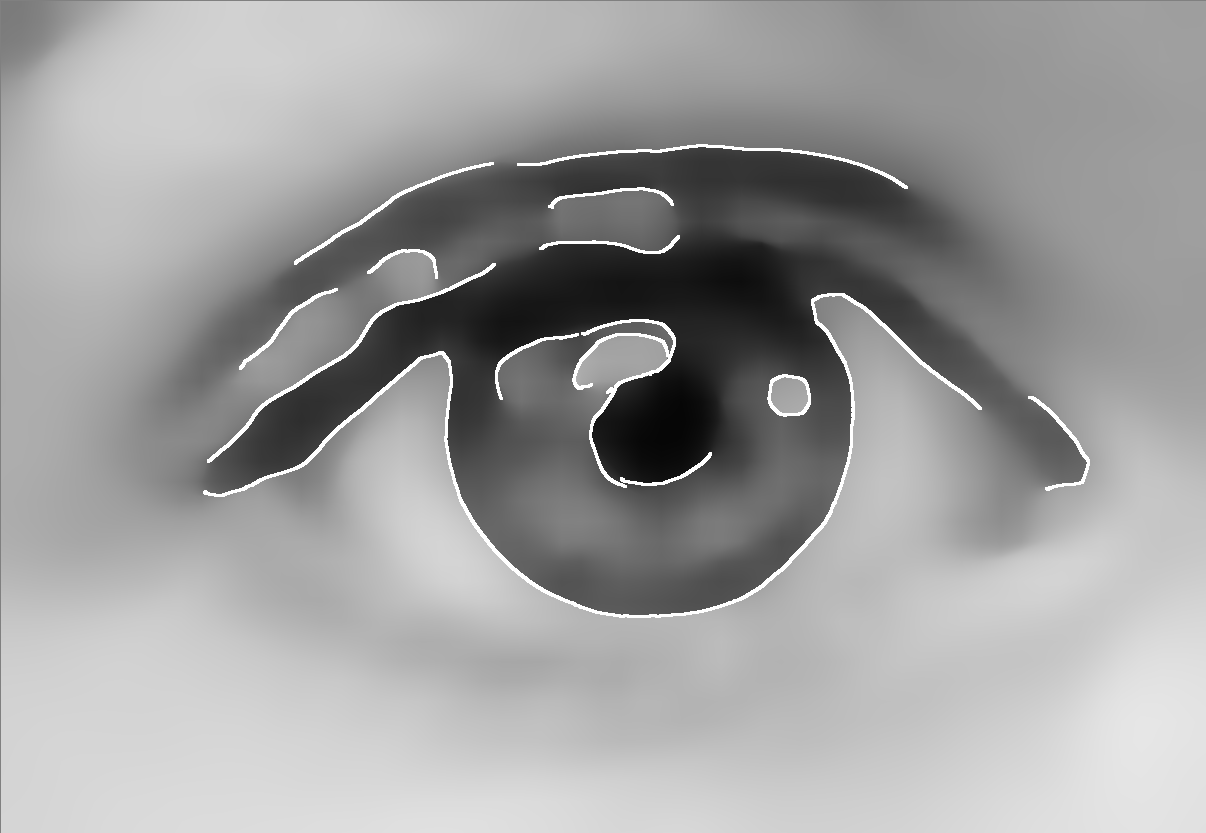}}
\subfigure[Bilateral filter]{\label{subfig:c}\includegraphics[width=0.195\textwidth]{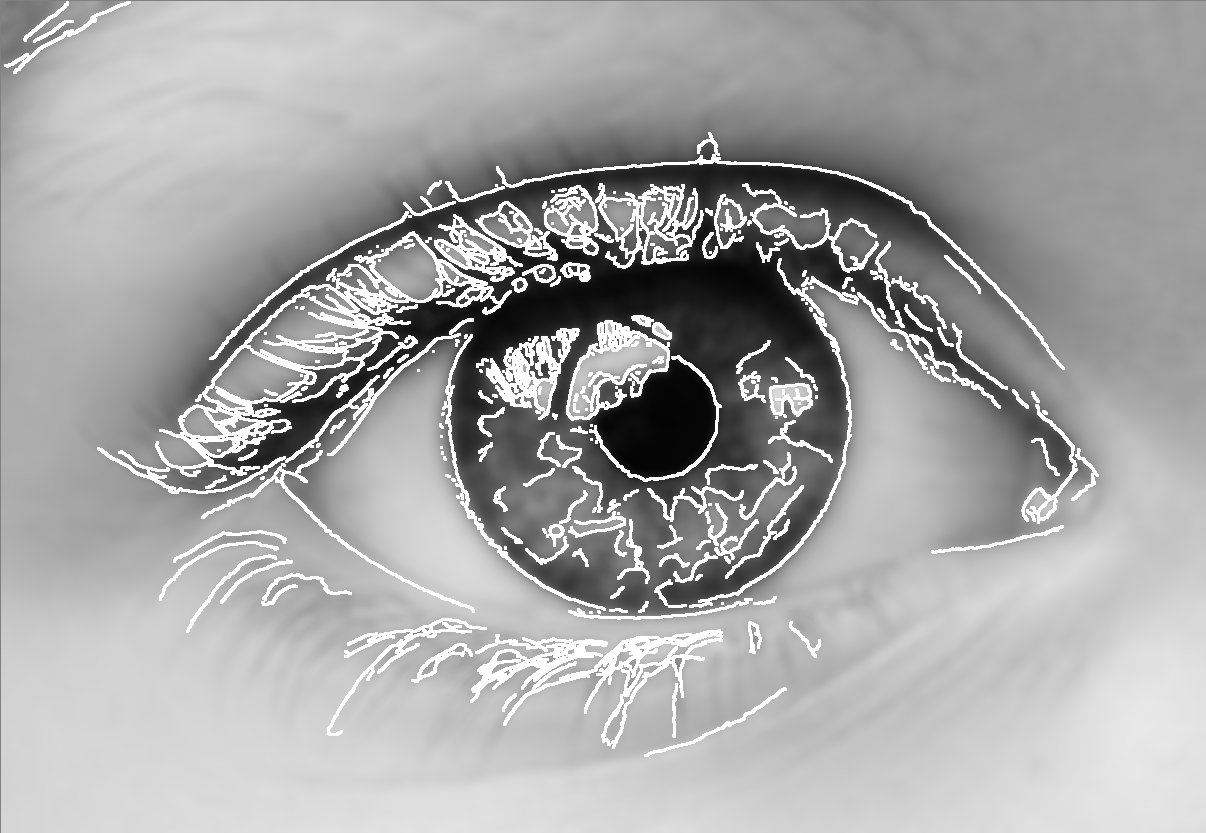}}
\subfigure[Potential field]{\label{subfig:d}\includegraphics[width=0.195\textwidth]{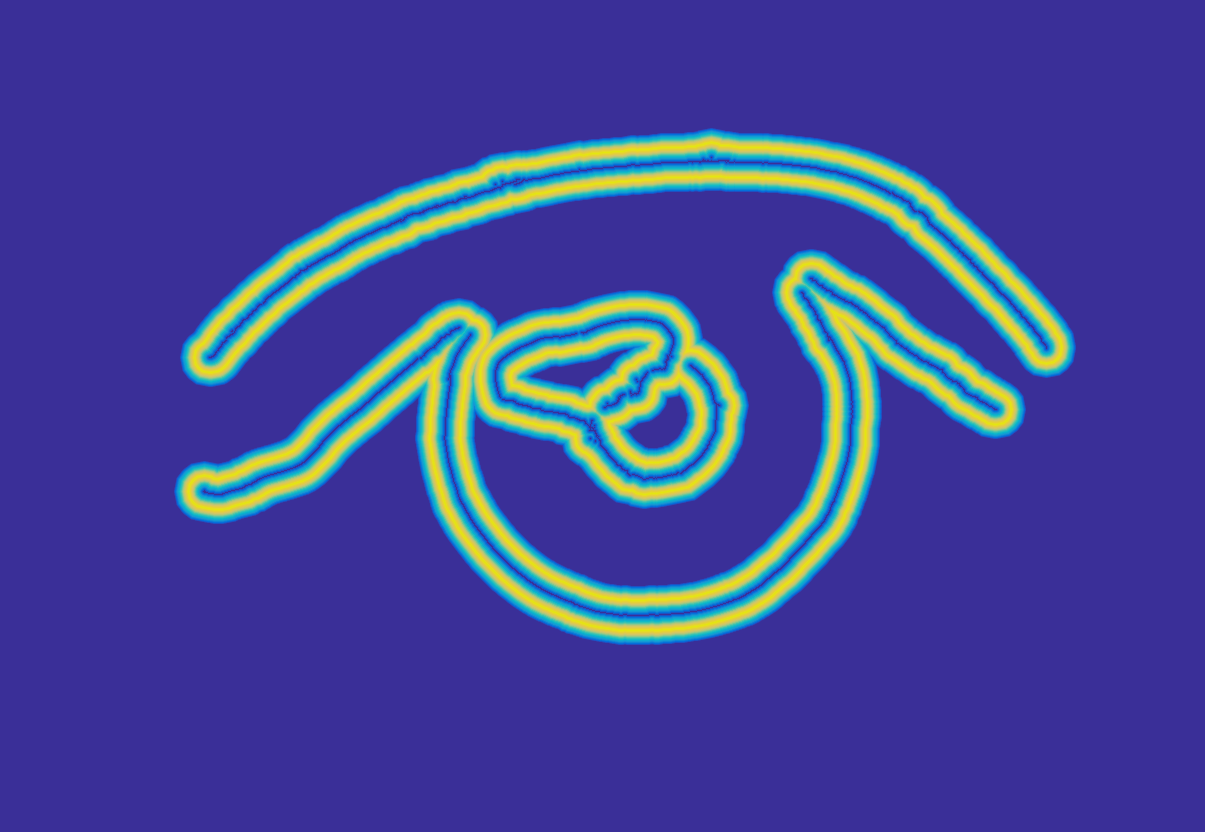}}
\subfigure[Resulting morphed grid]{\label{subfig:e}\includegraphics[width=0.195\textwidth]{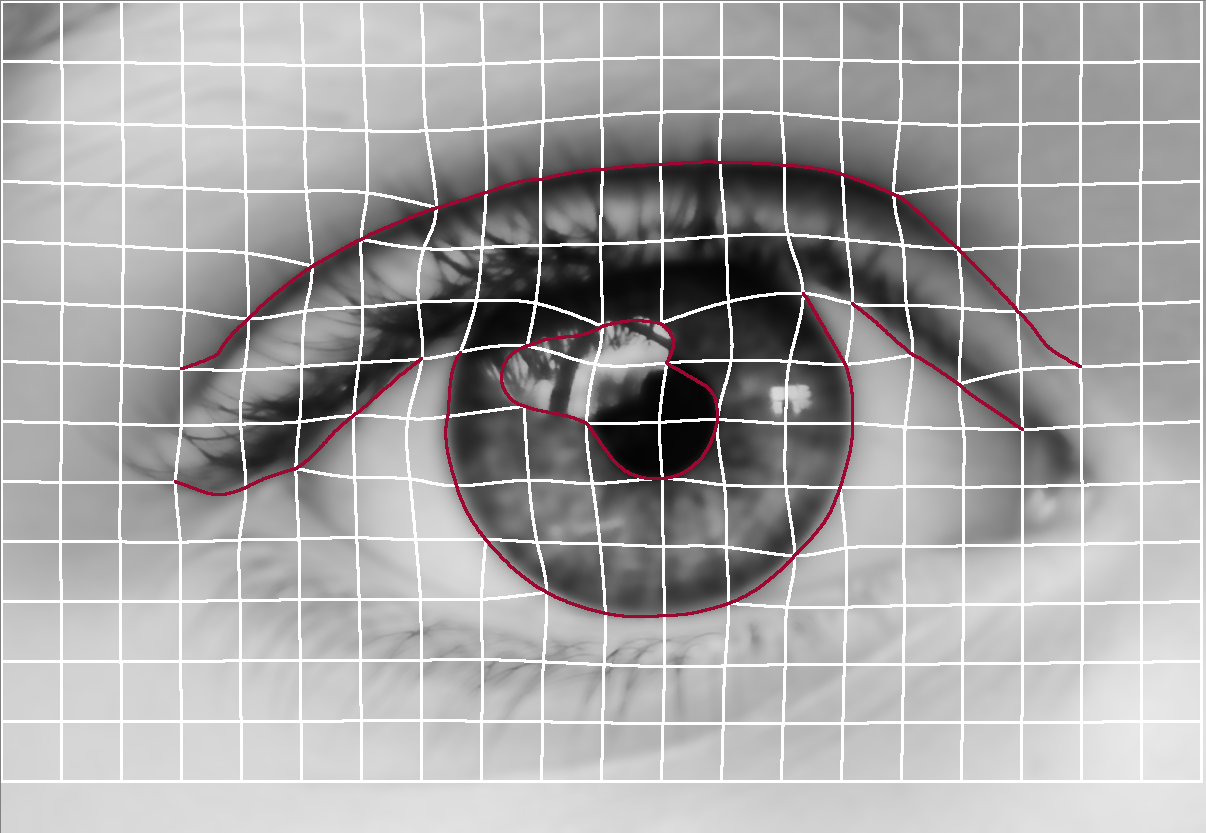}}
\caption{\changed{Selected stages of the grid morphing workflow. We filter the input image (a) using the rolling guidance filter (b) \cite{zhang:2014} to extract large-scale image features, as well as using the bilateral filter (c) \cite{tomasi:1998} to extract higher-frequency features. Canny edge detection \cite{canny:1986} is applied to extract edges, which can be selected using a binary mask. Finally, grid points are snapped to the edges under a fabricability constraint by iteratively solving a mass-spring-system with an additional edge-snapping term using a scalar potential field (d). The output is a smooth, curvilinear grid that follows the user-selected image edges (e).}}
\label{fig:morph_grids}
\end{figure*}

\changed{
\paragraph{A-priori, image-space grid morphing (optional)}
In order to increase the contrast of the fabricated output, it is beneficial to perform a feature-aligned segmentation of the target image that leads to better matching performance than regular grids.
We have implemented a semi-automatic, real-time grid morphing application that allows the user to generate a curvilinear, morphed grid that follows edge features, preserves fabricability, and reflects the user's artistic intent, see \Fig{fig:morph_grids}.}

\changed{The first step is to generate the edge features that the morphed grid should follow.
To this end, we filter the input image using two edge-preserving filters: 
we apply the rolling guidance filter \cite{zhang:2014} to filter out small structures (which should be reconstructed by the template matching, not the overall patch shape) and apply Canny edge detection \cite{canny:1986} to extract the skeletonized edge image $\vec{E}_\text{RG}$.
To extract higher-frequency edges, we smooth the input image using the bilateral filter \cite{tomasi:1998} and again use the Canny edge detector to generate an edge image $\vec{E}_\text{bilateral}$.
We combine the edge images using user-editable masks $\vec{M}_\text{RG}$, $\vec{M}_\text{bilateral}$ to generate the final edge image $\vec{E}$ that our cuts should follow,
\begin{equation}
\vec{E} = \max \left( \vec{M}_\text{RG} \circ \vec{E}_\text{RG}, \vec{M}_\text{bilateral} \circ \vec{E}_\text{bilateral} \right),
\end{equation}
where $\circ$ denotes the Hadamard product.
Given $\vec{E}$, we can compute a scalar potential field that allows us to snap grid vertices to the filtered edges,
\begin{equation}
\vec{P} = \max \left(0, \min \left( D(\vec{E}), r-D(\vec{E}) \right) \right),
\end{equation}
where $D(\vec{E})$ is the distance transform of the binary edge image \cite{felzenszwalb2012distance} and $r$ is the attraction radius of the edges, which is set to half the grid spacing.}

\changed{Next, we snap mesh vertices to the edge image by iteratively solving an energy-minimization problem.
We model the mesh as a damped spring-mass system, where each of the mesh vertices $\vec{x}_{i}$ is connected to its eight neighbors by linear springs, exerting the force
\begin{equation}
\begin{aligned}
\vec{F}(\vec{x}_{i}) &= \sum_{\vec{x}_j \in \mathcal{N}(\vec{x}_i)} m \left( \vec{x}_j - \vec{x}_i \right) - \gamma \frac{\text{d}}{\text{d}t} \vec{x}_i - w \nabla P(\vec{x}_i)\\
&= \vec{F}_{\text{spring}}(\vec{x}_i) + \vec{F}_{\text{damp}}(\vec{x}_i) + \vec{F}_{\text{edge}}(\vec{x}_i),
\end{aligned}
\label{eq:morph_grid_force}
\end{equation}
where $m$ denotes the vertices' mass, $\gamma$ the damping coefficient, and $w$ the weight for the edge-snapping force term.
Here, $\vec{F}_{\text{edge}}$ forces grid vertices to snap to the edges, while the term $\vec{F}_{\text{spring}}$ ensures that vertices adhere to a minimum distance to each other, which in turn enforces fabricability of the final grid.
We find an equilibrium to \Eq{eq:morph_grid_force} using a simple Euler solver.}

\changed{After the positions of the mesh vertices have been determined, we continue to generate the mesh edges, \ie the patch boundaries, by fitting cubic B\'{e}zier curves to the edge image $\vec{E}$.
Due to the edge snapping process, some edges do not follow patch outlines, but run along patch diagonals, in which case we fit an additional edge and split the respective patches into two triangular ones.
In order to generate a smooth appearance, we enforce {G2} continuity while filling in the remaining (not fitted) edges.}

\changed{Together, this preserves the pixelized look of the regular grid, while adding an almost 3D-like appearance due to the curved patch boundaries.}

\changed{
\paragraph{A-posteriori, cost-based patch refinement (optional)}
As an alternative approach to increase the visual fidelity of the final wood puzzle, we introduce an refinement step \emph{after} patch matching.
To this end, the initial patch size needs to be increased in order to produce an overlap between neighboring patches.
This area of overlap can be used to find optimal cuts according to the target image reproduction cost
\begin{equation}
\sum_{x,y} \left( \vec{F}(\vec{R}(x, y)) - \vec{F}(\vec{I}_T(x, y)) \right)^2,
\end{equation}
where $\vec{R}$ denotes the reconstructed wood parquetry image and $\vec{F}$ denotes the filter response from \Sec{sec:feature_extraction}.
Overlapping areas can be shared by either two or four individual source patches.}
For image regions with only two overlapping patches, we obtain an optimum solution using dynamic programming.
For details regarding the implementation of axis-aligned patch merging using dynamic programming see e.g.\@ \cite{Efros01}.
As we enforce  our cuts to be guided by features in the target image, the corresponding, local cost $c(x, y)$ for merging two horizontally neighboring patches $\vec{P}_{S,\{1,2\}}$ along pixel $x$ is given by
\begin{equation}
\begin{split}
c(x, y) = &\sum_{x^\prime=0}^{x-1} \left( \vec{F}(\vec{P}_{S,1}(x^\prime,y)) - \vec{F}(\vec{P}_T(x^\prime,y)) \right)^2 +\\
&\sum_{x^\prime=x}^{n-1} \left( \vec{F}(\vec{P}_{S,2}(x^\prime,y)) - \vec{F}(\vec{P}_T(x^\prime,y)) \right)^2,
\end{split}
\end{equation}
where $\vec{P}_T \subset \vec{I}_T$ and $n$ is the size of the overlap. We assign patch $\vec{P}_{S,1}$ to the region left of the cut and $\vec{P}_{S,2}$ to the remaining region.
By approaching this problem using dynamic programming, we enforce 6-connectivity of the cut and in turn physical fabricability.
Vertically neighboring patches can be aligned in an analogous manner.

In regions where four patches overlap, we have to find two intersecting cuts, one for the horizontal and one for the vertical direction.
This prevents cut optimization via dynamic programming and instead, we find an approximate solution by alternating optimizations for one cut direction while keeping the other direction fixed.
We experimentally observed two repetitions of this process to be sufficient.

In order to generate a representation that is laser-cuttable, we fit cubic B\'{e}zier curve segments to the cuts, where the user can choose between {G0} continuous and {G1} continuous curve segments.
Finally, the output of this step is a vector graphics file containing cut instructions which can be directly executed by the laser cutter.

\subsection{Fabrication}
In the next step, the optimized, still digital piece of parquetry is physically fabricated.
To this end, we use a laser cutter for cutting the veneer boards from the back side and for engraving IDs which facilitate the identification of individual patches during their assembly.
For other materials, this step could also be conducted using a CNC mill or a water jet cutter.
The patches are separated from the rest of the veneer and laid out in a frame.
To fix the patches, we attach a back plate using wood putty.
After the putty has dried, we sand the veneers and finish them with clear coat or hard wax oil.

\subsection{Implementation details}
The method was implemented in C\texttt{++} using the OpenCV library \cite{opencv_library} and parallelized with OpenMP.
Fitting a single patch typically takes \SIrange{0.5}{3}{s} on an Intel Core i7-5820K CPU, where the runtime is dominated by template matching.
Thus, the runtime primarily depends on the number of pixels per patch, and on the size of the wood samples tested.

During our experiments, we used a Plustek OpticPro A360 Plus flatbed scanner for A3-sized veneer boards, and a Cruse Synchron Table Scanner 4.0 for scanning larger panels.
The fabrication (cutting) was performed on a Trotec Rayjet with a \SI{12}{W} CO$_2$ laser and an Epilog Fusion 40 M2 engraver with a \SI{75}{W} CO$_2$ laser.
\begin{figure*}[t]
\includegraphics[width=0.19\linewidth,valign=t]{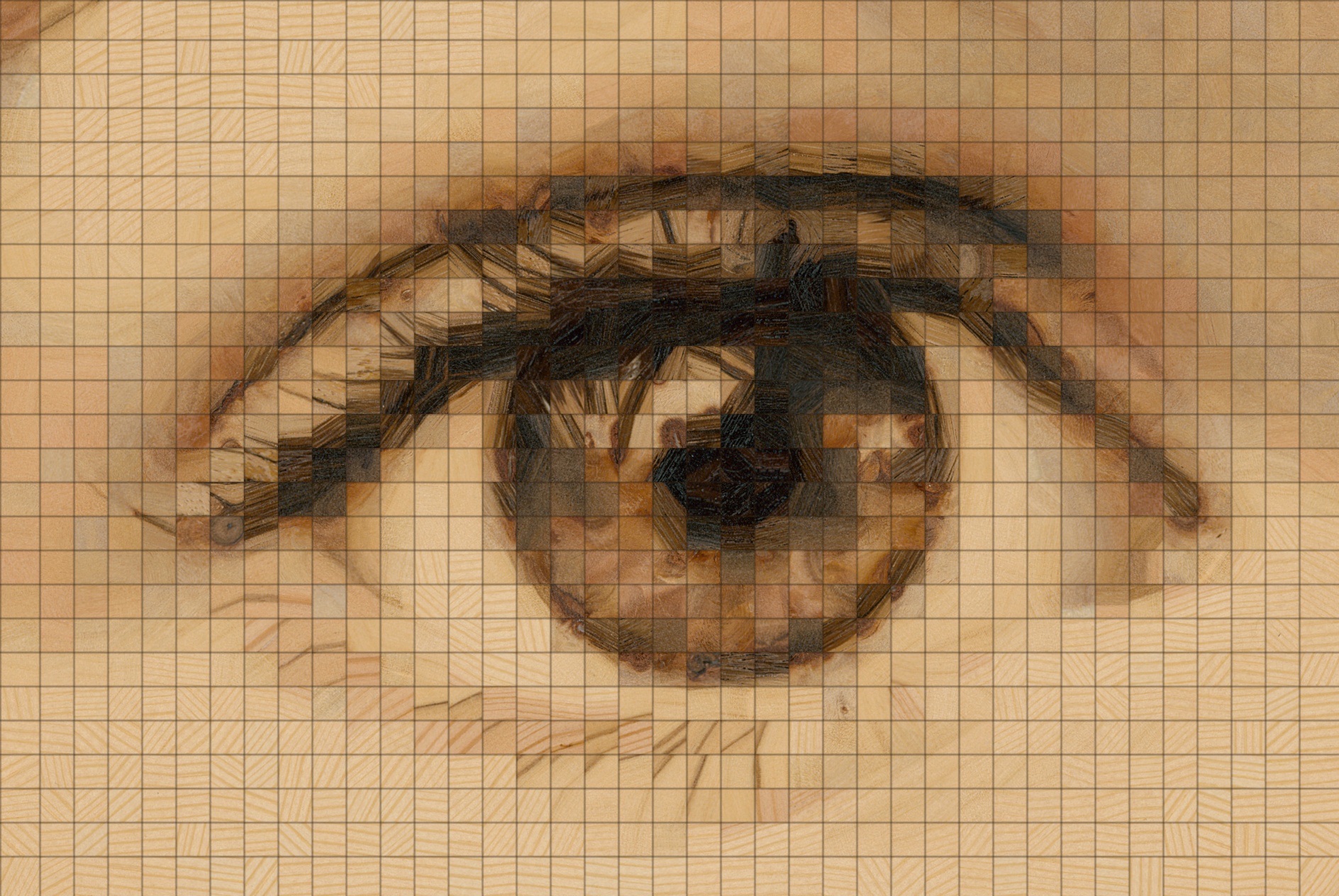}
\hfill
\includegraphics[width=0.19\linewidth,valign=t]{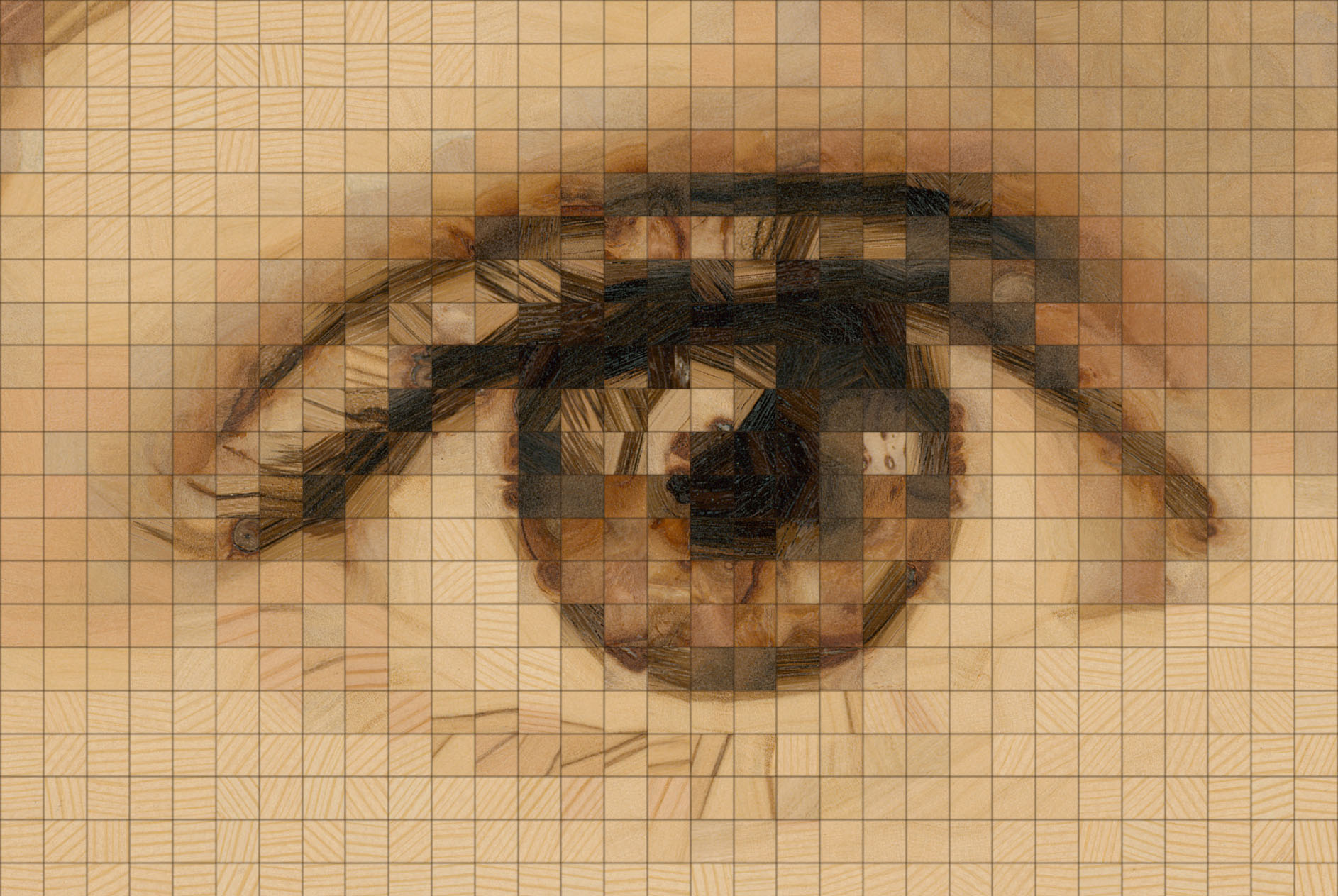}
\hfill
\includegraphics[width=0.19\linewidth,valign=t]{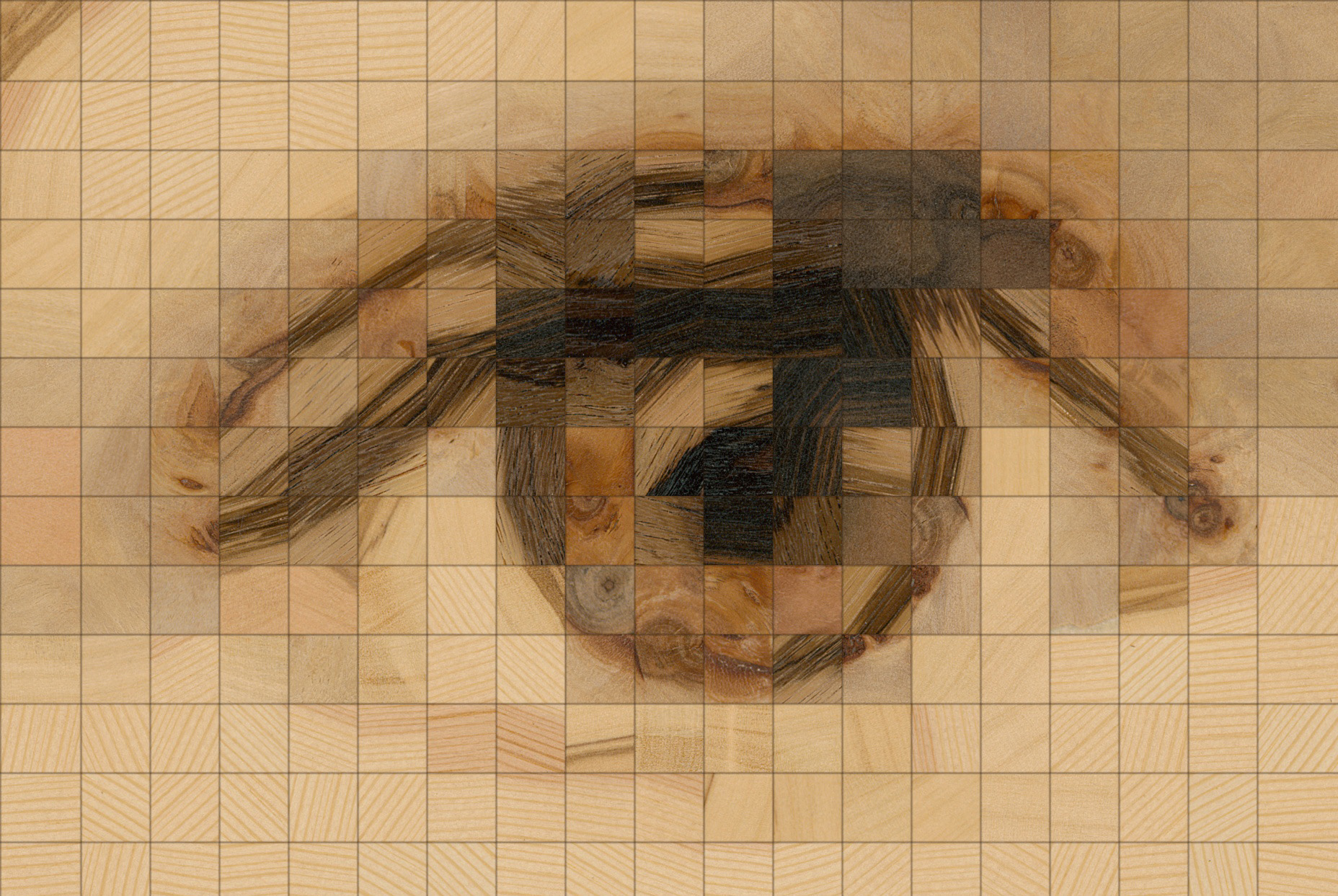}
\hfill
\includegraphics[width=0.19\linewidth,valign=t]{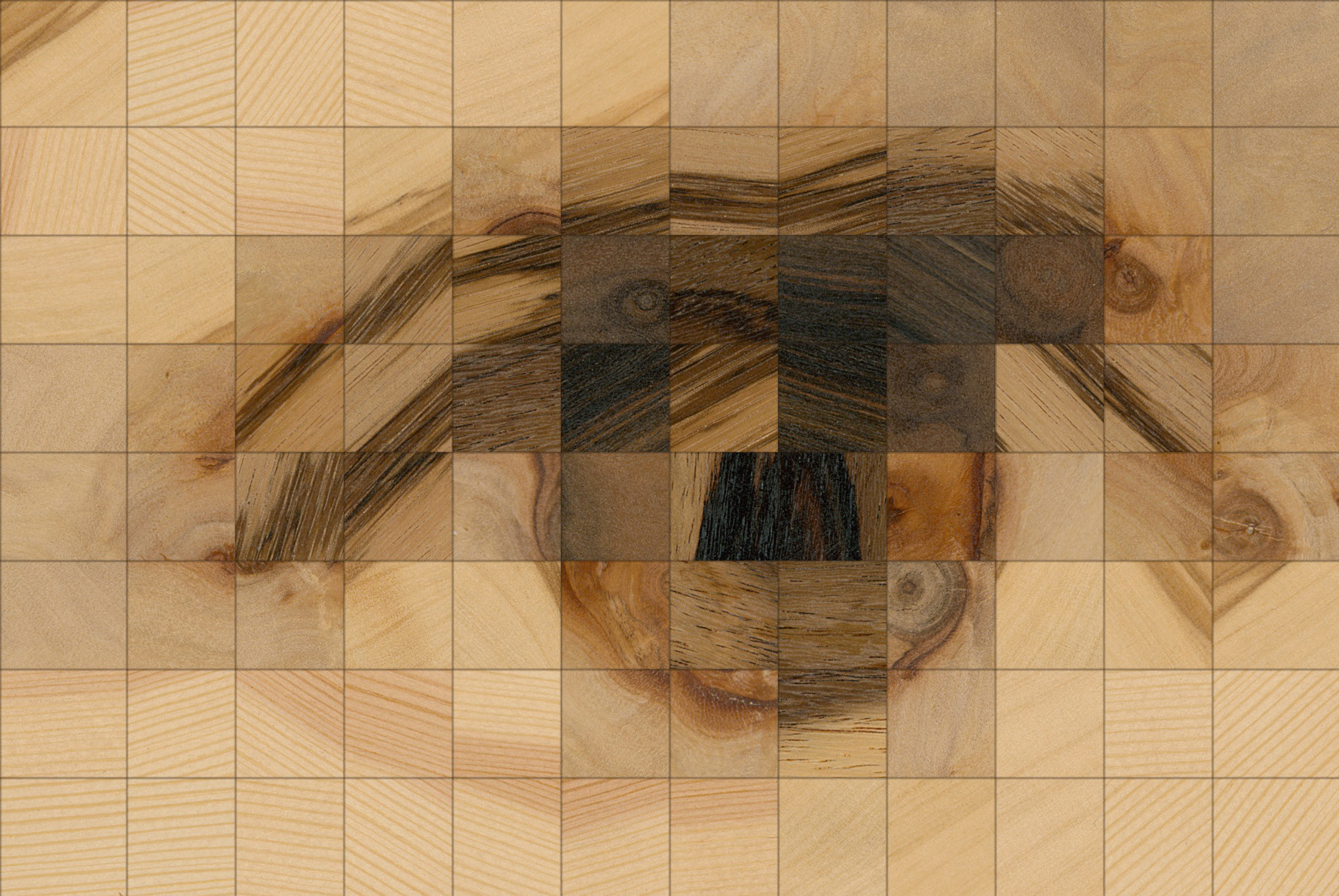}
\hfill
\includegraphics[width=0.19\linewidth,valign=t]{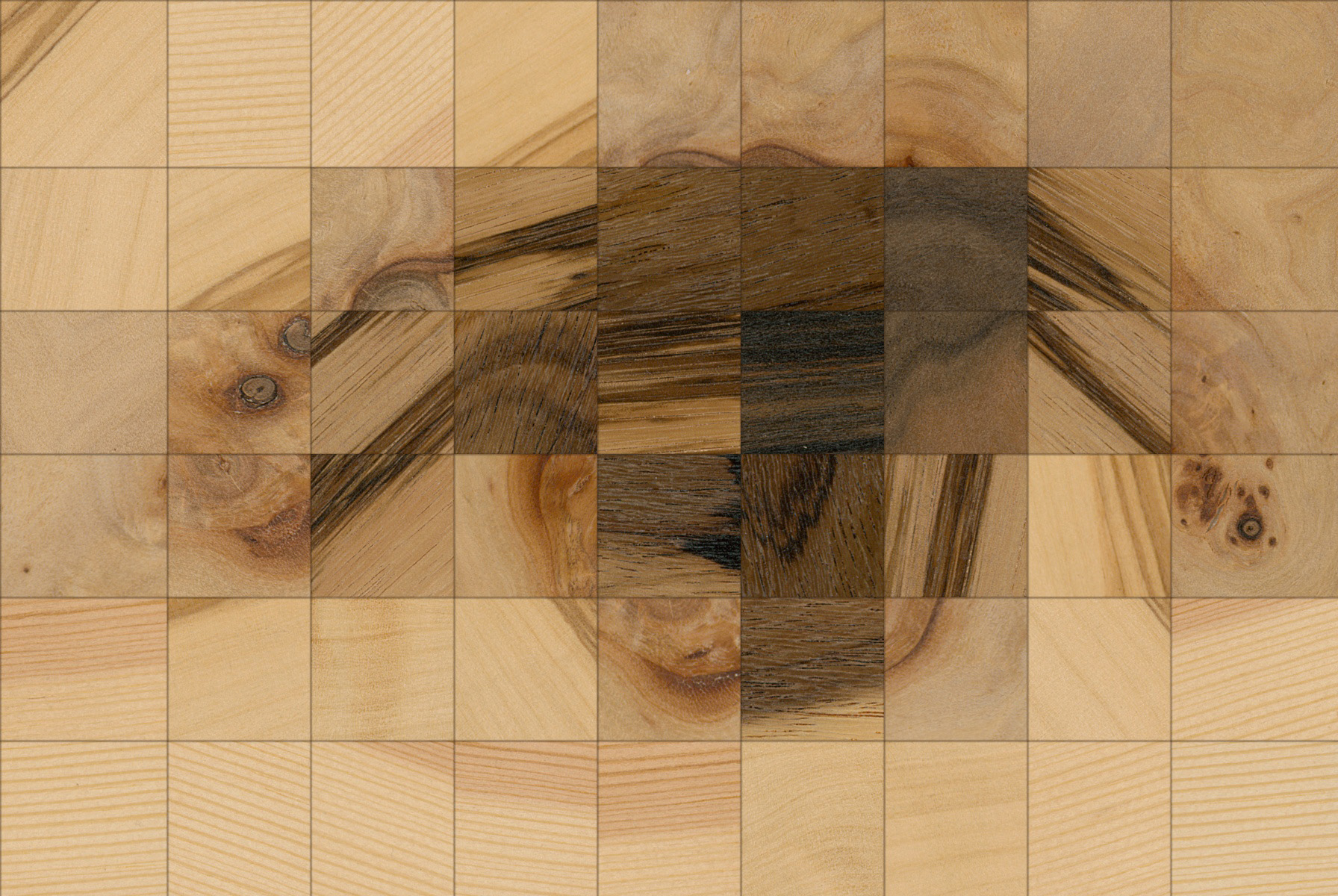}\\
\vspace{1.4mm}
\includegraphics[width=0.19\linewidth,valign=t]{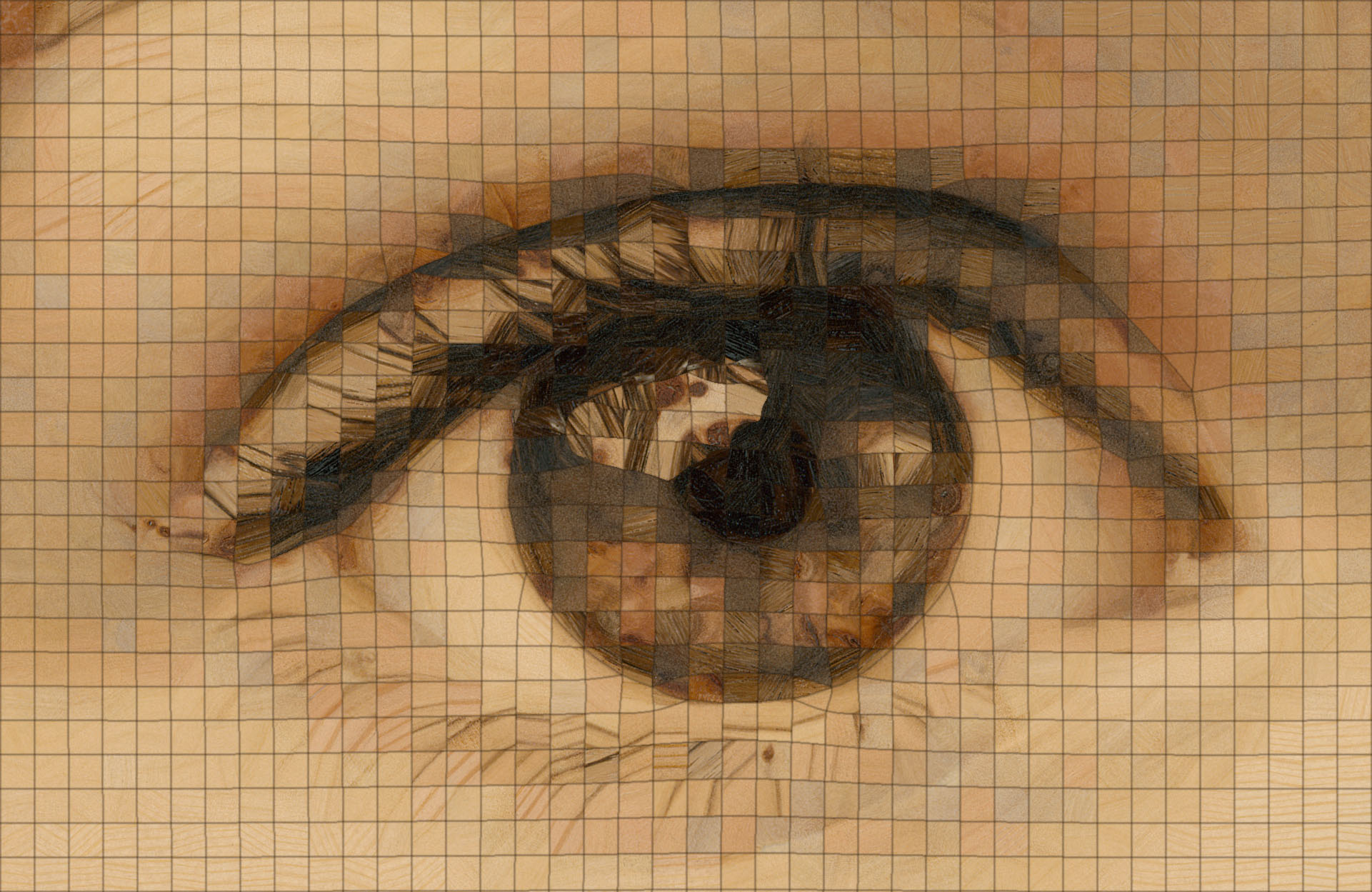}
\hfill
\includegraphics[width=0.19\linewidth,valign=t]{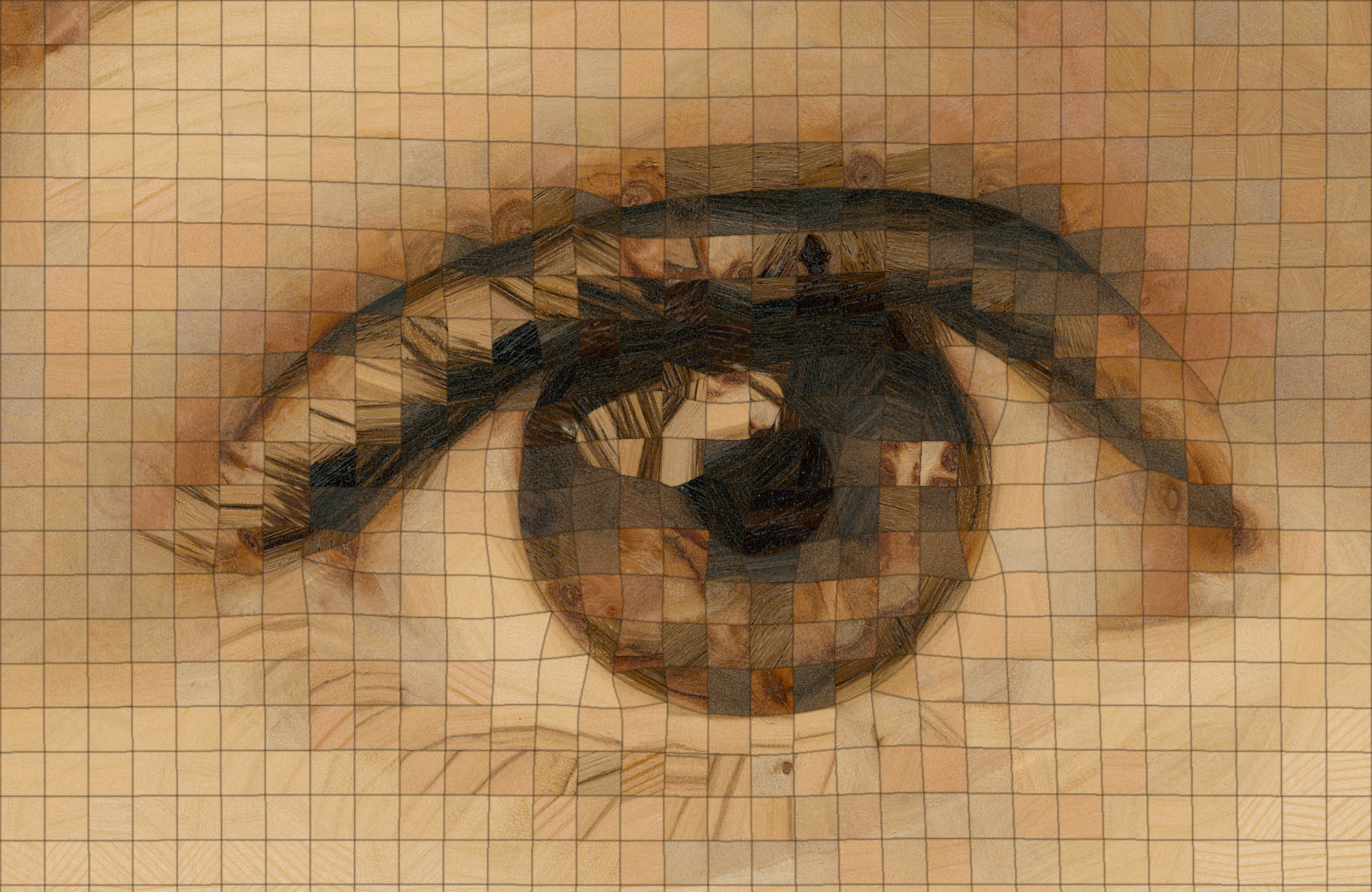}
\hfill
\includegraphics[width=0.19\linewidth,valign=t]{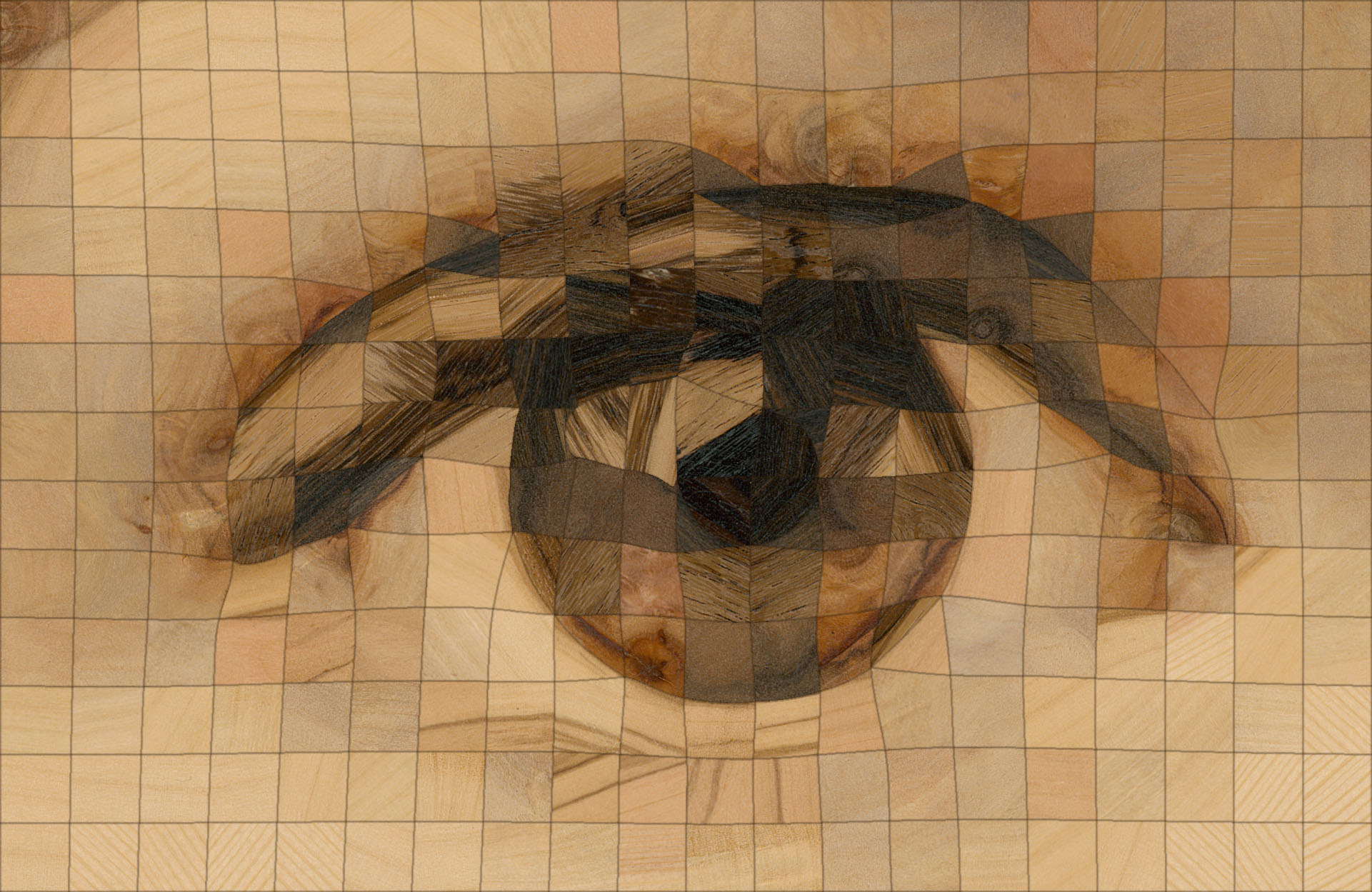}
\hfill
\includegraphics[width=0.19\linewidth,valign=t]{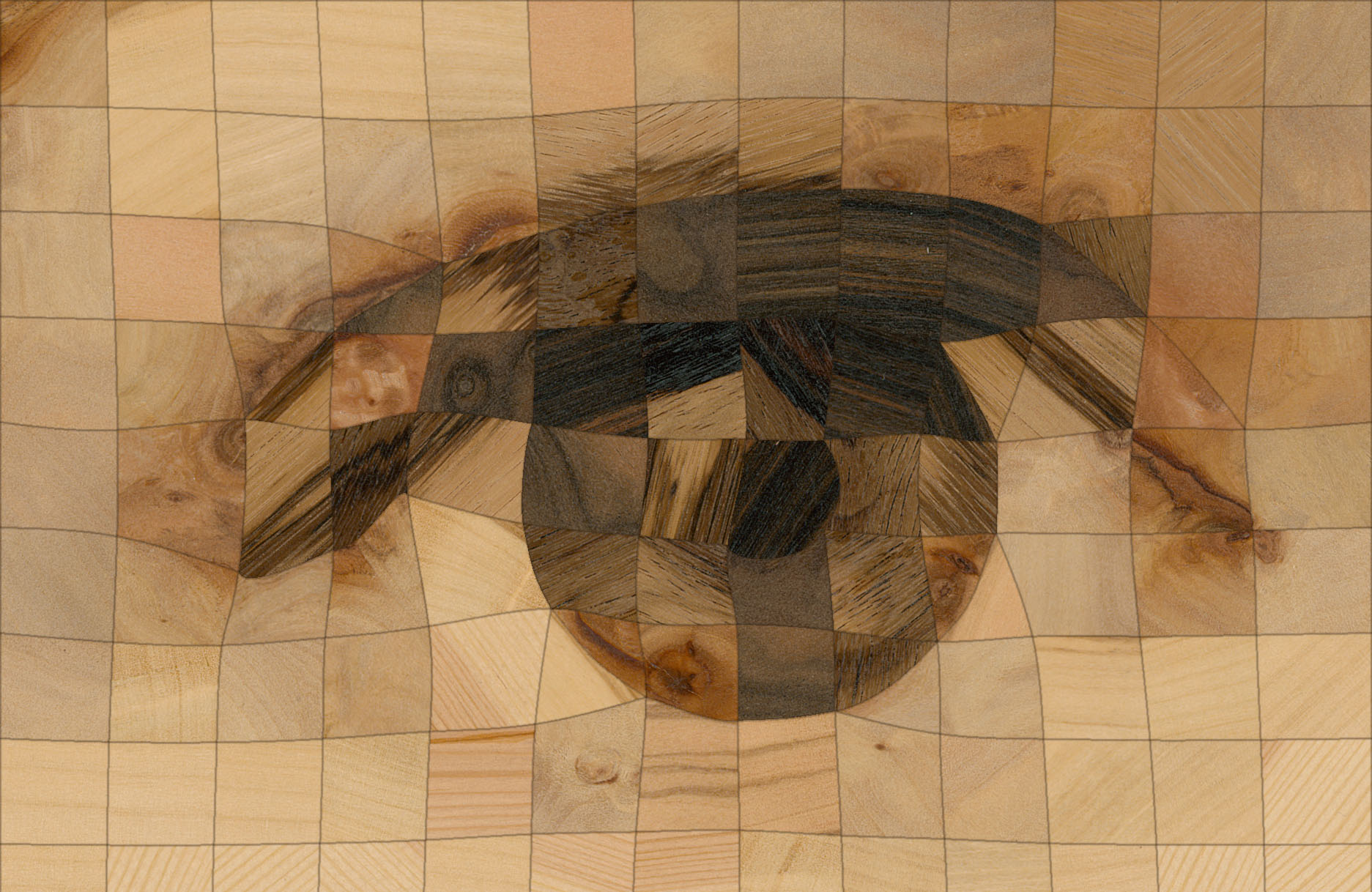}
\hfill
\includegraphics[width=0.19\linewidth,valign=t]{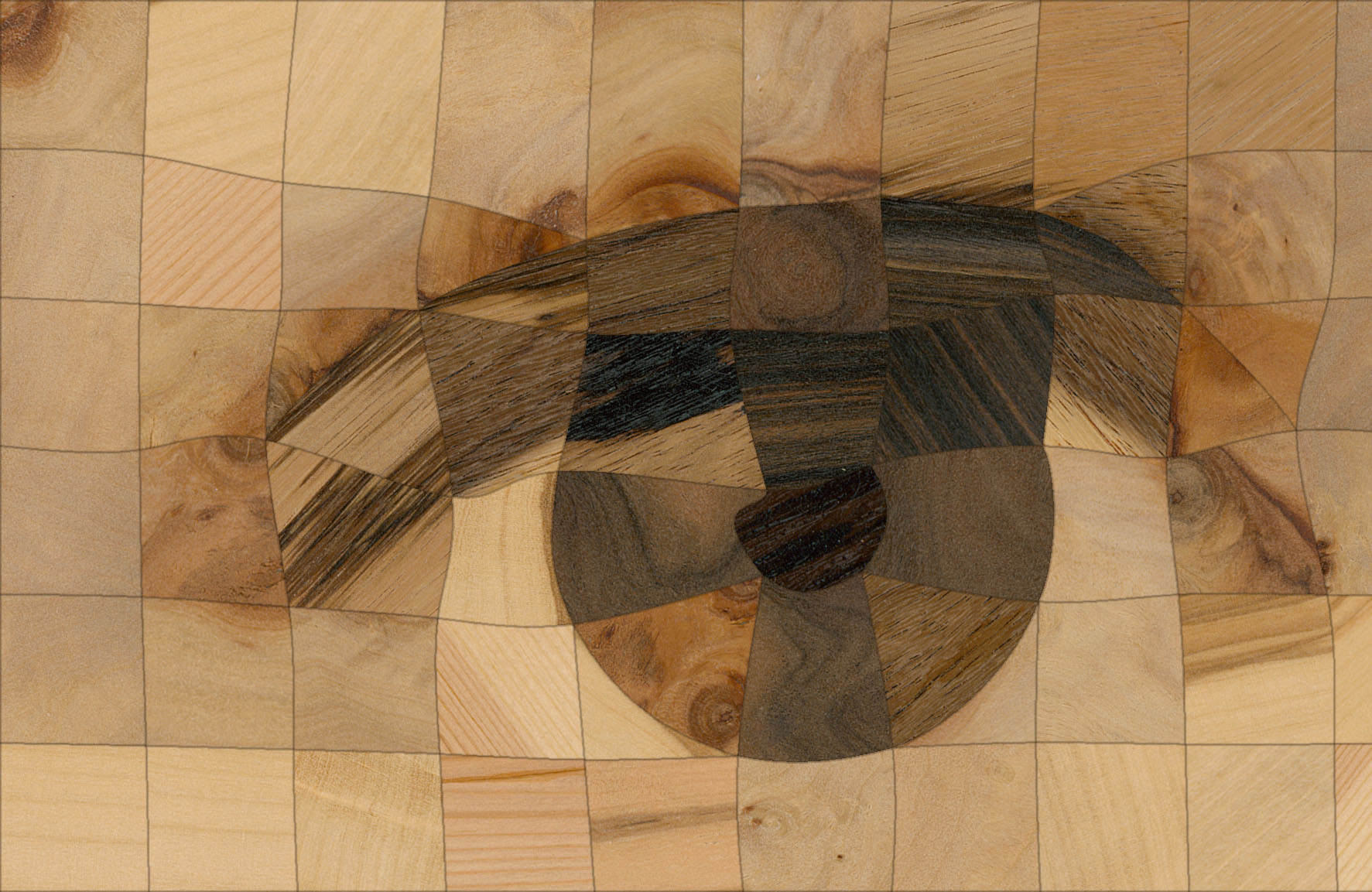}\\
\vspace{1.4mm}
\includegraphics[width=0.19\linewidth,valign=t]{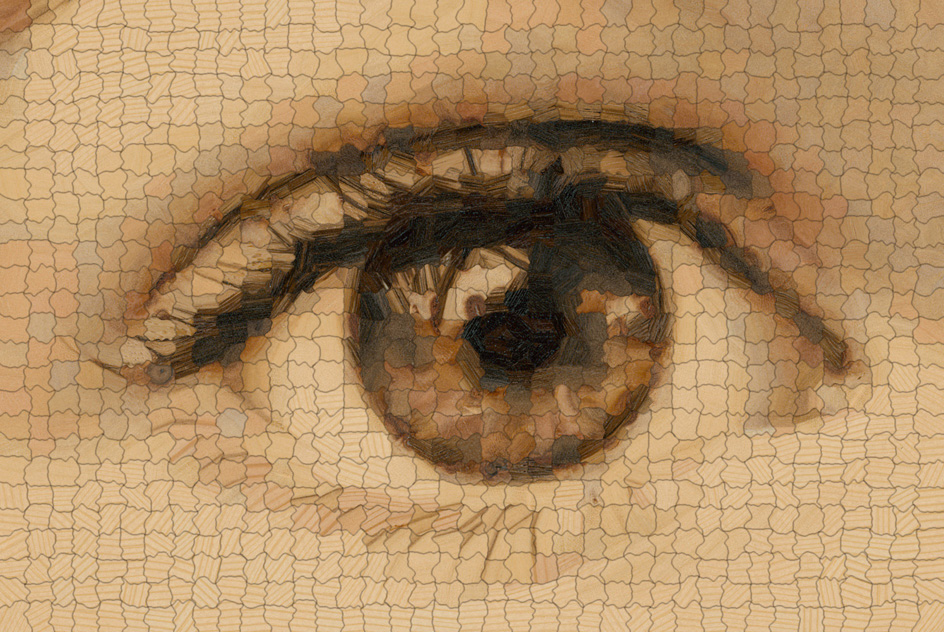}
\hfill
\includegraphics[width=0.19\linewidth,valign=t]{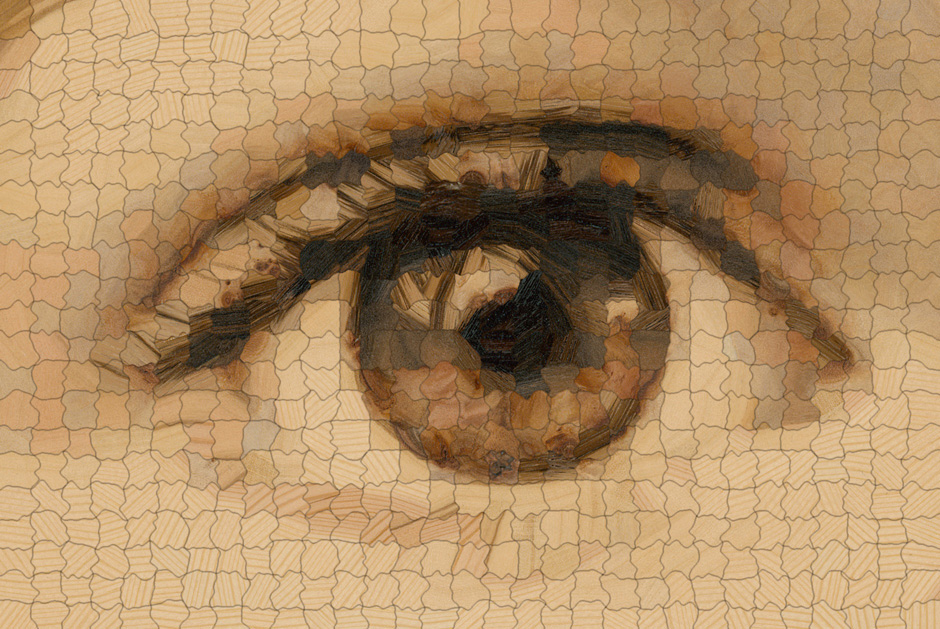}
\hfill
\includegraphics[width=0.19\linewidth,valign=t]{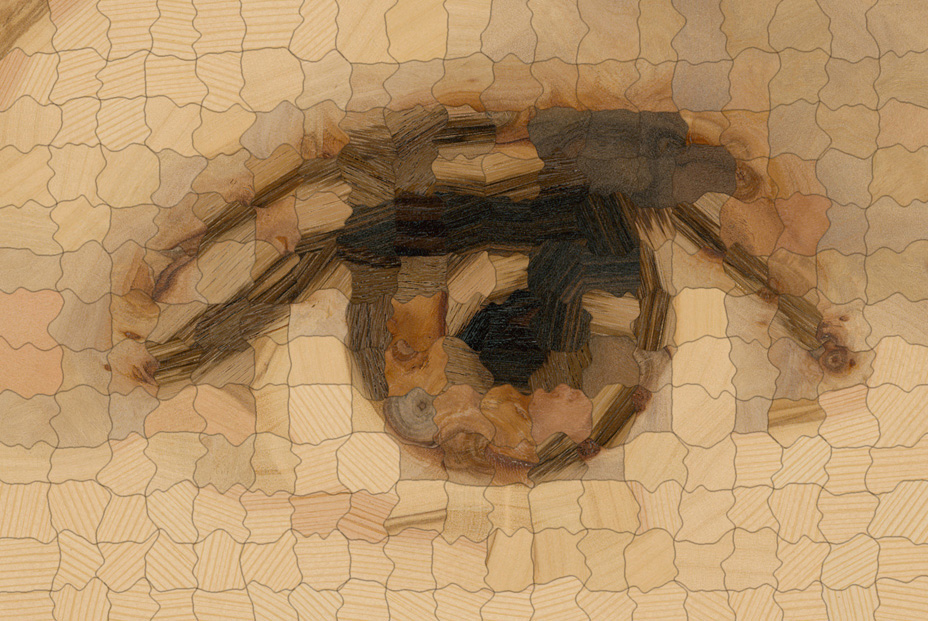}
\hfill
\includegraphics[width=0.19\linewidth,valign=t]{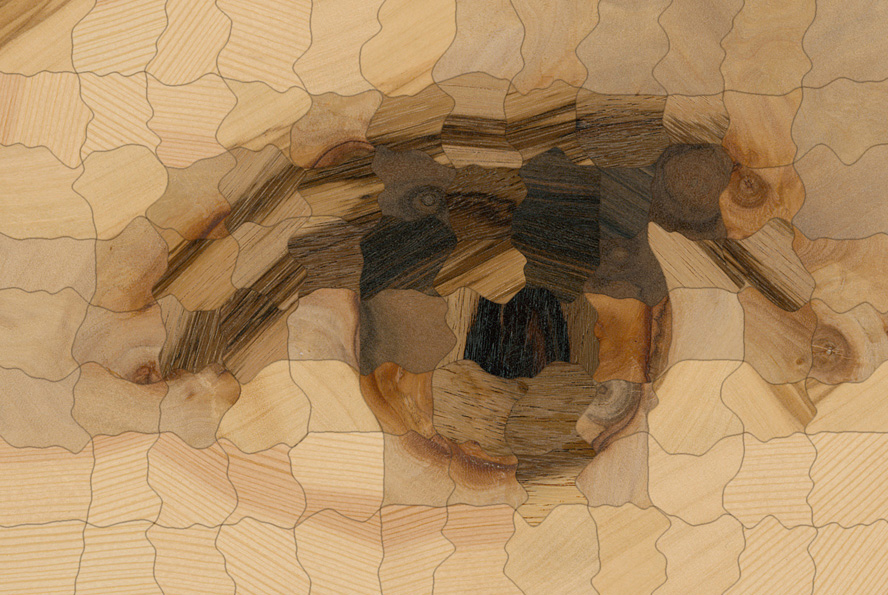}
\hfill
\includegraphics[width=0.19\linewidth,valign=t]{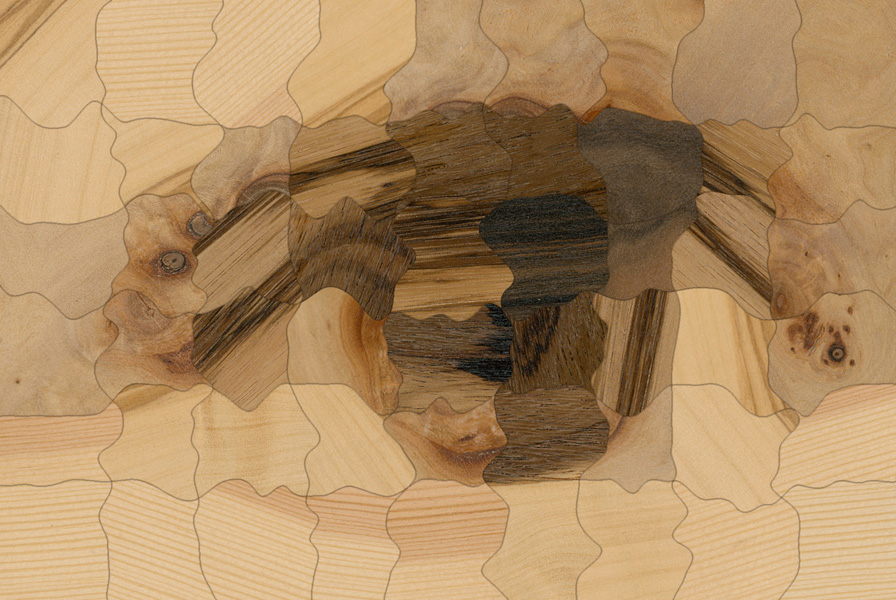}\\
\vspace{1.4mm}
\includegraphics[width=0.19\linewidth]{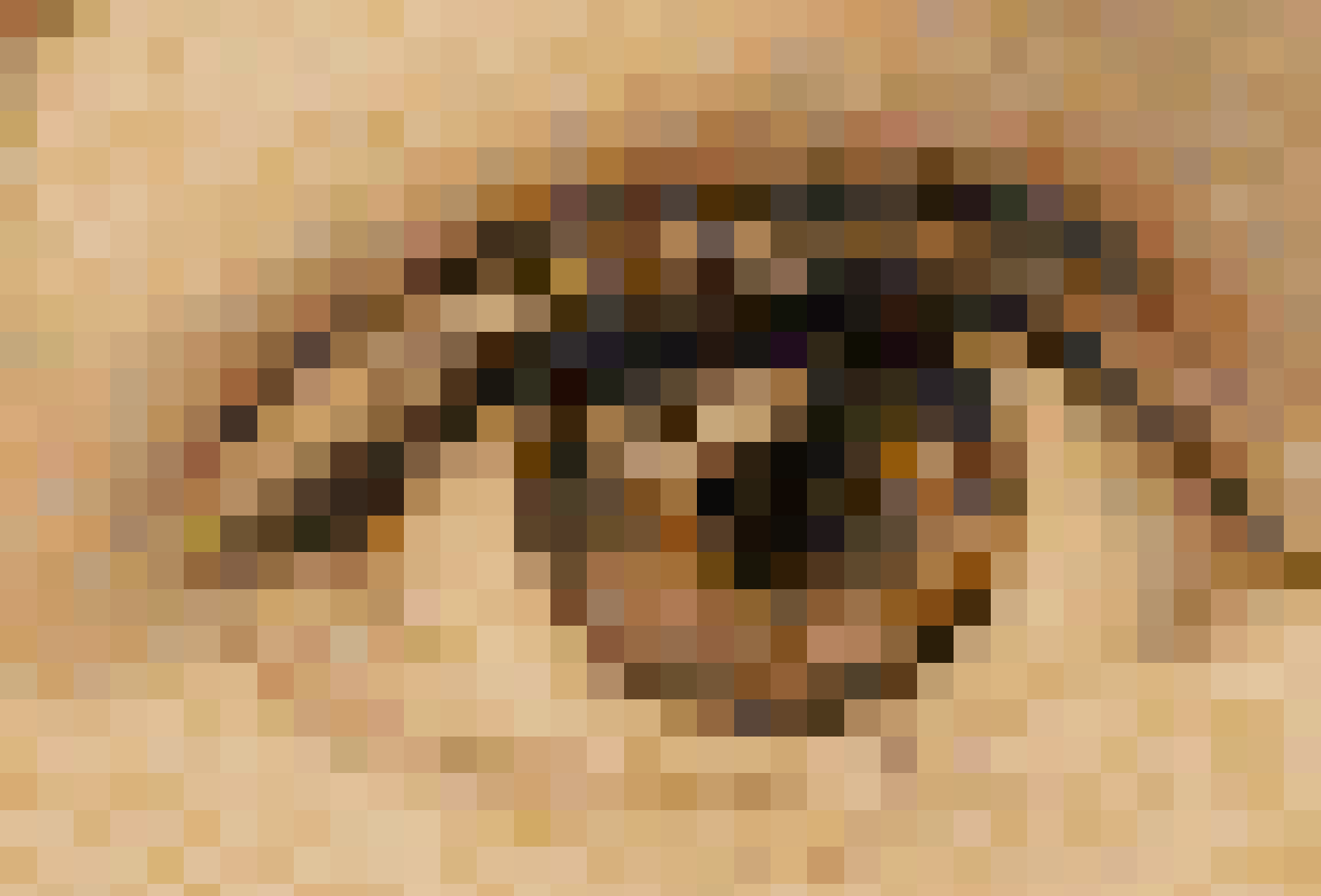}
\hfill
\includegraphics[width=0.19\linewidth]{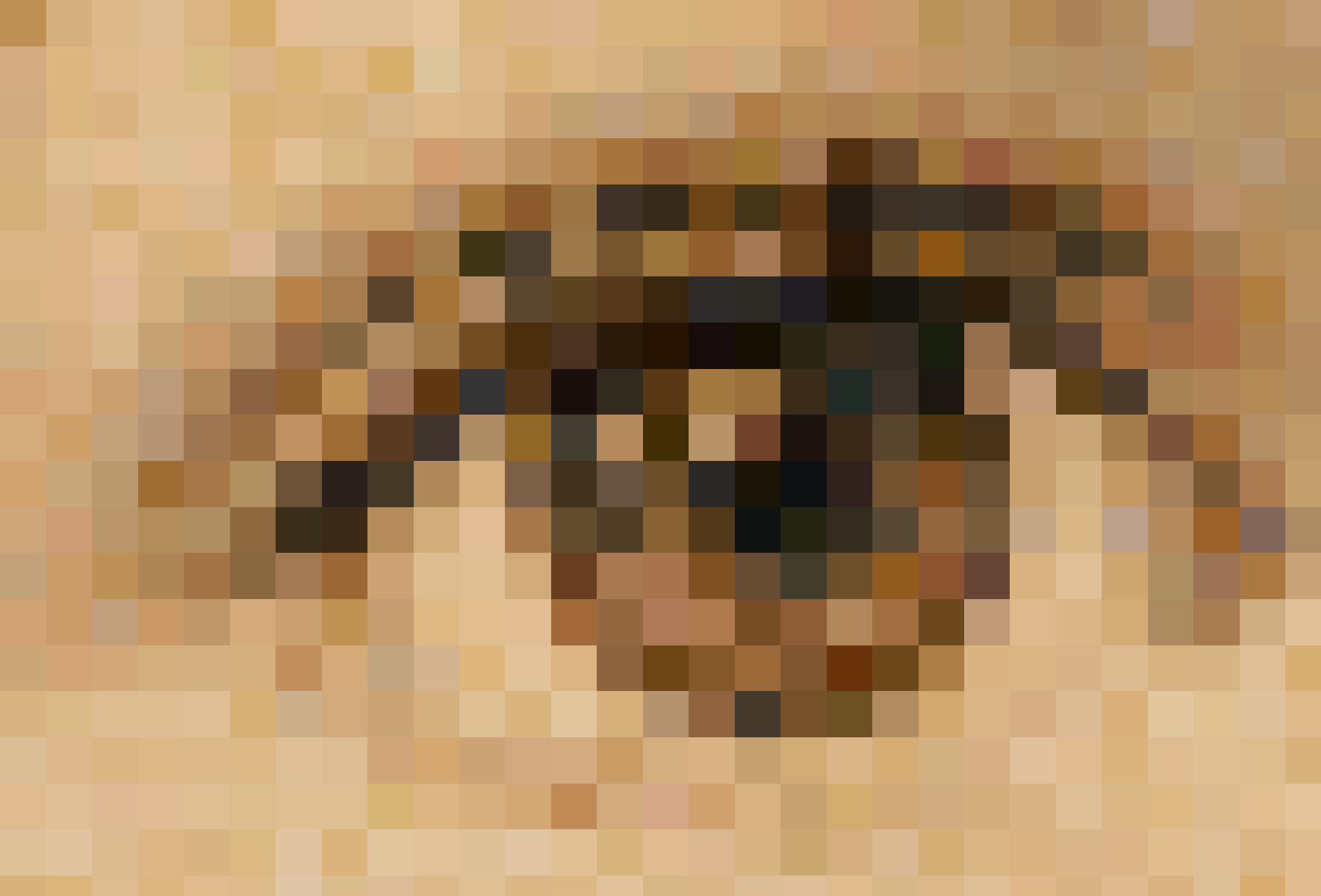}
\hfill
\includegraphics[width=0.19\linewidth]{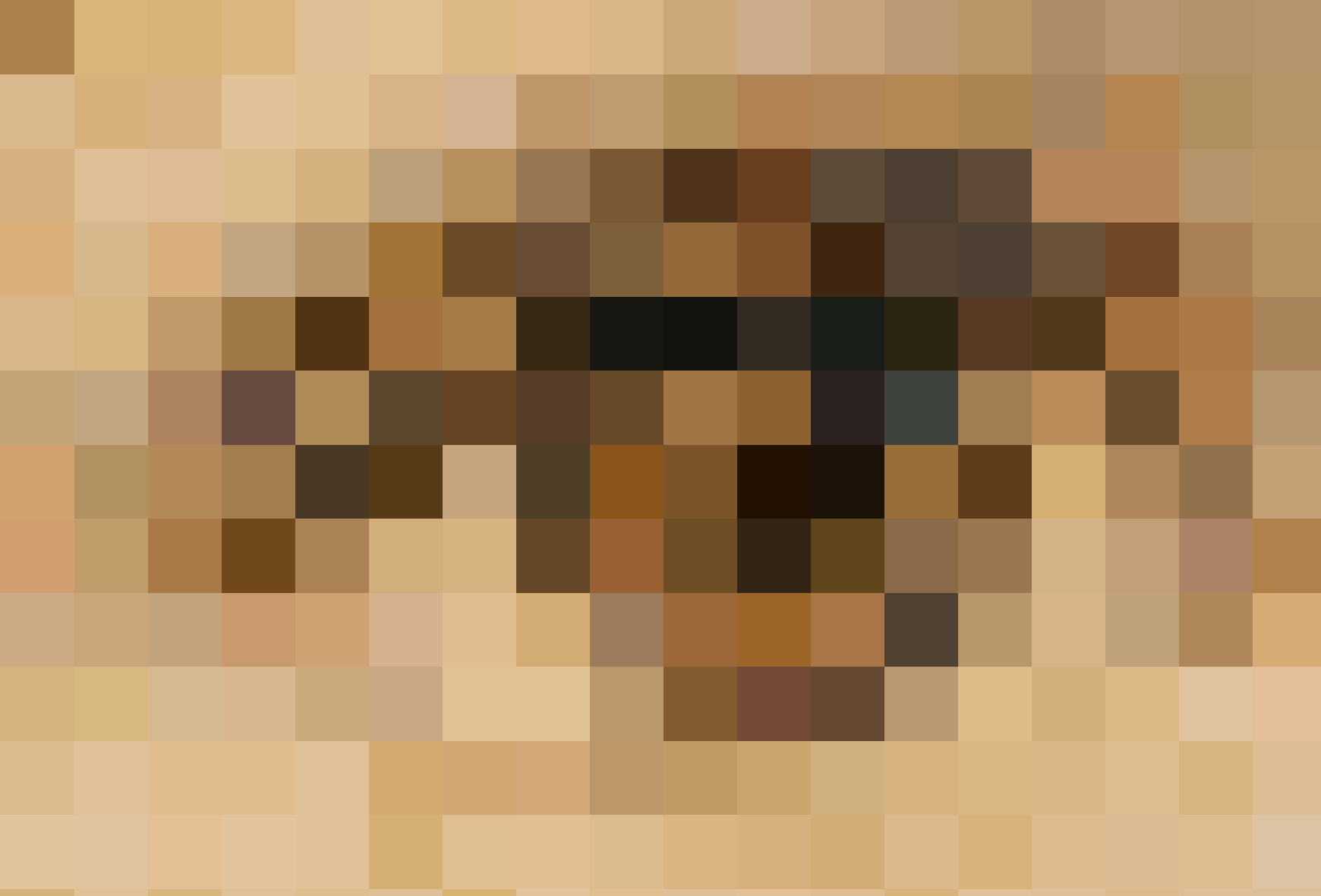}
\hfill
\includegraphics[width=0.19\linewidth]{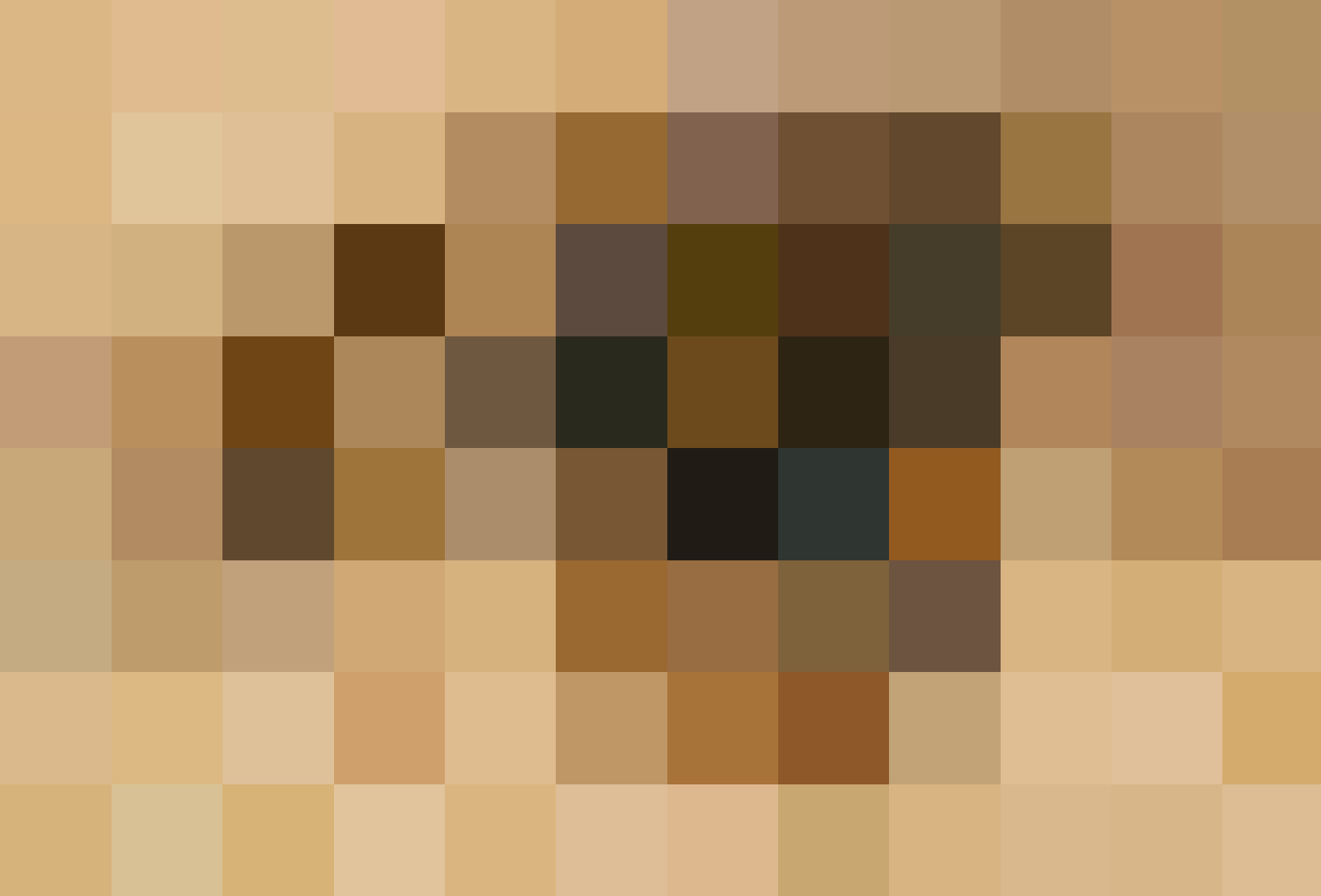}
\hfill
\includegraphics[width=0.19\linewidth]{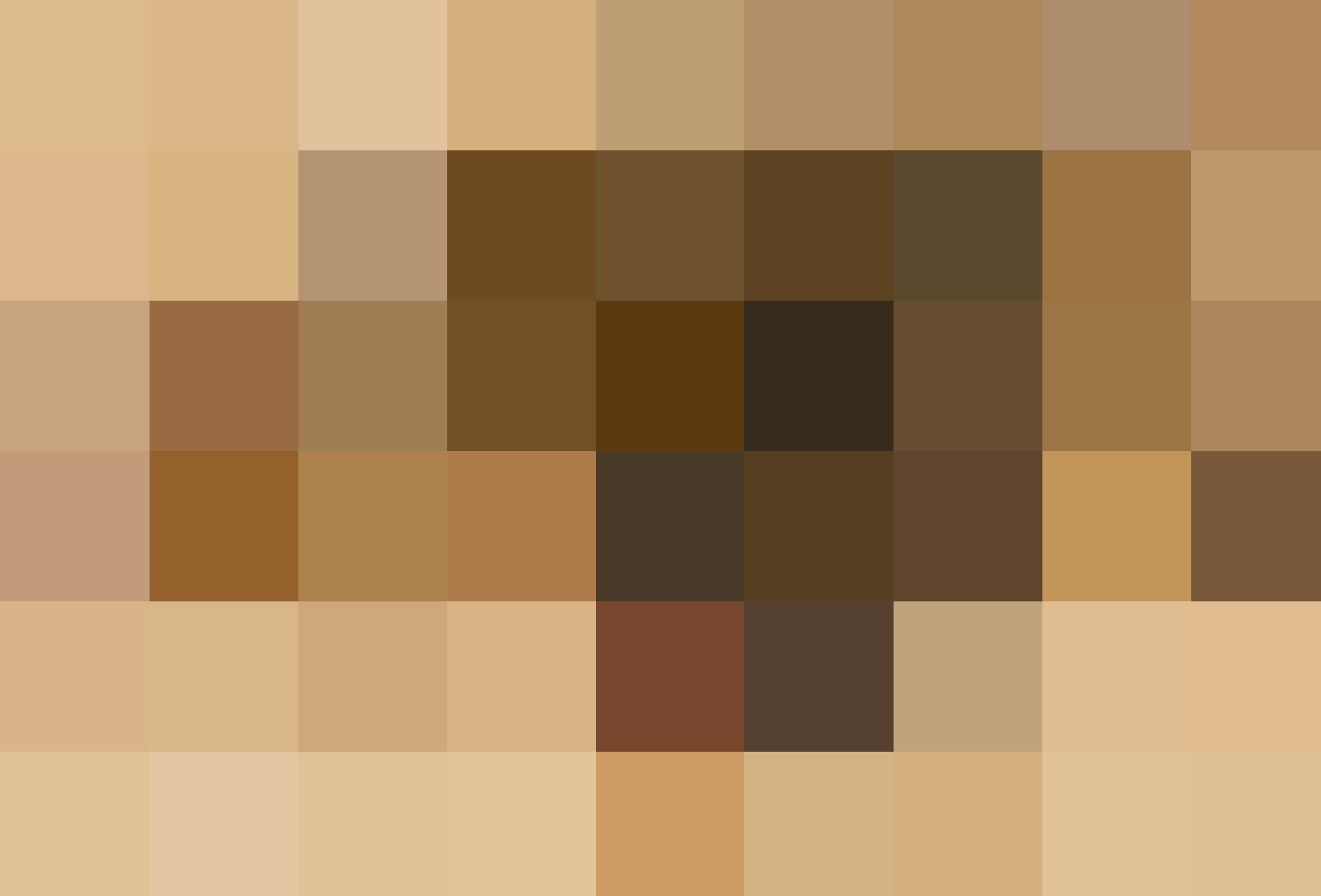}\\
\caption{\changed{\textbf{Synthetic} Effect of different resolutions and refinement strategies on the reconstruction quality. We show reconstructions obtained using regular grids (top row), the \emph{a-priori} grid morphing refinement (second row), the \emph{a-posteriori} cost-based patch refinement (third row), and a ``baseline'' where high frequency features are removed and each patch is replaced by its mean color (bottom row). With decreasing resolution (from left to right), we observe that the structurally aware filters, as well as the refinement schemes, are having an increased influence on reconstruction quality. The reconstruction quality obtained with our proposed technique gracefully declines with patch resolution and still produces visually pleasing results for very coarse patches.}}
\label{fig:resolution}
\end{figure*}

\section{Results}
\label{sec:results}
We begin our evaluation with the analysis of user-controllable design choices in the optimization, such as the effect of different energy terms, different sizes and shapes of the individual patches.
This is followed by an ablation study, where we investigate the gradual decline in quality that occurs when repeatedly producing the same target image from the same wood veneer panel.
We further demonstrate a few examples of fabricated parquetry obtained from different woods and under different conditions.
Finally, we show the robustness of our method with respect to different target images by presenting synthetic results for different targets, each optimized using the default parameter set.

\begin{table}[h]\small
\begin{tabular}{l|c|c}
Symbol & Parameter & Default\\
\hline
$w_\text{intens}$ & Intensity priority weight & $0.5$\\
$w_\text{edge}$ & Edge priority weight & $0.5$\\
$w_\text{hist}$ & Histogram matching weight & $0.5$\\
$s_\text{image}$ & Reconstructed image size (shorter axis) & $\SI{360}{mm}$\\
$s_\text{patch}$ & Patch size & $\SI{14}{mm}$\\
$n_\text{adaptive}$ & Adaptive patch levels & $0$\\
$w_\text{adaptive}$ & Adaptive patch quality factor & $1.2$
\end{tabular}
\caption{User-controllable stylization parameters and their default values.}
\label{tab:user_parameters}
\end{table}
\subsection{User-controlled stylization}
\begin{figure}[t]
\includegraphics[width=0.8\columnwidth]{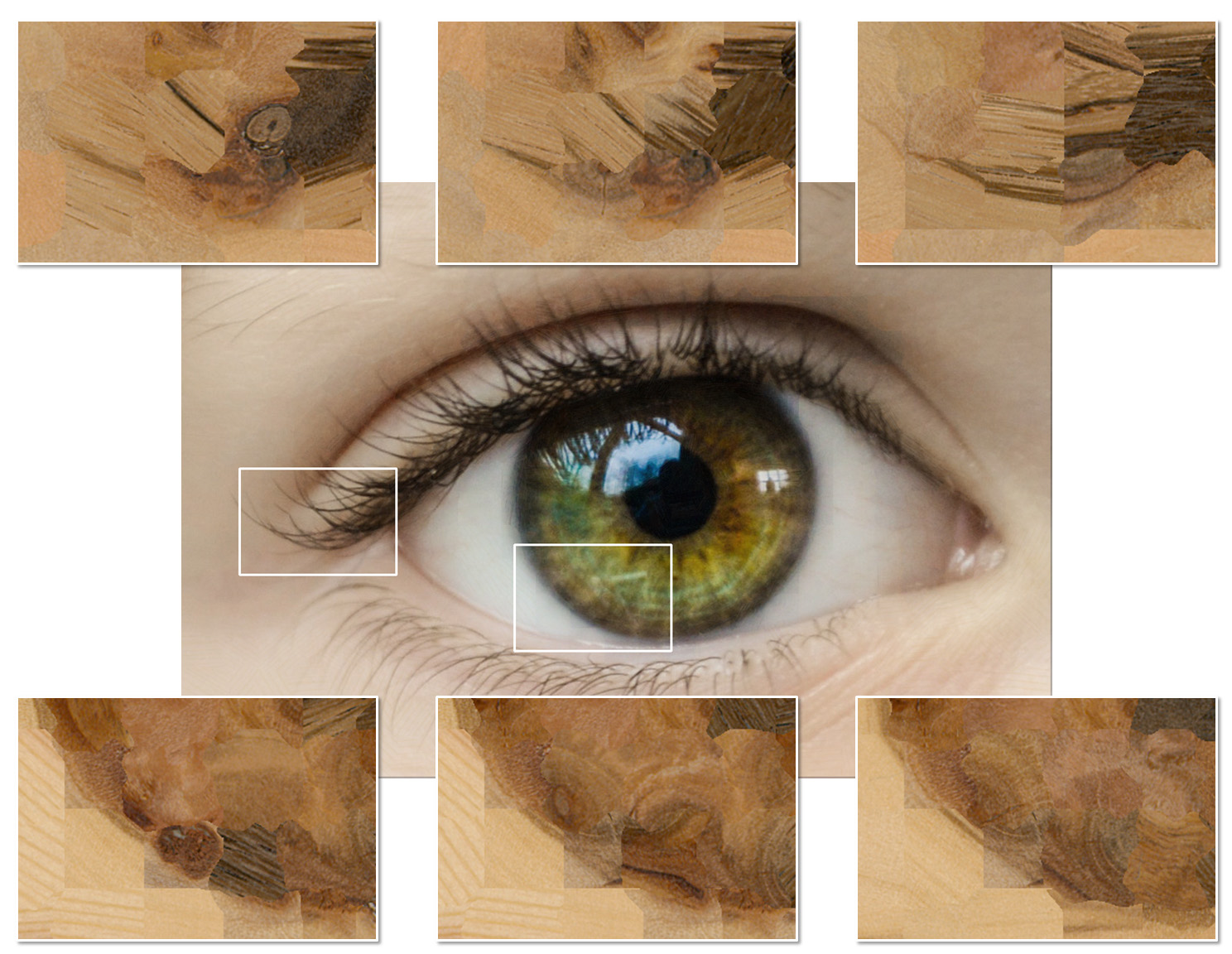}
\caption{\changed{\textbf{Synthetic}} Effect of intensity vs.\@ edge filter. The highlighted zoom-ins depict the respective reconstructed regions for weights $(w_\text{intens}, w_\text{edge})$: $(1.0, 0.0)$, $(0.5, 0.5)$, and $(0.2, 0.8)$ from left to right. Using only intensity penalty enforces the stylization to match intensity. Structural details become increasingly well preserved with an increasing weight of the edge term.}
\label{fig:sobel_effect}
\end{figure}

Our method allows the stylization of the generated renditions of target images based on user guidance.
Before discussing the effect of individual user-controllable parameter choices on the style of the generated renditions, we first provide insights regarding the involved physical materials.
We found  an image of a human eye (\Fig{fig:sobel_effect}) to be a good target for quality assessment, because it contains features with different frequencies, as well as rounded structures.
An overview over the user-controllable parameters related to stylization can be found in \Tab{tab:user_parameters} and a more detailed description in \Sec{sec:method}.

\paragraph{Materials}
\begin{figure}[t]
\includegraphics[width=\columnwidth]{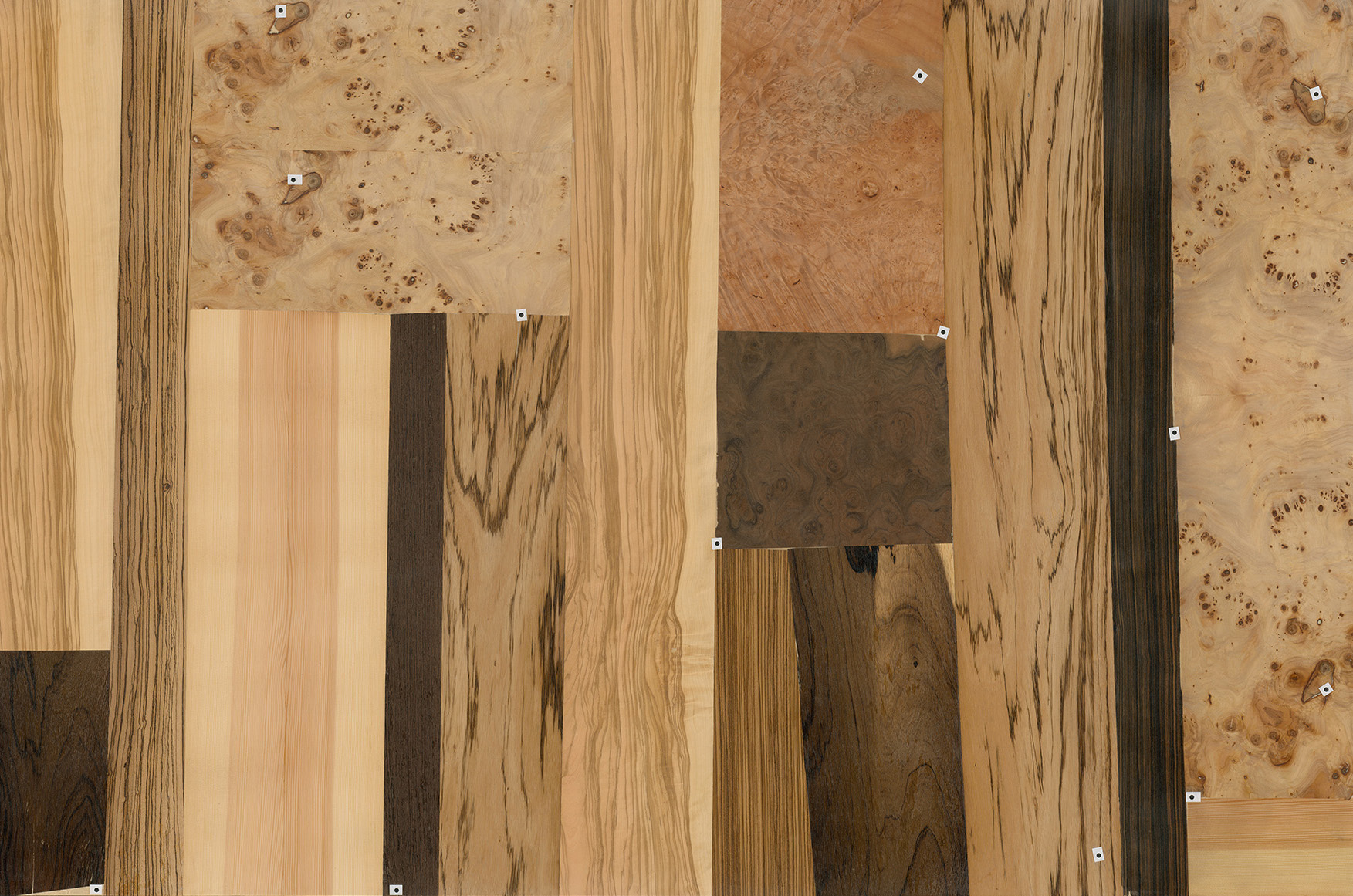}
\caption{Scan of the wooden veneer panel used for the results in \Sec{sec:results}. The panel has physical dimensions of \SI{1500 x 1000}{mm} and contains veneer samples from various wood types. The fiducial markers facilitate optical calibration on suitably equipped cutting systems.}
\label{fig:input_results}
\end{figure}
For the purpose of a better comparability, we generated synthetic renderings using the same scan of a wooden veneer panel as input for all results in this section (unless otherwise noted).
The panel has a size of \SI{1500 x 1000}{mm} and contains veneer samples from various wood types.
The woods used in our experiments are not protected under CITES. They include maple burl, ash burl, poplar burl, buckeye burl, elm burl, birch burl, walnut burl, pine, wenge, santos rosewood, olive tree, makassar ebony, apple tree, and zebrawood.
We sanded the panel and applied a layer of clear coat to enhance the contrast of the individual fiber strands.
The physical sample was scanned at \SI{300}{dpi} using a Cruse Synchron Table Scanner 4.0.
A downscaled version of the scan can be found in \Fig{fig:input_results}.

\paragraph{Histogram matching}
\begin{figure}[t]
\includegraphics[width=0.49\columnwidth]{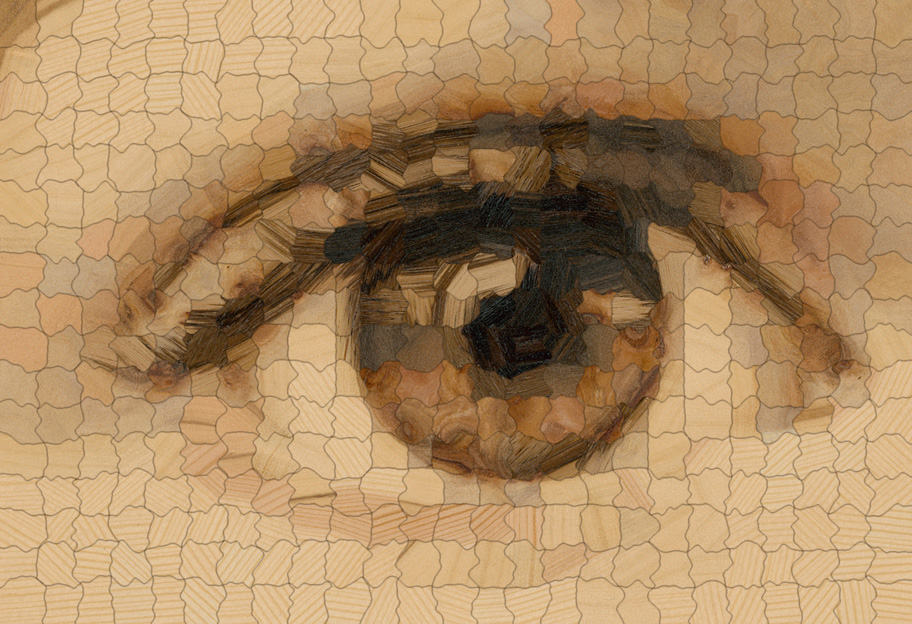}
\hfill
\includegraphics[width=0.49\columnwidth]{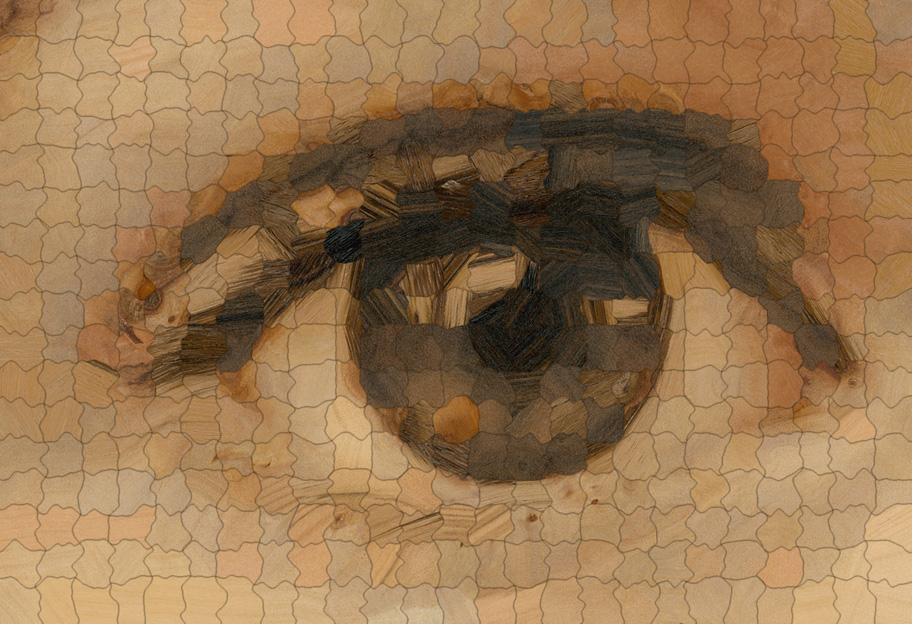}
\caption{\changed{\textbf{Synthetic}} The effect of histogram matching. Without histogram matching ($w_\text{hist}=0$, left), we obtain a higher contrast. With histogram matching ($w_\text{hist}=1$, right), the contrast is reduced, but the shading appears less flat.}
\label{fig:histogram_equalization_eye}
\end{figure}
The target image gamut is generally larger than the gamut of the wood textures. Without taking this into account, the template matching step will generally draw patches from the gamut boundaries, which results in reproductions with high contrasts, but flat shading.
By matching the target image histogram to the wood texture histogram, we compress the target image gamut to match the wood textures. This reduces the overall contrast, but puts more emphasis on shading nuances, see Figures~\ref{fig:histogram_equalization} and \ref{fig:histogram_equalization_eye}.
\hide{
In the opposite case, where the target image foreground exhibits a low contrast or where the foreground and the background are not separated well, the histogram matching appears to overfit, i.e.\ color variations tend to be exaggerated within the wooden rendition in comparison to the input target image.
In contrast, omitting histogram matching results in extremely flat reconstructions.
As a compromise, we apply a weighted average of the original input image (i.e.\ without histogram matching) and the image resulting after histogram matching to improve contrast while preserving the original image style.
Without histogram matching, on the other hand, the wood reconstruction looks extremely flat.
}
We found a simple interpolation between the matched and the unmatched input image to effectively improve contrast while preserving the original style of the image (\Fig{fig:histogram_equalization}). %

\begin{figure*}[t]
\includegraphics[width=0.32\linewidth]{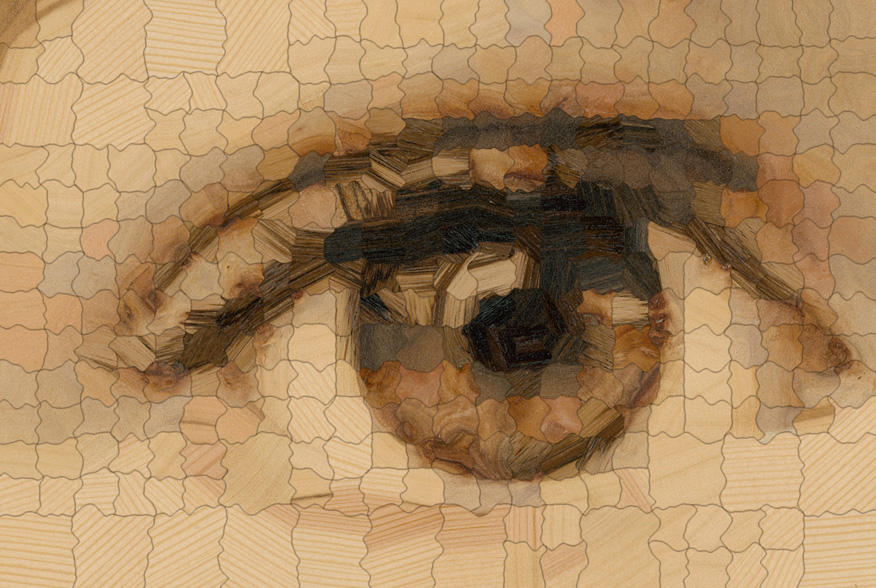}
\hfill
\includegraphics[width=0.32\linewidth]{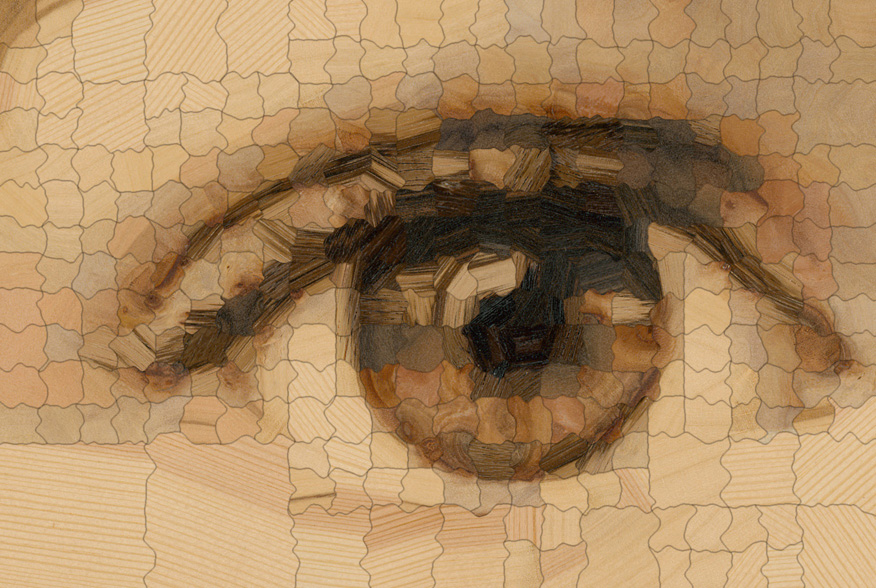}
\hfill
\includegraphics[width=0.32\linewidth]{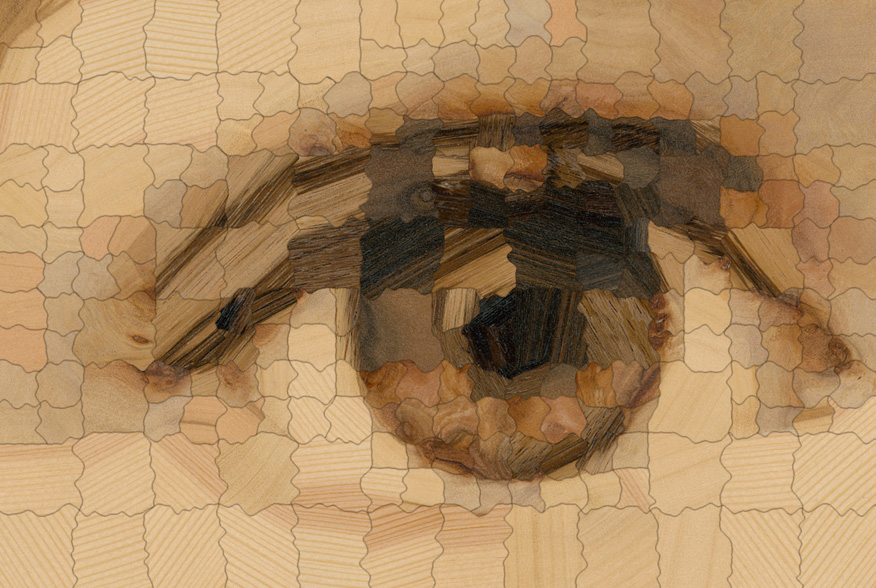}
\caption{\changed{\textbf{Synthetic}} Effect of different adaptive reconstruction parameters. From left to right: $(n_\text{adaptive}=1, w_\text{adaptive} = 1.2)$, $(n_\text{adaptive}=2, w_\text{adaptive}=1.2)$, $(n_\text{adaptive}=1, w_\text{adaptive} = 1.5)$. As expected, high-frequency image structures are only touched for large values of $w_\text{quality}$ (i.e.\ we accept a large decline in reconstruction quality). Nonetheless, we find the effect to be visually pleasing in all images and subject to personal preferences.}
\label{fig:adaptive}
\end{figure*}

\paragraph{Patch size}
We evaluated the influence of the patch size on the style of the resulting target image renditions.
\Fig{fig:resolution} shows rendered results for different patch sizes ranging from $7.7$ to \SI{31.0}{mm}.
Our experiments suggest that patches with \SI{5}{mm} edge length are the lower bound for physical producibility using our pipeline.
Smaller patches could easily get lost and would be difficult to assemble.
The reconstruction quality improves as the patch size decreases and approaches an almost photorealistic appearance for very small patches.
In contrast, reconstructions with coarse patch sizes exhibit a different, more sketch-like style.

As demonstrated in \Fig{fig:resolution}, exploiting the structures inherent to the wooden materials greatly enhances the visual quality on all resolutions, thereby providing evidence for the effectiveness of our structurally aware template matching step.
The perceived resolution of any image depends on the image size, resolution, and viewing distance.
In order to give the reader an impression about the amount of additional perceived resolution introduced by the wood pixels, we include a comparison to a ``baseline'' that discards the wood structure and instead replaces each patch by its mean color. 

Finally, we evaluate the effect of adaptive patch sizes in \Fig{fig:adaptive}.
Analogous to adaptive grid methods, this allows us to reduce the total number of wood patches without sacrificing reconstruction quality.
Regarding stylization, the larger patches result in an overall smoother appearance with fewer cuts.

\paragraph{Feature vector weights}
To analyze the effect of differently weighted feature vectors in the template matching step (\Eq{eq:template_matching}) on the wood puzzle appearance, we show results obtained for various parameter choices in \Fig{fig:sobel_effect}.
The obtained renditions for the highlighted regions of the eyelid (top row) and the iris (lower row) show that high weights for the intensity penalty $w_\text{intens}$ enforce the matching regarding the intensity features.
Finer structures, such as eyelashes, become better preserved by increasing the penalty $w_\text{edge}$ on the edge filter responses.

\paragraph{Boundary shape optimization}
\begin{figure}[t]
\includegraphics[width=0.49\columnwidth]{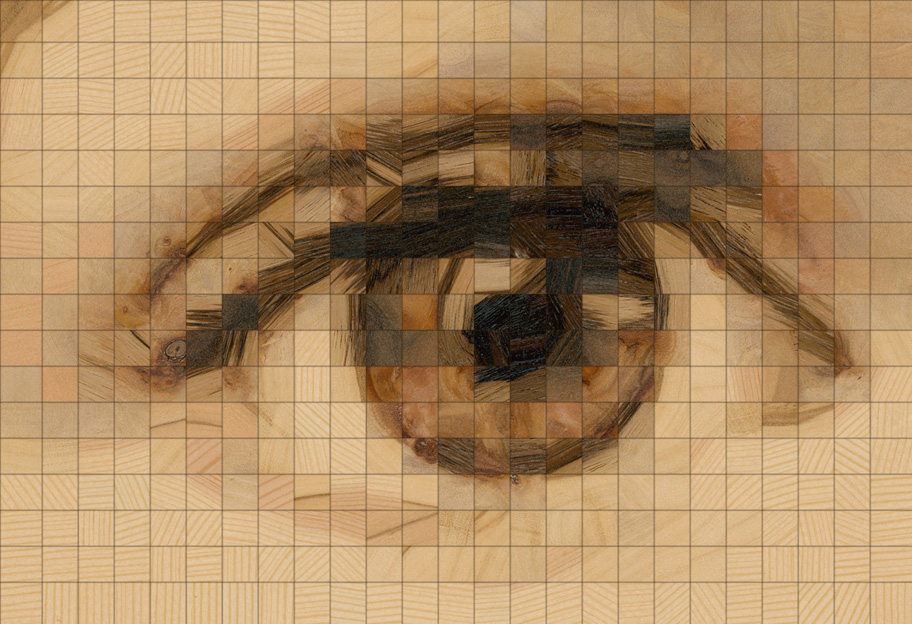}
\hfill
\includegraphics[width=0.49\columnwidth]{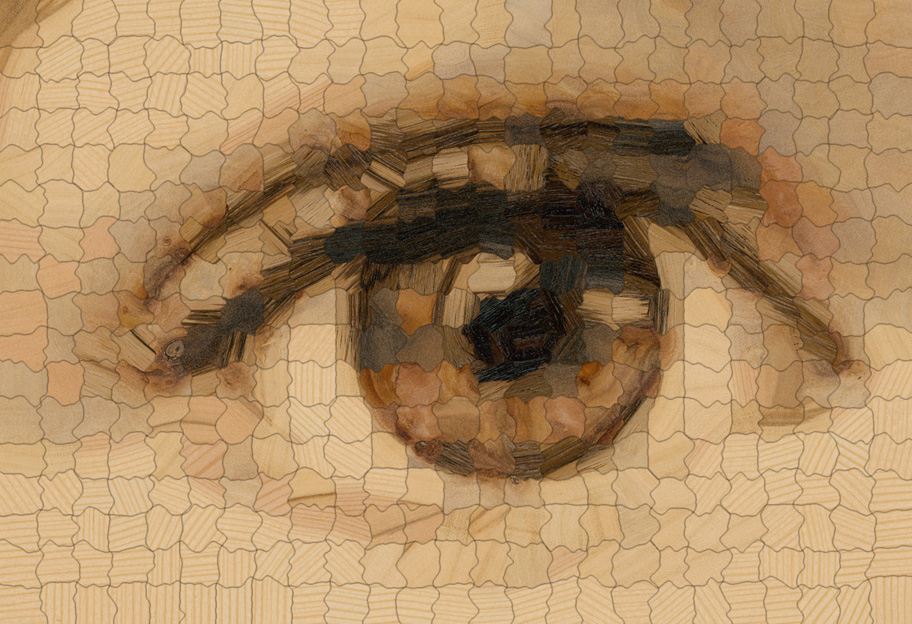}
\caption{\changed{\textbf{Synthetic}} The effect of the boundary shape optimization using dynamic programming. Without dynamic programming (left), the generated rendition of the target image has a pixelized style. With dynamic programming (right), the cuts are optimized according to the underlying data term and the rendition exhibits a smoother, more organic style.}
\label{fig:dynamic}
\end{figure}
We also show the respective results before and after cut optimization.
As demonstrated in \Fig{fig:dynamic}, the use of square patches on a regular grid results in a pixel-like rendition of the target image.
Merging neighboring patches according to the data term reduces the pixelation effect, thereby putting more emphasis onto the underlying image structures.
We found that the representation of rounded, high-contrast image features specifically benefits from the dynamic programming step.

\subsection{Ablation study}
\label{sec:ablation_study}
\begin{figure*}[ht]%
\includegraphics[width=0.12\textwidth]{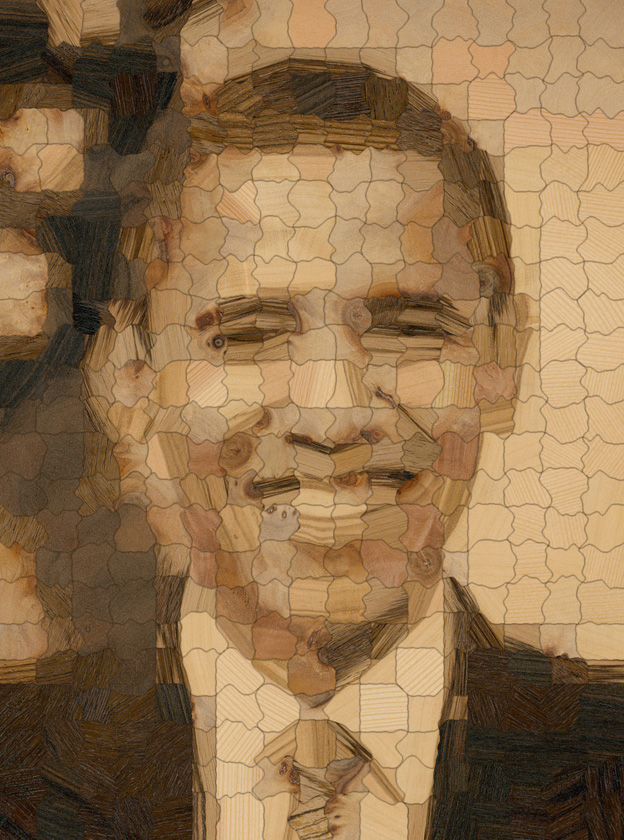}\hfill%
\includegraphics[width=0.12\textwidth]{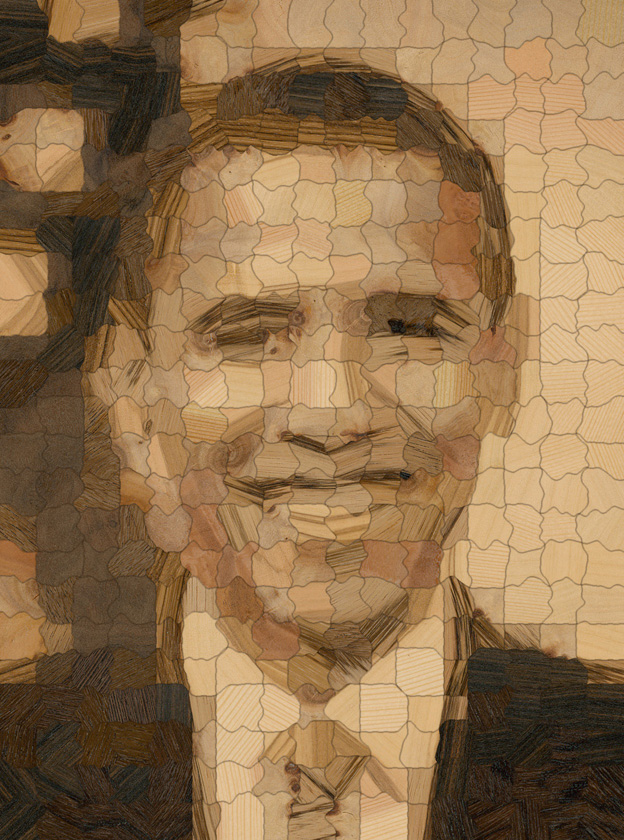}\hfill%
\includegraphics[width=0.12\textwidth]{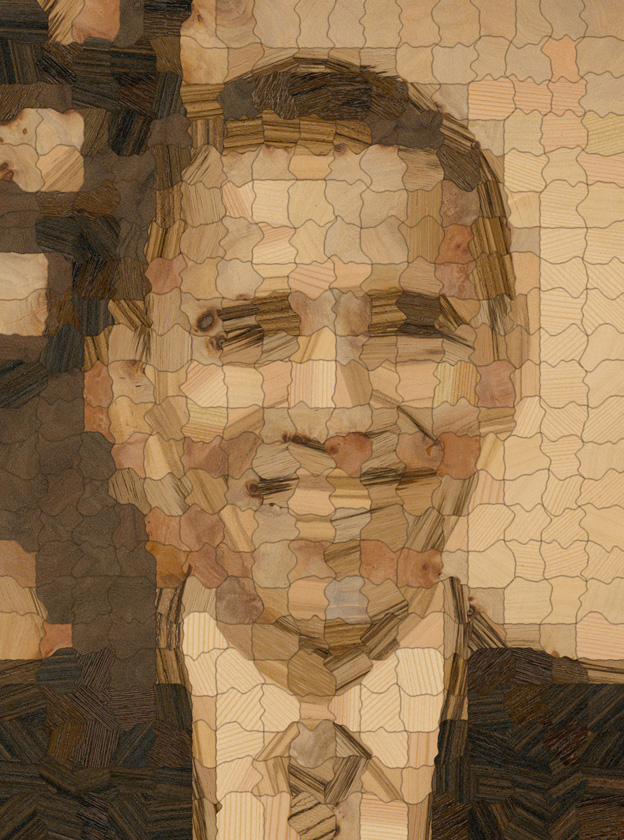}\hfill%
\includegraphics[width=0.12\textwidth]{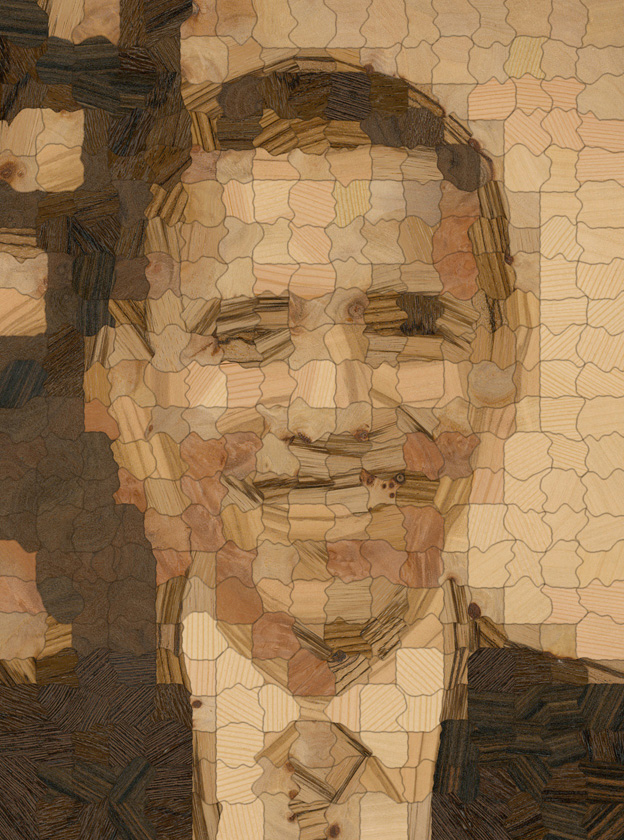}\hfill%
\includegraphics[width=0.12\textwidth]{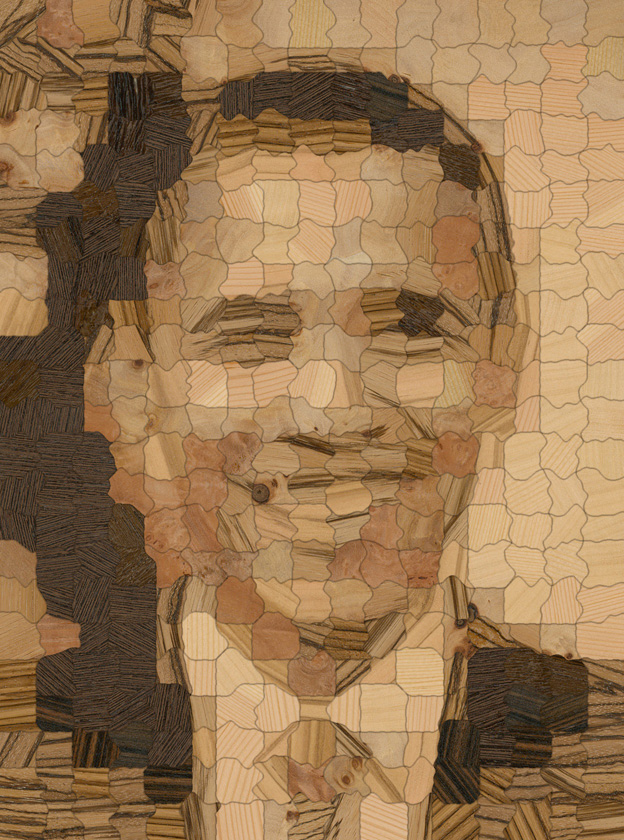}\hfill%
\includegraphics[width=0.12\textwidth]{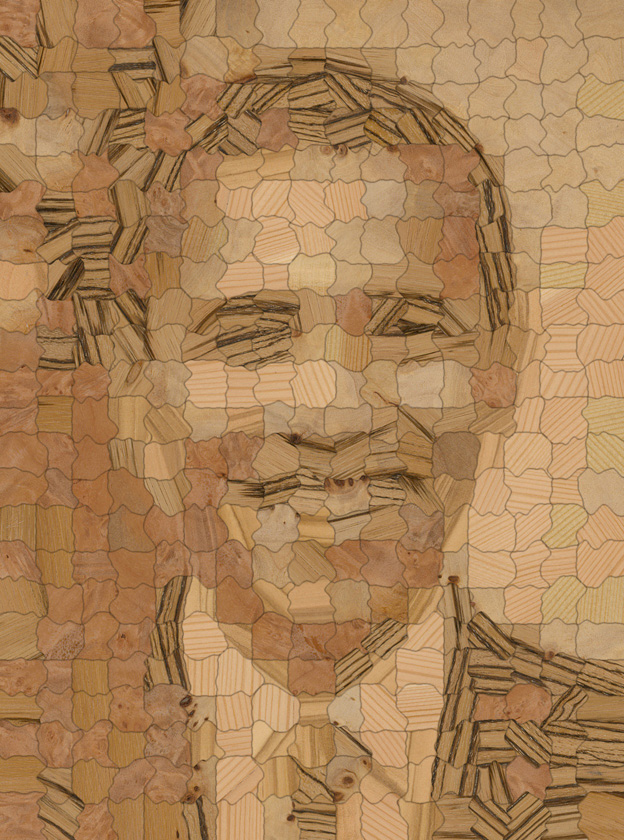}\hfill%
\includegraphics[width=0.12\textwidth]{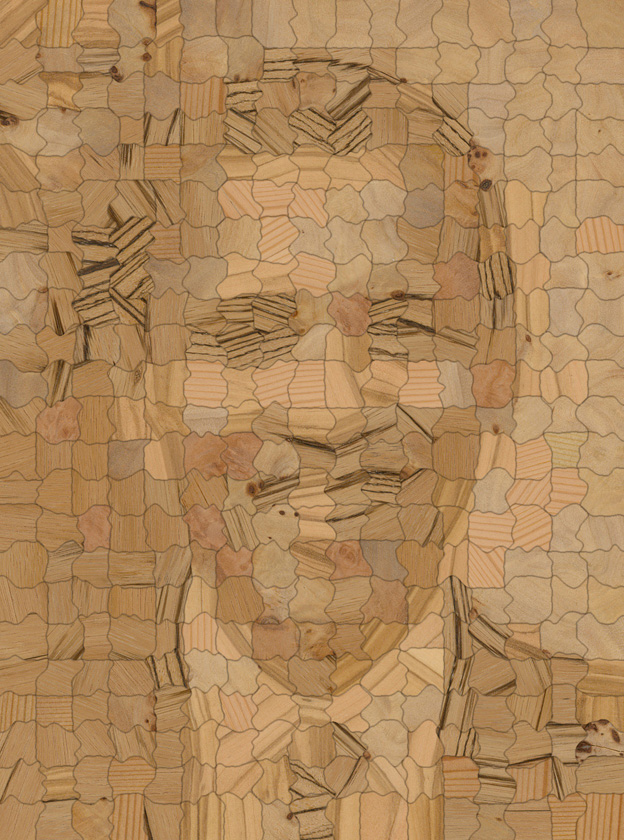}\hfill%
\includegraphics[width=0.12\textwidth]{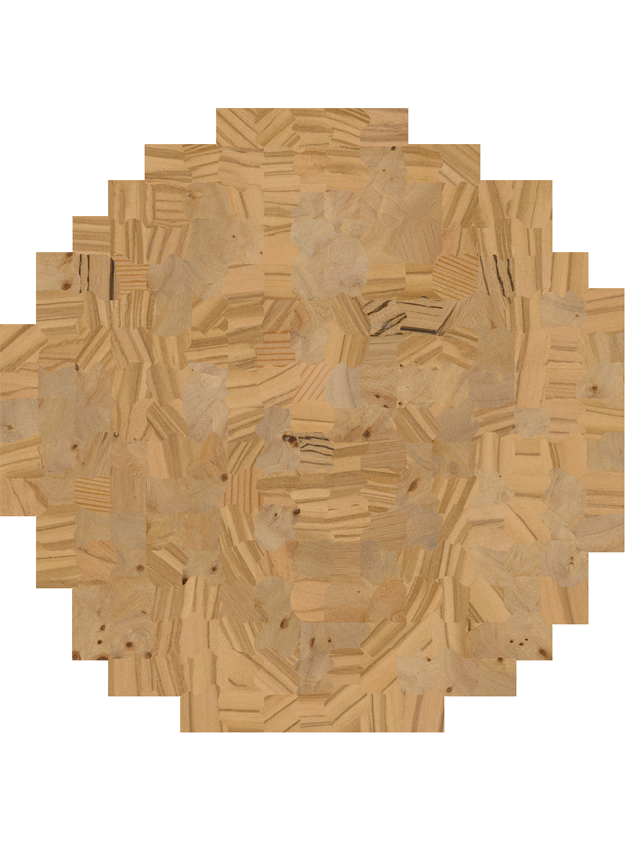}\\[0.5mm]%
\includegraphics[width=0.12\textwidth]{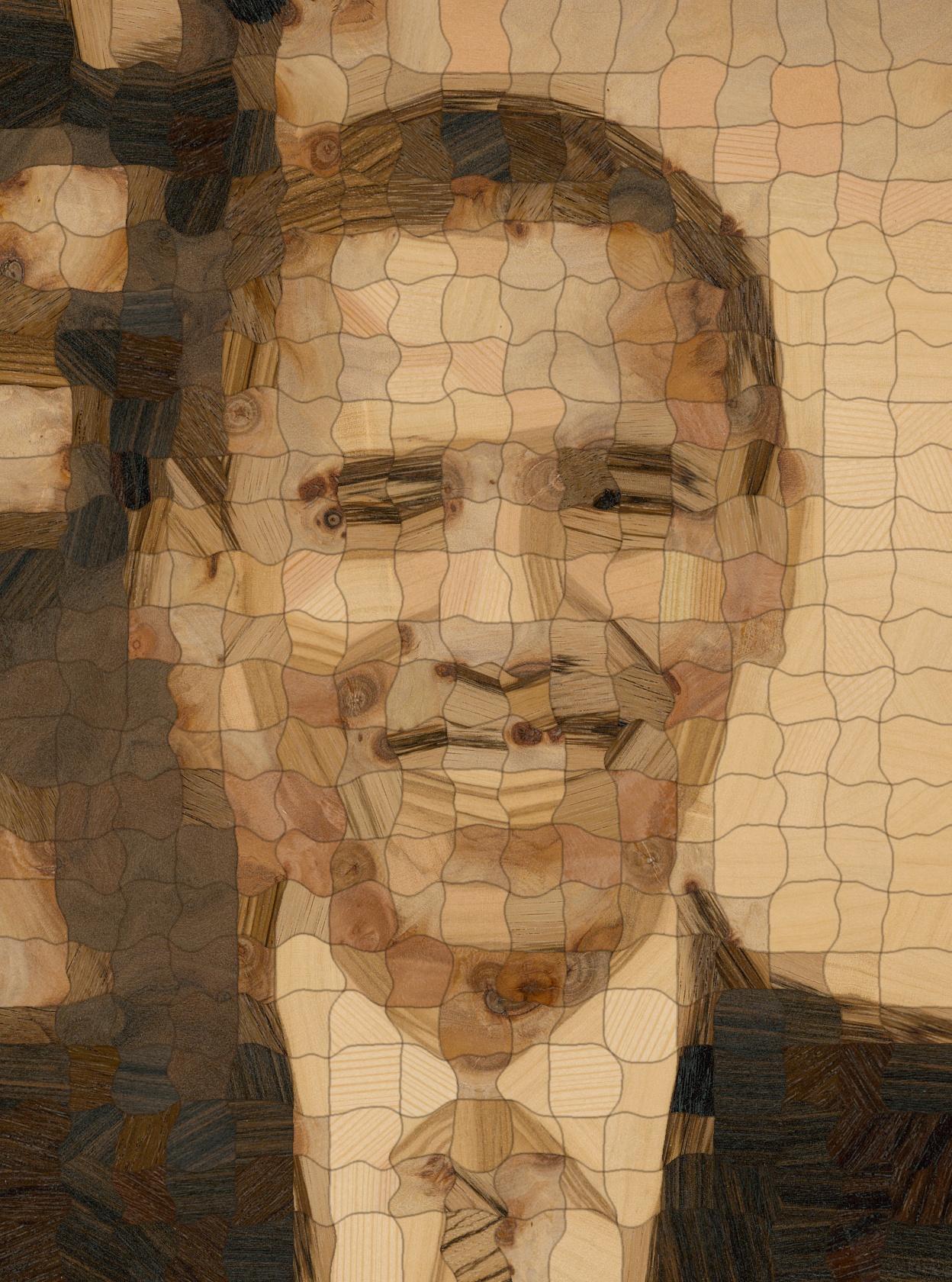}\hfill%
\includegraphics[width=0.12\textwidth]{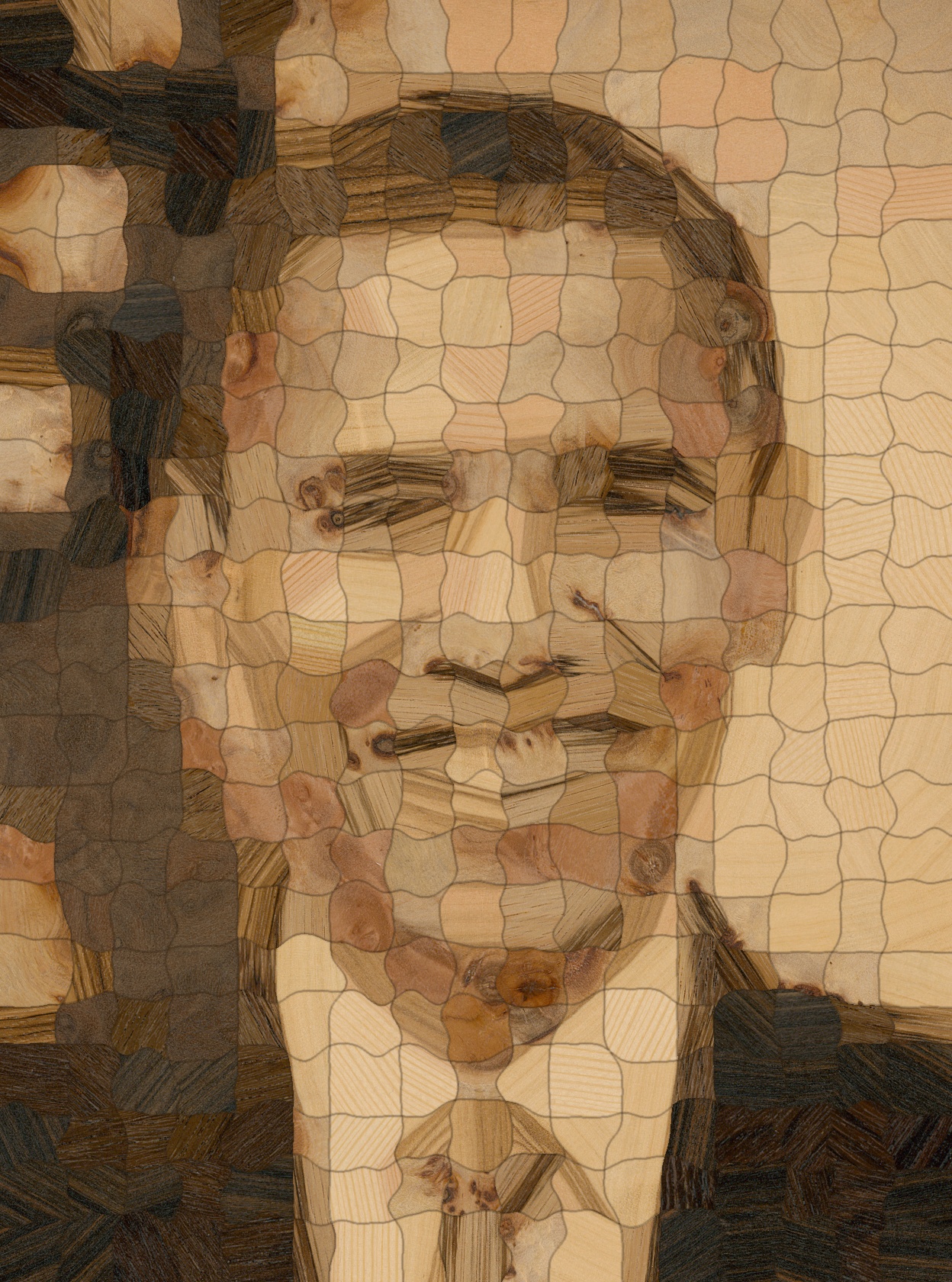}\hfill%
\includegraphics[width=0.12\textwidth]{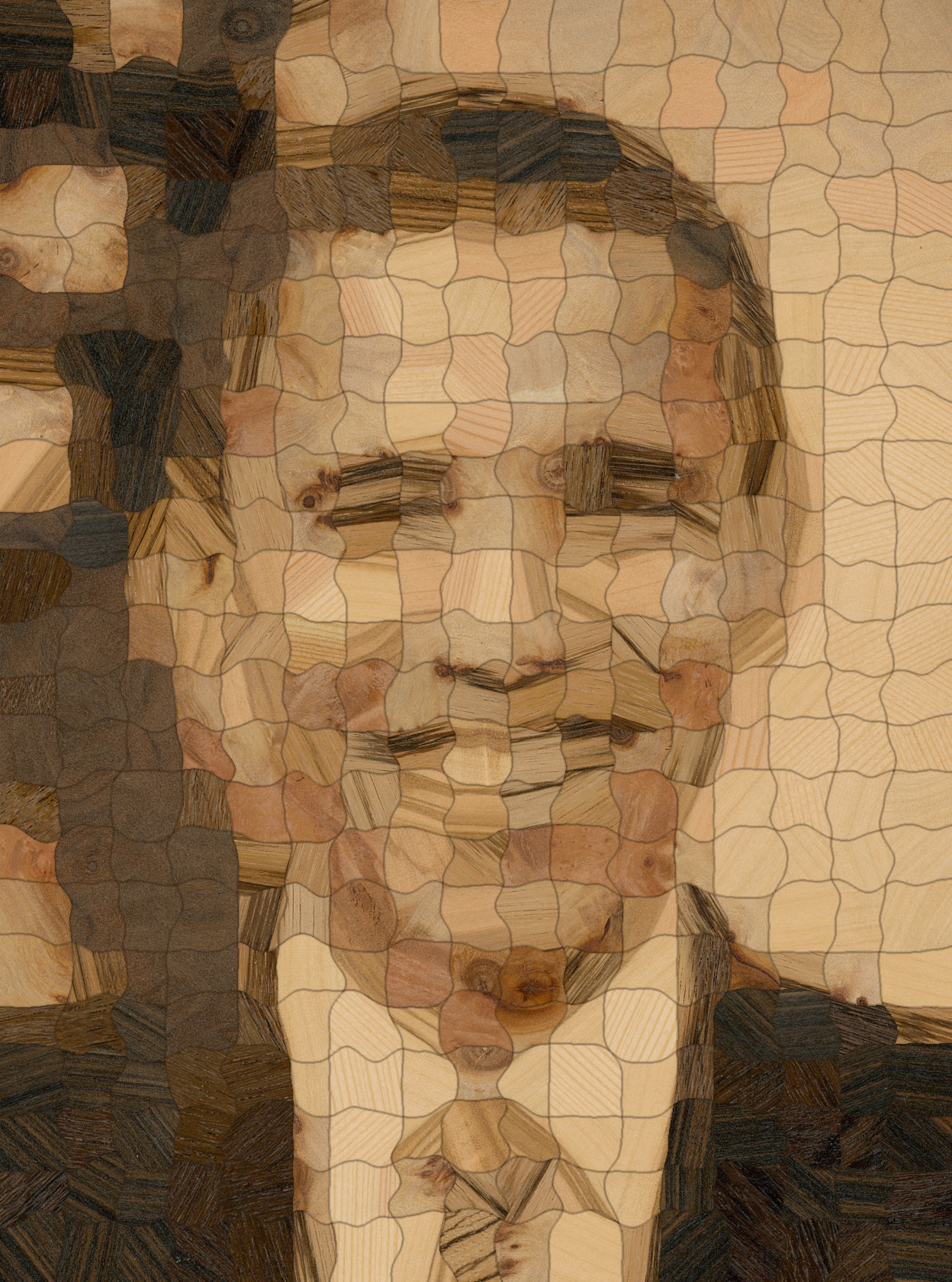}\hfill%
\includegraphics[width=0.12\textwidth]{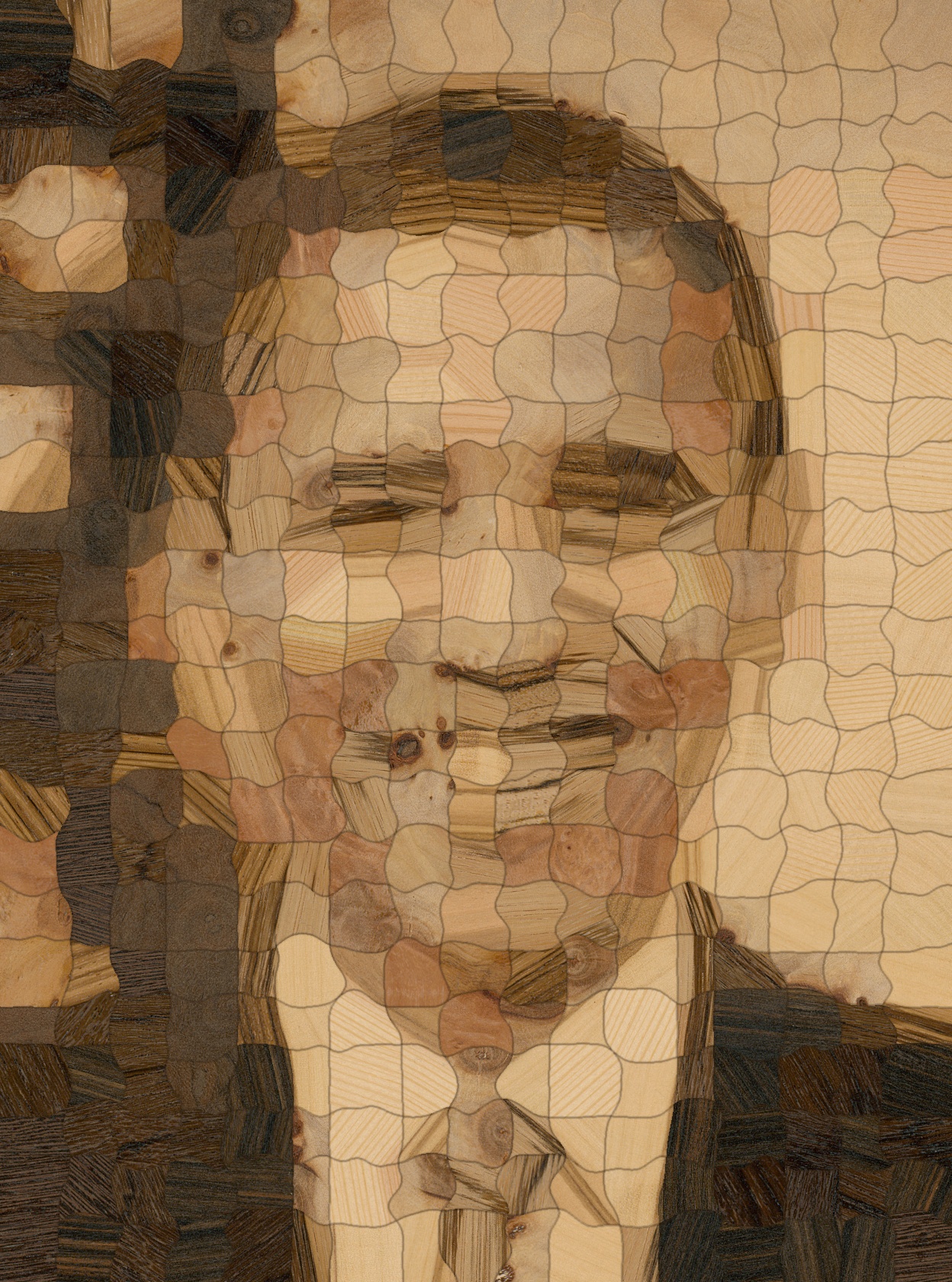}\hfill%
\includegraphics[width=0.12\textwidth]{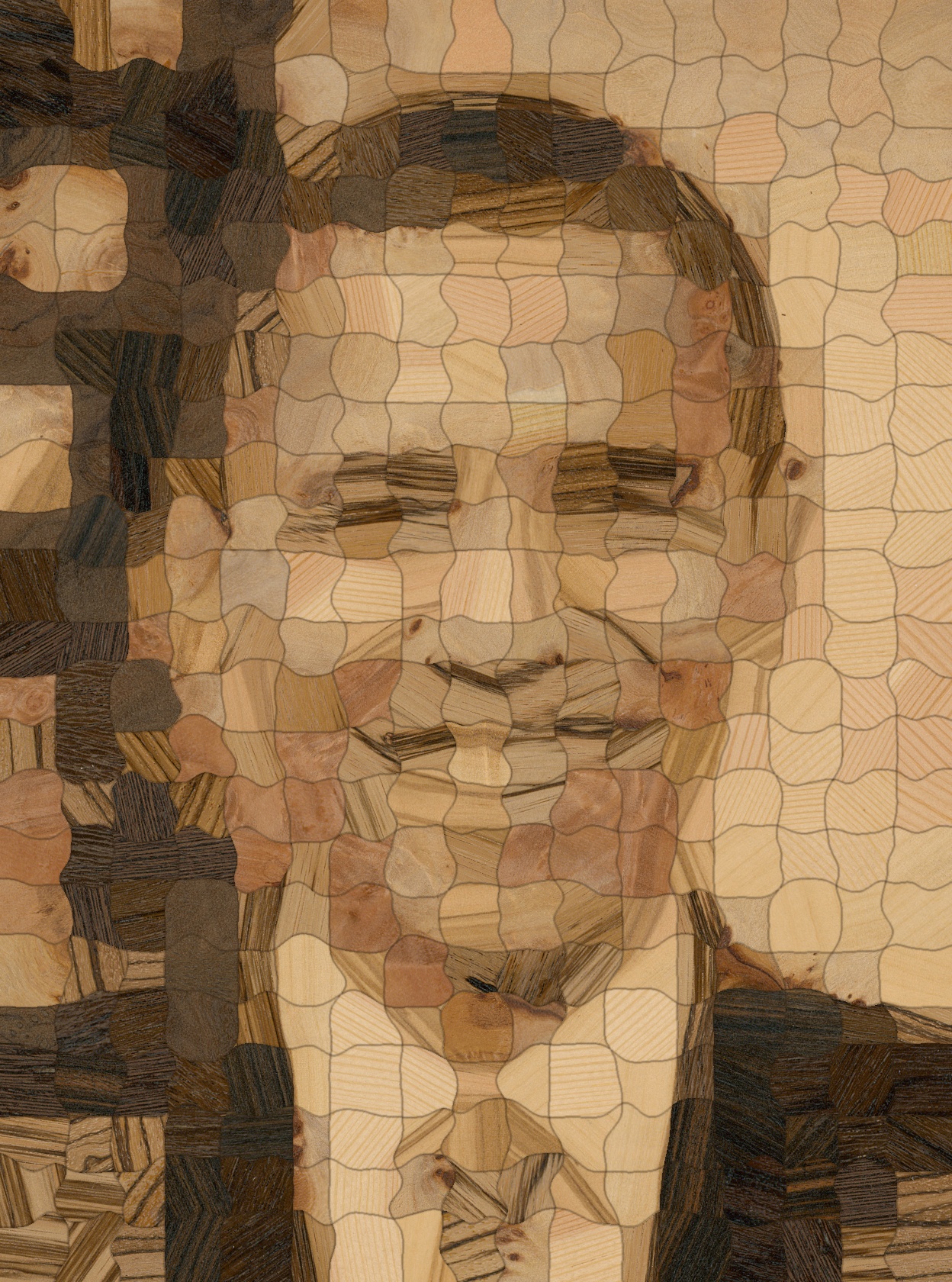}\hfill%
\includegraphics[width=0.12\textwidth]{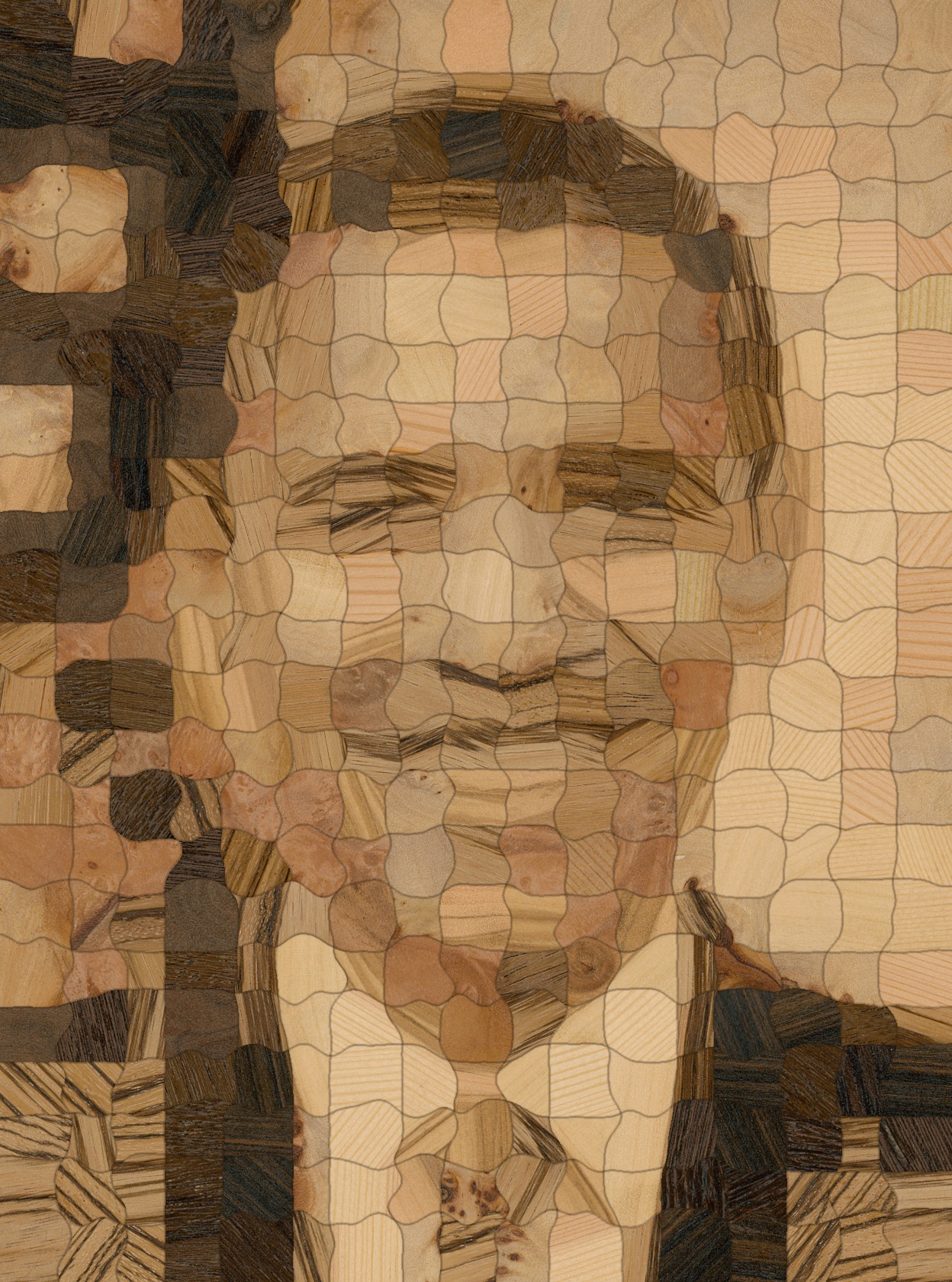}\hfill%
\includegraphics[width=0.12\textwidth]{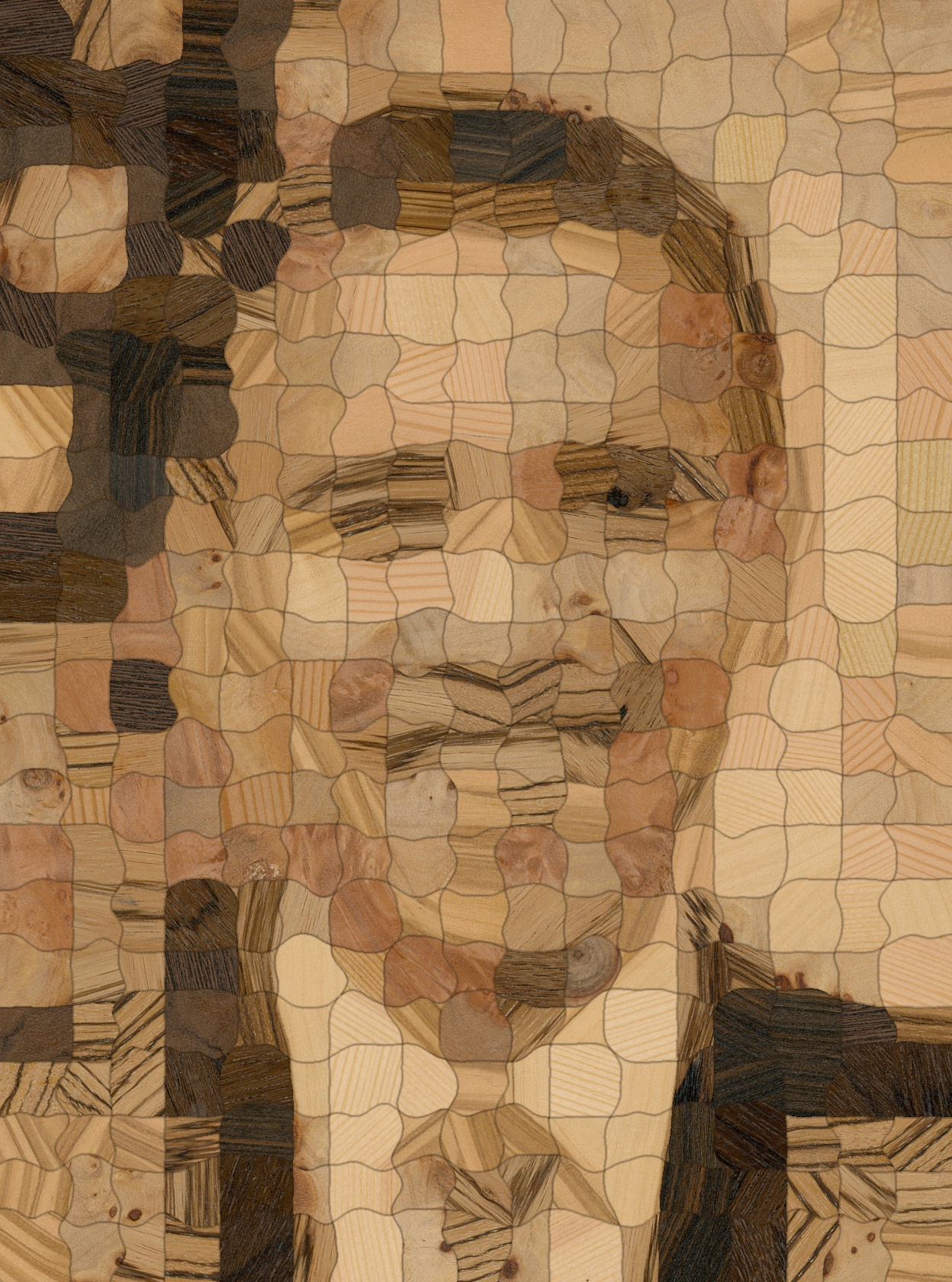}\hfill%
\includegraphics[width=0.12\textwidth]{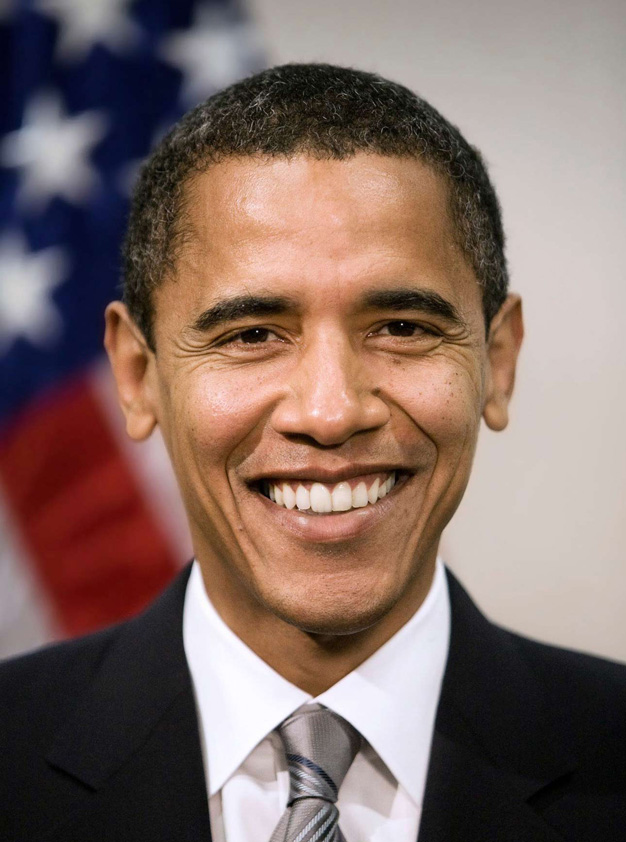}\\%
\caption{\changed{\textbf{Synthetic} Renditions of a target image generated under a decreasing amount, and quality, of available patches from a single wood sample. Top row: Patches are sorted by their distance to the image center and reconstructions are conducted sequentially from left to right, \ie matching of the second column starts after all patches of the first column have been matched. The last reconstruction did not complete because there were no valid patches left on the wood sample. Bottom row: Patches from multiple copies of the same image are sorted by their image saliency score and reconstructions are carried out in an interleaved manner. The first column shows the result for a single reconstructed target image, the second column for two simultaneously reconstructed target images, and so on. For the first four columns, the differences are negligible. With less suitable source patches available, the interleaved reconstruction yields higher quality results compared to the sequential reconstruction. Please zoom in to see image details.}}%
\label{fig:AblationStudy}%
\end{figure*}
Our approach is inherently resource constrained.
Thus we expect the reconstruction quality to scale with the area of available wood samples.
To evaluate this effect, we applied our pipeline several times to generate renditions of the same target image under a decreasing availability (and quality) of source patches.
\changed{We have evaluated the effect of a sequential reconstruction using a circularly sorted priority queue and an interleaved, simultaneous reconstruction using a queue sorted by image saliency.}
The respective results are shown in \Fig{fig:AblationStudy}.
We observe that the reconstruction quality decreases gracefully and the target image stays recognizable until the very last reconstruction.
After the last reconstruction (partially) finished, there was no space left on the veneer panel that was large enough for another patch.%

We noticed two types of degradation: intensity and high-frequency detail degradation.
Most noticeable is the degradation in overall intensity matching after the panel runs out of dark patches (iteration 5).
Less noticeable is the degradation of high-frequency content, e.g.\@ around the eyes after iteration 3.
\changed{We note that the interleaved, simultaneous reconstruction using image saliency produces generally favorable results over the circularly sorted reconstructions.
The degradations could be alleviated by reconstructing target images with different intensity distributions or, preferably, enforcing sufficient resources during reconstruction.}

\subsection{Fabricated results}
\begin{figure}[t]
\includegraphics[width=\columnwidth]{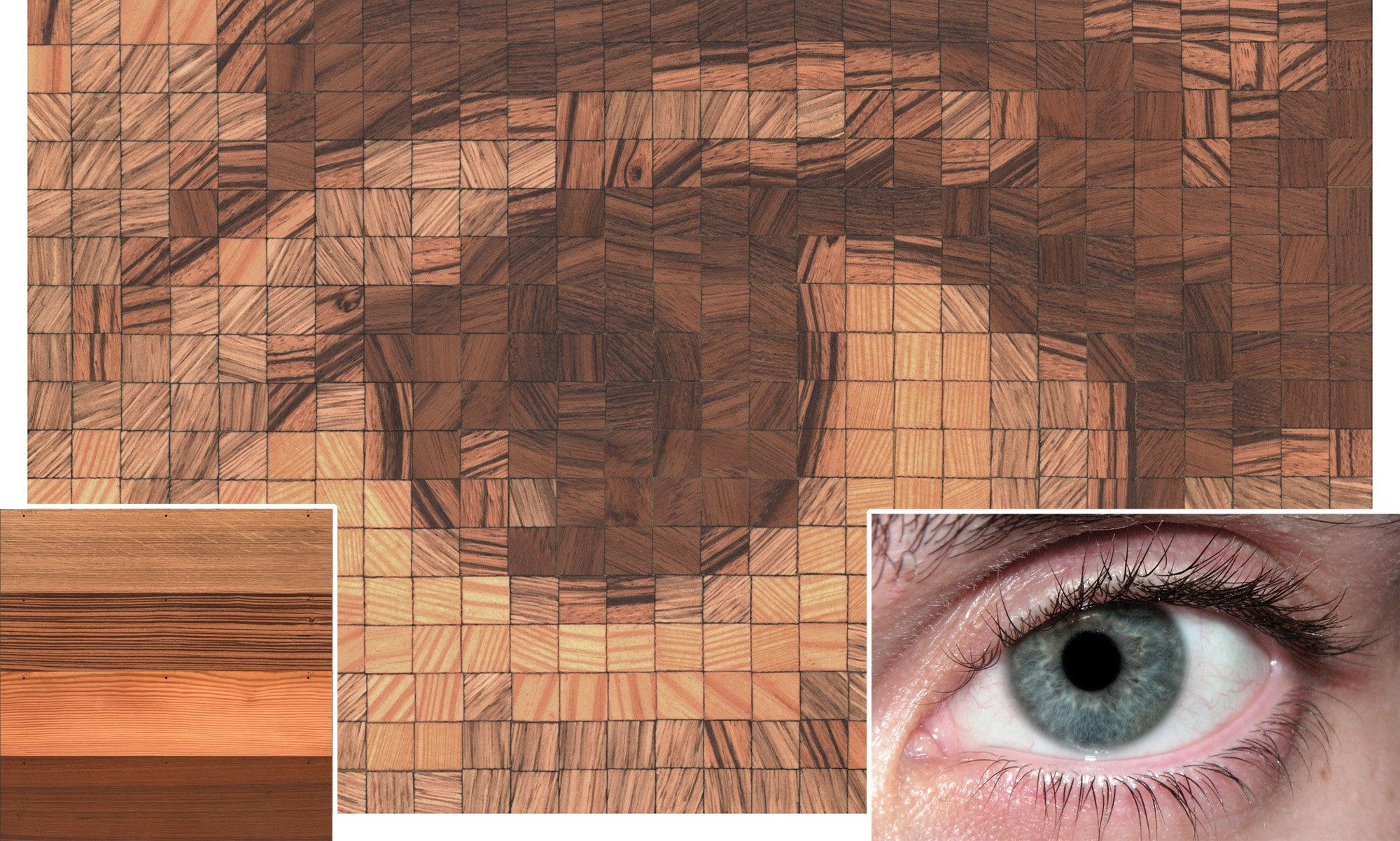}\\[0.5em]
\caption{\changed{\textbf{Fabricated}} A fabricated piece of wood parquetry made from four different quarter-cut thick veneers (bottom left corner, from top to bottom: oak, zebrawood, fir, American walnut). The target image is a human eye (bottom right corner). The veneer puzzle consists of \num{28 x 17} wooden pixels and has a total size of approx.\@ \SI{28 x 17}{cm}.}
\label{fig:teaser}
\end{figure}

\begin{figure}[t]%
\includegraphics[width=\columnwidth]{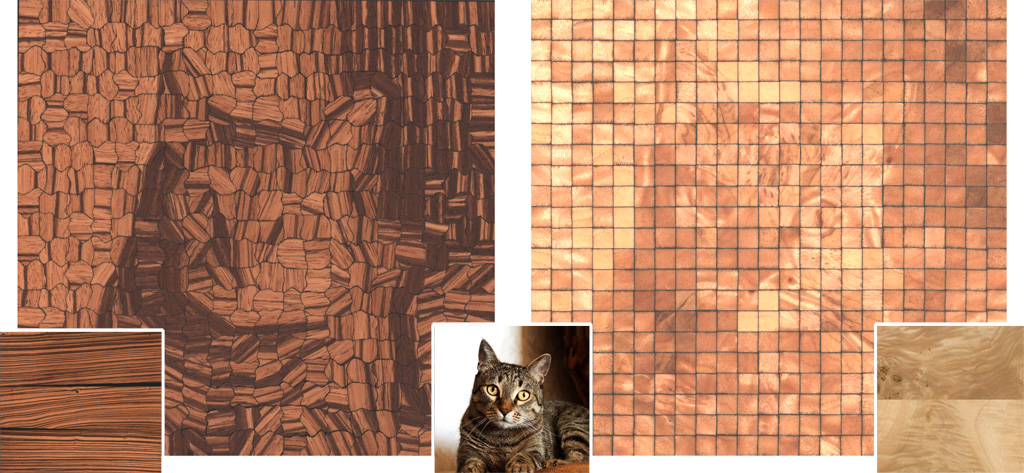}
\\%
\caption{\changed{\textbf{Fabricated}} Exemplary results of fabricated parquetries using the same target image (bottom center), but different wood types and finish. The left image was fabricated using zebrawood with an oil finish. The right image was produced using poplar burl veneer with clear coating, resulting in a highly specular appearance with limited contrast. The samples consist of \num{20 x 19} and \num{23 x 22} wooden pixels respectively and their physical dimensions are about \SI{15 x 15}{cm}. The left puzzle has optimized patch boundaries, the right puzzle consists of square patches.}%
\label{fig:RealResults_cats}%
\end{figure}

We present exemplary results of physically produced veneer puzzles in \cref{fig:RealResults_beethoven,,fig:teaser,,fig:RealResults_cats}.
The veneer puzzles in \cref{fig:RealResults_beethoven,,fig:teaser} have been fabricated using multiple wood types.
Since different wood types can differ vastly in color and grain structure, these results show a high contrast and perceived resolution.
Fine details, such as hair, eyebrows, or eyelashes are faithfully reproduced.

The results in \Fig{fig:RealResults_cats} have each been produced using a different single wood type.
The amount and quality of detail within a pixel is inherently limited to the features present in the original material.
Woods with a limited feature gamut thus lead to a strongly stylized outcome, which we imagine could also be utilized as an artistic tool.

We decided to finish most of the pieces using hard wax oil in order to accomplish a natural look.
A clear coat finish (\Fig{fig:RealResults_cats}, right) results in a highly specular appearance.

With row/column labels engraved on the back side, 
it takes about \SIrange{1}{2}{h} for a single person to assemble a 500-piece parquetry inside a suitably dimensioned frame.
Although somewhat repetitive, the authors found this activity to be satisfying and relaxing.
For thin veneers that are laminated onto a plywood substrate, the final image remains hidden until the finished composition is turned around.

\subsection{Synthetic results}
\begin{figure*}[ht]
\includegraphics[width=\linewidth]{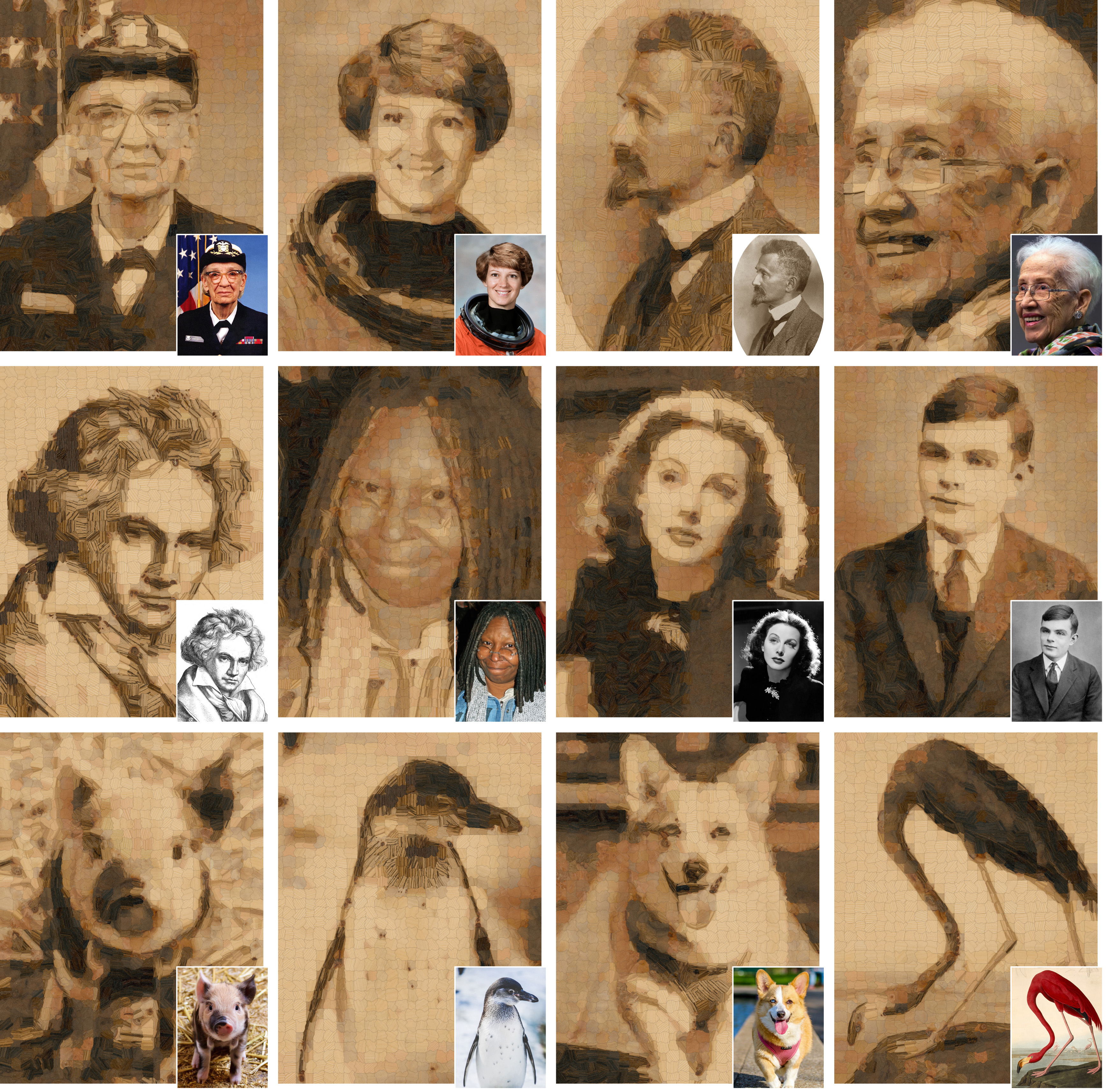}
\caption{\changed{\textbf{Synthetic}} Exemplary synthetic renditions of portraits and animals. Each of these results has been composed using the veneer sample panel shown in \Fig{fig:input_results} and the default parameters listed in \Tab{tab:user_parameters}. Our algorithm is able to handle a wide range of input including color photographs, black and white photographs, drawings, and paintings. The images show, from left to right, top to bottom: Grace Hopper, Eileen Collins, Felix Hausdorff, Katherine Johnson, Ludwig van Beethoven, Whoopi Goldberg, Hedy Lamarr, Alan Turing, a piglet, a penguin, a Corgi, and a flamingo.}%
\label{fig:ExamplesPortraitsAnimals}%
\end{figure*}

In addition to the evaluation of different parameter choices, we show renditions for several target images depicting portraits and animals in \Fig{fig:ExamplesPortraitsAnimals}.
To demonstrate the robustness of our approach with respect to different target images, each of these results has been produced using the default parameters shown in \Tab{tab:user_parameters}.
The depicted results demonstrate the potential of computational parquetry for fine arts.
Portraits and animal pictures can be easily recognized as their characteristic appearance is preserved in the stylized result.
Please see the supplemental material for additional results.

\hide{
\begin{figure}%
\includegraphics[width=0.31\columnwidth]{inputdata/bobby2.jpg}\hfill%
\includegraphics[width=0.31\columnwidth]{figures/bobby-finished.jpg}\\%
\caption{Stylized parquetry of a photo of Bobby McFerrin using intensity, Gabor and Laplacian penalties and the same wood samples as used throughout the paper. Left: input image. Right: finished piece.}%
\label{fig:bobby}%
\end{figure}
\begin{figure}%
\includegraphics[width=0.31\columnwidth]{inputdata/cat.jpg}\hfill%
\includegraphics[width=0.31\columnwidth]{figures/results/cat-square.jpg}\hfill%
\includegraphics[width=0.31\columnwidth]{figures/results/cat-fab.jpg}\\%
\caption{``Photo-realistic'' parquetry of a cat photo using intensity, Gabor and Laplacian penalties, square cuts and a single wood sample. From left to right: input image, simulation, finished piece. }%
\label{fig:cat}%
\end{figure}
}
\section{Discussion and future work}

\begin{figure}[t]
\includegraphics[width=0.49\columnwidth]{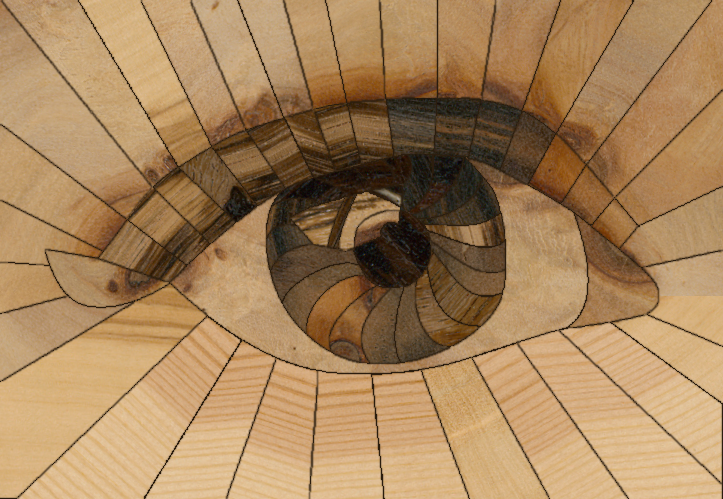}
\hfill
\includegraphics[width=0.49\columnwidth]{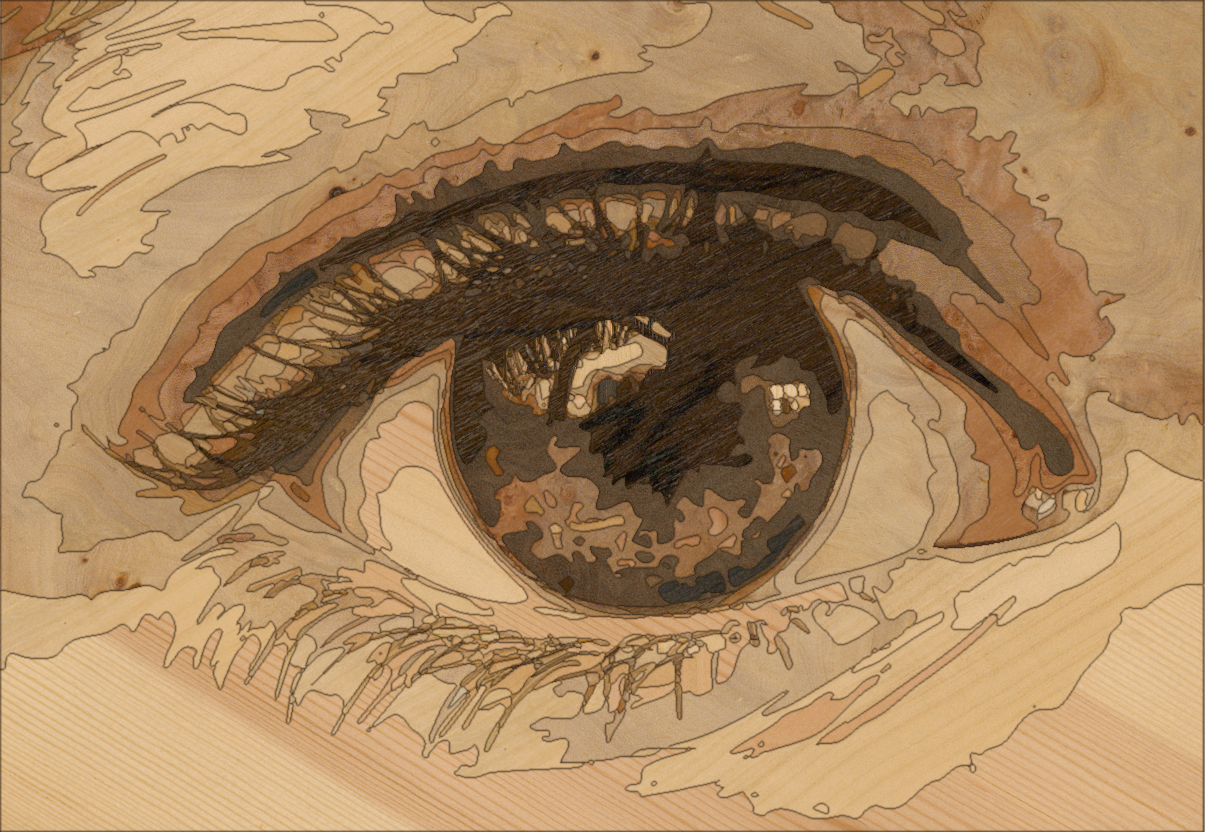}
\caption{\changed{\textbf{Synthetic} Two results using custom segmentations of highly varying complexity. The left reconstruction is based on a hand-drawn segmentation, while the right reconstruction is based on a \emph{posterized} version of the input image. Depending on the segmentation, cutting and assembly can be complex and might require a skillful artist to execute.}}
\label{fig:segmented_result}
\end{figure}

\changed{In this work, we have focused on the generation of cut patterns that are fabricable and can easily be assembled even by untrained users, which has led us to a solution based on regularly or semi-regularly placed patches.
We would like to note though that our pipeline is not inherently restricted to these kinds of segmentations and instead custom, user-defined segmentations can be provided as well, \eg to produce marquetry art comparable to the one presented in \Fig{fig:examples}.
Due to the structure-aware patch matching step, our approach could be able to produce art pieces of even higher fidelity than by manual matching.
However, in this case, the complexity of the wood puzzle art strongly depends on the provided segmentation and the assembly might require a skillful artist to execute.
See \Fig{fig:segmented_result} for two synthetic results using user-defined segmentations with highly varying style and complexity.}

\begin{figure}[t]
\includegraphics[width=0.32\columnwidth]{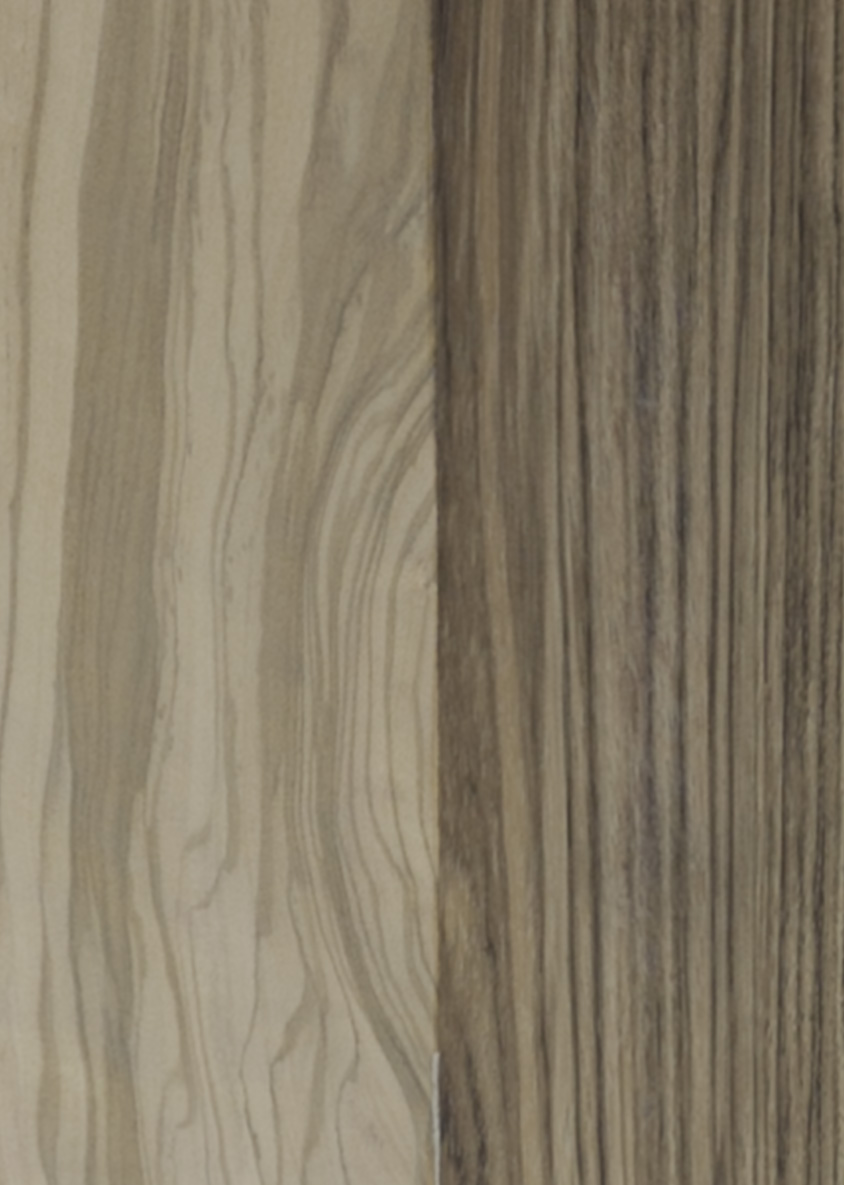}\hfill%
\includegraphics[width=0.32\columnwidth]{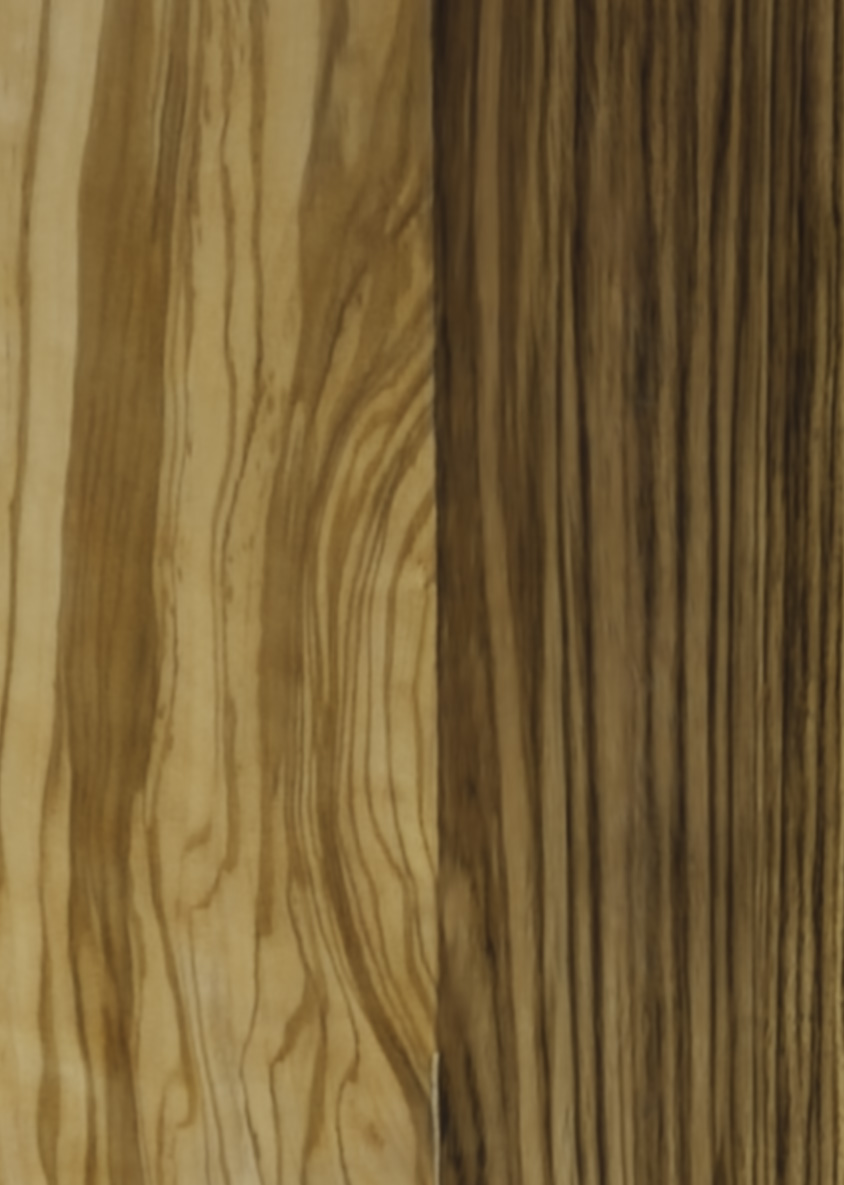}\hfill%
\includegraphics[width=0.32\columnwidth]{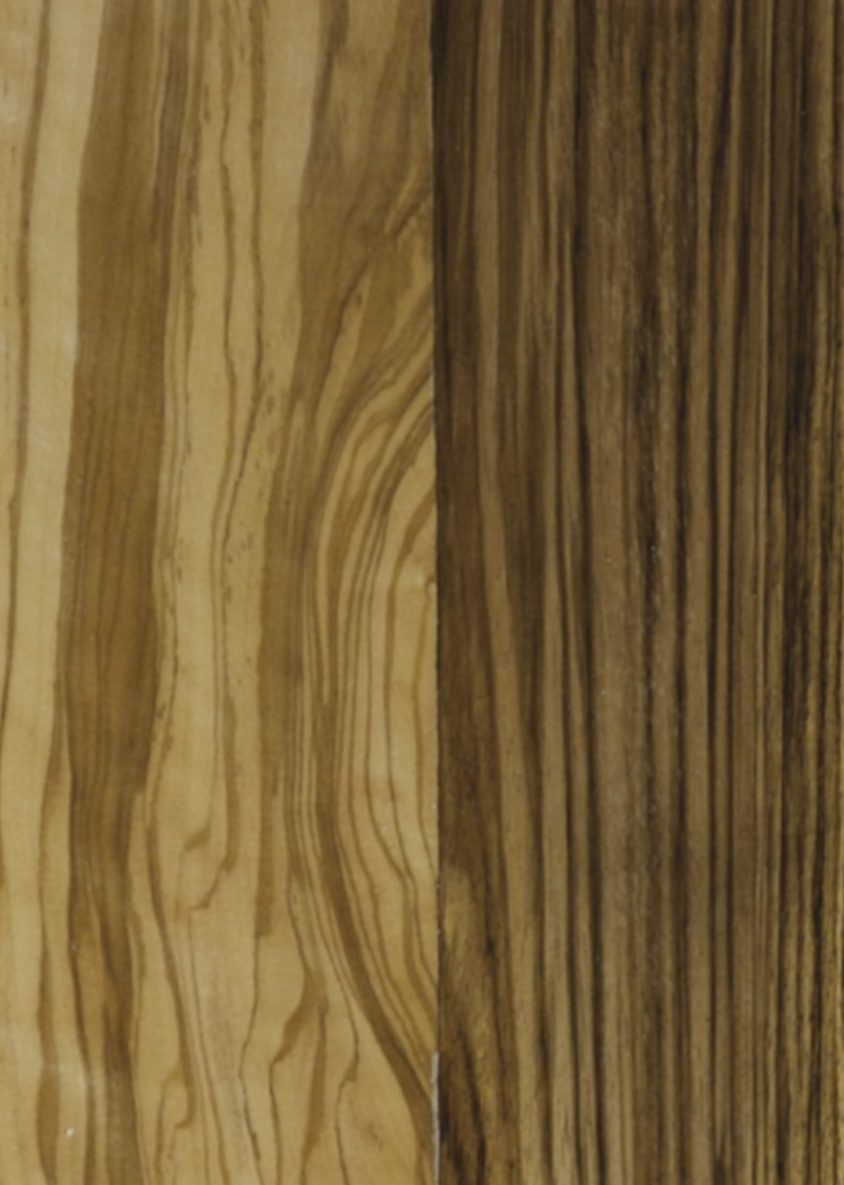}%
\caption{We envision using deep learning to predict the change in surface finish induced by a layer of oil or clear coat. Being able to do so would alleviate the need for a pre-finishing step prior to texture acquisition. From left to right: input image, surface finish appearance predicted by our preliminary model, ground truth image.}
\label{fig:deep_learning}
\end{figure}

A practical drawback of our method is that it requires a surface finish to be applied to the wood two times, once before scanning and then again after the final assembly of the finished puzzle.
The first application is important, since this step changes the appearance of the wood samples significantly. For the algorithm, it is crucial to choose suitable patches based on their final appearance.
We apply the sanding/finishing procedure a second time in order to flatten out small height variations, which are inevitable after puzzling.
For a large-scale, automatic production of custom, wooden parquetry puzzles, we would like to minimize the amount of manual interaction.
Thus, we conducted initial experiments on training a model to predict the change of appearance from unfinished to finished veneers.
Using these predictions, it might become possible to defer the application of surface finish until after the final puzzle has been assembled.
To this end, we trained a U-Net \cite{DBLP:journals/corr/RonnebergerFB15} on image pairs before and after applying the finish.
Based on the preliminary results in \Fig{fig:deep_learning}, we believe that this would be a good direction for future work.

Our approach allowed us to produce visually pleasing pieces of wood parquetry, even without having a professional wood-working background.
However, we expect that certain technical imprecisions (such as sub-perfectly applied clear coating) would be mitigated with more experience.
Also, we expect that cut clearances and discolorations will be improved with further fine tuning of the cutting equipment.

Here, we treat wood as being a diffuse reflector and ignore any directional effects.
Real wood exhibits anisotropic BRDF characteristics, which means that rotation of a part could be used to modulate its intensity.
This might also enable the generation of new types of puzzles, where a hidden pattern is revealed by the right permutation and rotation of some parts\changed{, comparable to the work of Sakurai et al.~\shortcite{Sakurai:2018:FRD:3197517.3201400}}.

In our experiments, we restricted ourselves to fabricating parquetry based on wood veneers, since they are commonly available and can be cut using a laser cutter.
Generally, our pipeline is not restricted to this type of material.
Using a water jet cutter, other materials like marble or brushed metal could be processed as well.
The process could also be extended to multi-material parquetry.

Parquetry generation is inherently resource-constrained and in the scope of our work, the amount of available source samples was limited.
Having access to a larger database of veneers (either by increasing the number of samples per wood type, or by introducing new wood types) would certainly improve the reconstruction quality.
However, since this is an artistic process, reaching the highest reconstruction quality might not always be the goal.
Using only a single type of wood, or a selection of wood samples with a particular structure, can lead to equally interesting and fascinating results, see e.g.\@ \Fig{fig:RealResults_cats}.

When preparing our puzzle for assembly as a game, various degrees of difficulty could be imagined.
As all pieces are made from wood, semantic labels are not immediately accessible as they sometimes are in regular puzzles (water, buildings, skin, foliage, sky/clouds, etc.).
Given a bag of identically-shaped (square) pieces, it would seem extremely challenging to arrive at the one ``correct'' solution; at the same time, there would be numerous mechanically valid ``approximate'' solutions, or permutations between sets of similar-looking parts.
Here, the \changed{irregularly-shaped pieces generated by our refinement steps offer welcome visual and tactile cues for assembly.}

\section{Conclusions}
We approached the fabrication of structure-aware parquetry based on a novel end-to-end pipeline that takes wood samples and a target image as inputs and generates a cut pattern for parquetry puzzles.
To the best of our knowledge, there is no prior work that addresses the challenges inherent to the task of producing a physical sample of wood parquetry using commodity hardware from minimal input (a target image).
The challenges include the single use of individual pieces of input material without being deformed, scaled, blended, or filtered, as well as keeping track of resource use in order to prevent source patches from colliding with each other, while still faithfully reproducing the target image. Practical aspects regarding the fabricability have also been taken into account.
The varying structural details within the wood samples lead to unique and fascinating artworks, and the design of the overall process allows even users without a particular woodworking background to experience producing pieces of this new type of art.

 \bibliographystyle{ACM-Reference-Format}

\subsubsection*{Attribution of source materials}
\footnotesize
\begin{itemize}
\item
\urlcolor{\href{https://commons.wikimedia.org/wiki/File:Left_Blue_Eye.jpg}{Left Blue Eye}} (\Fig{fig:teaser}): public domain.
\item
\urlcolor{\href{https://commons.wikimedia.org/wiki/File:Marquetry_portrait_by_Laszlo_Sandor.jpg}{Marquetry Self Portrait}} (\Fig{fig:examples}): \copyright~2008 Laszlo Sandor, \urlcolor{\href{https://creativecommons.org/licenses/by/4.0/}{CC BY 4.0}}.
\item
\urlcolor{\href{http://veneerimages.tumblr.com/post/134920852655/girl-1-marquetry-21-x-21-this-piece-includes}{Marquetry portrait ``Girl 1''}} (\Fig{fig:examples}): \copyright~2015 Rob Milam, included with permission of the artist.
\item
\urlcolor{\href{https://commons.wikimedia.org/wiki/File:Intarsienbild_Roentgen_Zick_makffm_6889.jpg}
{Intarsia image, Workshop David Roentgen}} (\Fig{fig:examples2}): 2011, public domain.
\item
\urlcolor{\href{https://commons.wikimedia.org/wiki/File:Mosa\%C3\%AFque_d\%27Ulysse_et_les_sir\%C3\%A8nes.jpg}
{Mosa\"ique d'Ulysse et les sir\`{e}nes, Bardo Museum in Tunis}} (\Fig{fig:examples2}): public domain.
\item
\urlcolor{\href{https://www.pexels.com/photo/adult-brown-tabby-cat-747795/}{Adult brown tabby cat}} (\cref{fig:pipeline,fig:RealResults_cats}): \copyright~Tomas Andreopoulos, \urlcolor{\href{https://www.pexels.com/photo-license/}{Pexels license}}.
\item
\urlcolor{\href{https://www.pexels.com/photo/close-up-photo-of-dog-wearing-golden-crown-1663421/}{Close-up Photo of Dog Wearing Golden Crown}}: \copyright~rawpixel.com, \urlcolor{\href{https://www.pexels.com/photo-license/}{Pexels license}}.
\item
\urlcolor{\href{https://www.pexels.com/photo/closeup-photo-of-human-eye-862122/}{Closeup Photo of Human Eye}} (\cref{fig:sobel_effect}): \copyright~Skitterphoto, \urlcolor{\href{https://creativecommons.org/publicdomain/zero/1.0/}{CC0 1.0}}.
\item
\urlcolor{\href{https://commons.wikimedia.org/wiki/File:Poster-sized_portrait_of_Barack_Obama.jpg}{Official presidential transitional photo of then-President-elect Barack Obama}} (\cref{fig:AblationStudy}): \copyright~2008 The Obama-Biden Transition Project, \urlcolor{\href{https://creativecommons.org/licenses/by/3.0/}{CC BY 3.0}}.
\item
\urlcolor{\href{https://commons.wikimedia.org/wiki/File:Beethovensmall.jpg}{Ludwig van Beethoven, oil on canvas}} (\cref{fig:RealResults_beethoven}): 1820 Joseph Karl Stieler, public domain.
\item
\urlcolor{\href{https://commons.wikimedia.org/wiki/File:Grace_Hopper.jpg}{Commodore Grace M. Hopper, {{USNR}} Official portrait photograph}} (\cref{fig:ExamplesPortraitsAnimals}): 1984 Naval History and Heritage Command, public domain.
\item
\urlcolor{\href{https://commons.wikimedia.org/wiki/File:Commander_Eileen_Collins_-_GPN-2000-001177.jpg}{STS-93 Commander Eileen M. Collins}} (\cref{fig:ExamplesPortraitsAnimals}): 1998 NASA, Robert Markowitz, public domain.
\item
\urlcolor{\href{https://commons.wikimedia.org/wiki/File:Hausdorff_1913-1921.jpg}{Felix Hausdorff}} (\cref{fig:ExamplesPortraitsAnimals}): 1913-1921 Universit\"atsbibliothek Bonn, public domain.
\item
\urlcolor{\href{https://share.america.gov/katherine-johnson-celebrates-100th-birthday/}{Katherine G. Johnson}} (\cref{fig:ExamplesPortraitsAnimals}): 2018 NASA, public domain.
\item
\urlcolor{\href{https://commons.wikimedia.org/wiki/File:Ludwig_van_Beethoven.jpeg}{Ludwig van Beethoven}} (\cref{fig:ExamplesPortraitsAnimals}): 1854 Emil Eugen Sachse, public domain.
\item
\urlcolor{\href{https://commons.wikimedia.org/wiki/File:Whoopi_Goldberg_at_a_NYC_No_on_Proposition_8_Rally.jpg}{Whoopi Goldberg in New York City, protesting California Proposition 8}} (\cref{fig:ExamplesPortraitsAnimals}): \copyright~2008 David Shankbone, \urlcolor{\href{https://creativecommons.org/licenses/by/3.0/}{CC BY 3.0}}.
\item
\urlcolor{\href{https://commons.wikimedia.org/wiki/File:Hedy_Lamarr_in_The_Heavenly_Body_1944.jpg}{Hedy Lamarr in ``The Heavenly Body''}} (\cref{fig:ExamplesPortraitsAnimals}): 1944 Employees of MGM, public domain.
\item
\urlcolor{\href{https://commons.wikimedia.org/wiki/File:Alan_Turing_Aged_16.jpg}{Passport photo of Alan Turing at age 16}} (\cref{fig:ExamplesPortraitsAnimals}): 1928-1929 unknown author, public domain.
\item
\urlcolor{\href{https://commons.wikimedia.org/wiki/File:Cute_Piglet.jpg}{Tiny cute piglet looking at the photographer}} (\cref{fig:ExamplesPortraitsAnimals}): 2012 Petr Kratochvil, public domain.
\item
\urlcolor{\href{https://pixabay.com/photos/animal-avian-bird-cold-nature-1867125/}{Penguin}} (\cref{fig:ExamplesPortraitsAnimals}): \copyright~2016 Pexels, \urlcolor{\href{https://pixabay.com/service/license/}{Pixabay license}}.
\item
\urlcolor{\href{https://www.pexels.com/photo/photographer-animal-photography-dog-58997/}{Adult Brown and White Pembroke Welsh Corgi Near the Body of Water}} (\cref{fig:ExamplesPortraitsAnimals}): \copyright~Muhannad Alatawi, \urlcolor{\href{https://www.pexels.com/photo-license/}{Pexels license}}.
\item
\urlcolor{\href{https://commons.wikimedia.org/wiki/File:Audubon-Flamingo.jpg}{Phoenicopterus ruber, the Greater Flamingo}} (\cref{fig:ExamplesPortraitsAnimals}): 1827-1838 John James Audubon, public domain.
\end{itemize}
 \end{document}